\documentclass{cmspaper}
\pdfoutput=1
\usepackage{graphicx}
\usepackage{amsmath}

\begin{document}

\newcommand{\degree}{\ensuremath{^\circ}}
\newcommand{\degC}{\ensuremath{^\circ\,\mathrm{C}}}
\newcommand{\TeVc}{\ensuremath{\mathrm{TeV}/c}}
\newcommand{\GeVc}{\ensuremath{\mathrm{GeV}/c}}
\newcommand{\MeVc}{\ensuremath{\mathrm{MeV}/c}}


\begin{titlepage}


\begin{center}\includegraphics{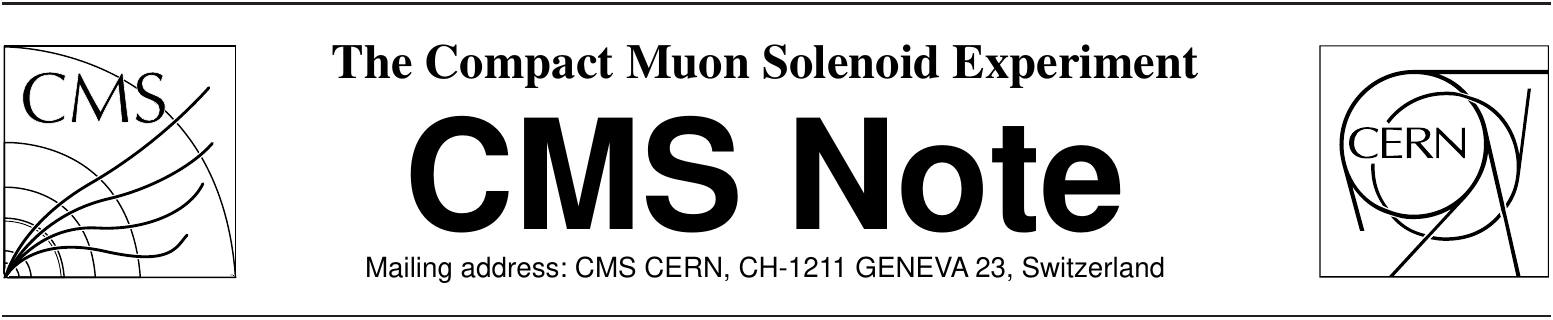}\end{center}\vspace*{0mm}

 \title{Stand-alone Cosmic Muon Reconstruction Before Installation of the CMS Silicon Strip Tracker}





\begin{Authlist}

W.~Adam, T.~Bergauer, M.~Dragicevic, M.~Friedl, R.~Fr\"{u}hwirth, S.~H\"{a}nsel, J.~Hrubec, M.~Krammer, M.Oberegger, M.~Pernicka, S.~Schmid, R.~Stark, H.~Steininger, D.~Uhl, W.~Waltenberger, E.~Widl
\Instfoot{vienna}{Institut f\"{u}r Hochenergiephysik der \"{O}sterreichischen Akademie der Wissenschaften (HEPHY), Vienna, Austria}
% ...
   \Anotfoot{a}{Also at CERN, European Organization for Nuclear Research, Geneva, Switzerland}
   \Anotfoot{b}{Now at University of Hamburg, Institute for Experimental Physics, Hamburg, Germany}
   \Anotfoot{c}{On leave from IFIN-HH, Bucharest, Romania}
   \Anotfoot{d}{Now at LLR-Ecole Polytechnique, France}
   \Anotfoot{e}{Now at Northeastern University, Boston,  USA}
   \Anotfoot{f}{Now at IFIC, Centro mixto U. Valencia/CSIC, Valencia, Spain}
   \Anotfoot{g}{Now at Universiteit Antwerpen, Antwerpen, Belgium}
  \Anotfoot{h}{Now at ETH Zurich, Zurich, Switzerland}
   \Anotfoot{i}{Also Peking University, China}
   \Anotfoot{j}{Now at Imperial College, London, UK}
   \Anotfoot{k}{Now at Universit\'{e} catholique de Louvain, UCL, Louvain-la-Neuve, Belgium}
   \Anotfoot{l}{Now at Eidgen\"{o}ssische Technische Hochschule, Z\"{u}rich, Switzerland}
   \Anotfoot{m}{Now at University of California, Davis, California, USA}
   \Anotfoot{n}{Now at Kansas University, USA}
   \Anotfoot{o}{Also at Research Institute for Particle and Nuclear Physics, Budapest, Hungary}
   \Anotfoot{p}{Now at University of Liverpool, UK}
   \Anotfoot{q}{Now at Massachusetts Institute of Technology, Cambridge, Massachusetts, USA}
   \Anotfoot{r}{Now at Stanford University, Stanford, California, USA}
   \Anotfoot{s}{Now at Harvard University, Cambridge, Massachusetts, USA}  
   \Anotfoot{t}{Now at University of California, Los Angeles, California, USA}


 

\end{Authlist}

  \begin{abstract}
    The subsystems of the CMS silicon strip tracker were integrated and commissioned at the 
Tracker Integration Facility (TIF) in the period from November 2006 to July 2007. 
    As part of the commissioning, large samples of cosmic ray data were recorded under 
various running conditions in the absence of a magnetic field. 
    Cosmic rays detected by scintillation counters were used to trigger the readout of up to 
15\,\% of the final silicon strip detector, and over 4.7~million events were recorded. 
This document describes the cosmic track reconstruction and presents results on the 
performance of track and hit reconstruction as from dedicated analyses.
	\end{abstract}

\end{titlepage}

\setcounter{page}{2}

\section{Introduction}

The CMS tracking system, composed of silicon pixel and micro-strip detectors, is designed to provide a precise and efficient measurement of the trajectories of charged particles emerging from the LHC collisions. With over 70 million electronic channels and an active area of about $\rm 200\,m^2$ it is the largest silicon tracker ever built. 

First experience with tracker operations and track reconstruction was gained during summer 2006, when elements of the silicon strip tracker were operated at room temperature in a comprehensive slice test involving various CMS subdetectors.
The tracker setup was limited and represented only 1\% of the total electronic channels and an active area of $\rm 0.75\,m^2$. Cosmic rays detected in the muon chambers were used to trigger the readout of all CMS subdetectors. 
The CMS superconducting solenoid provided a magnetic field of up to 4\,T. 
Over 25 million events were recorded, and the tracking performance was studied using both a dedicated algorithm for cosmic ray tracking and a general algorithm for track reconstruction in LHC collisions. In addition, tracks reconstructed in the silicon strip tracker were compared with tracks detected by the muon chambers. The results are summarized in Ref.~\cite{MTCCNote}.

In the period from November 2006 to July 2007 the different subsystems of the silicon strip tracker were integrated and commissioned in a large clean room at CERN, the Tracker Integration Facility (TIF). As part of the commissioning large samples of cosmic ray data were recorded under different running conditions. No magnetic field was present, and the tracker setup consisted of up to 15\% of the electronic channels. Over 4.7~million events were taken while operating the detector at five different temperature points. The data were used to verify the reconstruction and calibration algorithms for low- and high-level objects and for comparison with simulated events. The tracking performance was studied using a dedicated cosmic track reconstruction algorithm and two standard tracking algorithms for LHC collisions modified for cosmic ray tracking.

The results from tracker commissioning at the TIF are summarized in three publications. 
The operational aspects, commissioning studies and simulation tuning are described in Ref.~\cite{TIFPerformanceNote}. 
The alignment of the silicon strip tracker using cosmic tracks, survey information and a laser alignment system are described in Ref.~\cite{TIFAlignmentNote}. 
In this paper the results of the track reconstruction are reported.

The setup used for cosmic ray reconstruction is described in Section~\ref{sec:Setup}, followed by an overview of the data sets and the Monte Carlo simulation in Section~\ref{sec:Datasets}. A brief introduction to the local reconstruction is presented in Section~\ref{sec:LocalReconstruction}. The tracking algorithms are summarized in Section~\ref{sec:TrackReconstruction}. Track performance results are presented in Section~\ref{sec:TrackPerformance}. The validation of the track reconstruction is the subject of Section~\ref{sec:TrkValidation}. The hit performance studies related to tracking are described in Section~\ref{sec:HitPerformance} and conclusions are drawn in Section~\ref{sec:Conclusions}.

\section{\label{sec:Setup}Experimental Setup}

\begin{figure}[htb]
	\begin{center}
		\includegraphics[scale=0.9]{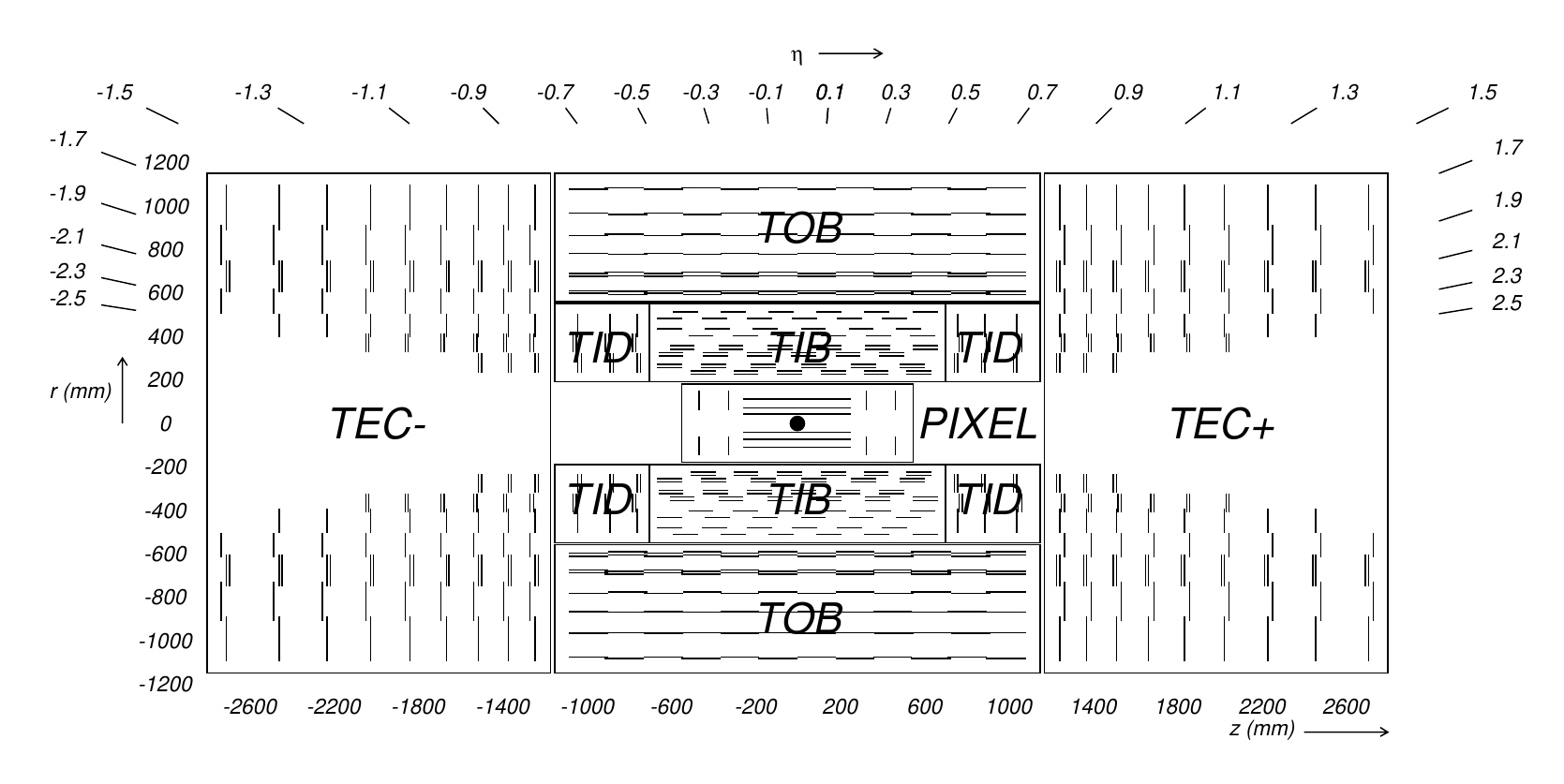}
		\caption{Schematic layout of the CMS tracker. Each line represents a detector module.}
		\label{fig:general_tracker_layout}
	\end{center}
\end{figure}

The CMS tracker occupies a cylindrical volume around the interaction point with a length of 5.8\,m and a diameter of 2.5\,m. The region closest to the interaction point is equipped with a pixel system, while the bulk consists of layers of silicon strip detectors. A schematic overview of the CMS tracker is shown in Figure~\ref{fig:general_tracker_layout}. Throughout this note the standard CMS reference system is used. It has its origin in the center of the detector, with the $z$-axis along the beam line in the anti-clockwise direction for an observer standing in the middle of the LHC ring. The $x$-axis points to the LHC center and the $y$-axis points upward. The azimuthal angle $\phi$ is measured starting from the $x$-axis towards the $y$-axis. The polar radius $r$ is defined as the distance from the $z$ axis in the transverse $x$-$y$ plane.

The pixel detector is the innermost part of the tracking system. Three cylindrical layers of pixel detector modules are complemented by two disks of pixel modules on each side. The strip detector surrounds the pixel detector and is composed of four subsystems. The central region, up to a pseudorapidity of $|\eta|\approx 1$, is covered by the Tracker Inner Barrel (TIB) and the Tracker Outer Barrel (TOB). At each side of the TIB the remaining volume inside the TOB is filled by the Tracker Inner Disks (TID). The silicon strip system is completed by two Tracker End Caps (TEC), extending the acceptance of the tracker up to a pseudorapidity of $|\eta|<2.5$. 

The TIB is composed of four layers using $\rm 320\,\mu m$ thick silicon micro-strip sensors. The strip pitch is $\rm 80\,\mu m$ on layers~1 and~2 and $\rm 120\,\mu m$ on layers~3 and~4. The TID consists of three disks on each side, also employing $\rm 320\,\mu m$ thick silicon micro-strip sensors. Its mean pitch varies between $\rm 100\,\mu m$ and $\rm 141\,\mu m$. The TOB encompasses the TIB/TID and consists of six layers of $\rm 500\,\mu m$ thick sensors with strip pitches of $\rm 183\,\mu m$ on the first four layers and $\rm 122\,\mu m$ on layers 5 and 6. The TEC is composed of nine disks on each side, carrying up to seven rings of silicon micro-strip detectors. The sensor thickness is $\rm 320\,\mu m$ on the inner four rings and $\rm 500\,\mu m$ on the outer three rings, and the mean pitch varies from $\rm 97\,\mu m$ to $\rm 184\,\mu m$.

All silicon strip subsystems are equipped with $r\phi$ modules. These modules have their strips parallel to the beam axis in the barrel and radial on the disks. In addition, the modules in the first two layers and rings of TIB, TID and TOB as well as rings 1, 2 and 5 in the TECs carry a second micro-strip detector module, generally referred to as stereo module. The stereo modules are mounted back-to-back to the $r\phi$ modules with a stereo angle of 100\,mrad, resulting in a measurement of $z$ in the barrel and $r$ on the disks. A detailed description of the CMS tracker can be found in Ref.~\cite{PTDR1}.

\begin{figure}[p]
	\begin{center}
		(a)
		\includegraphics[width=5cm]{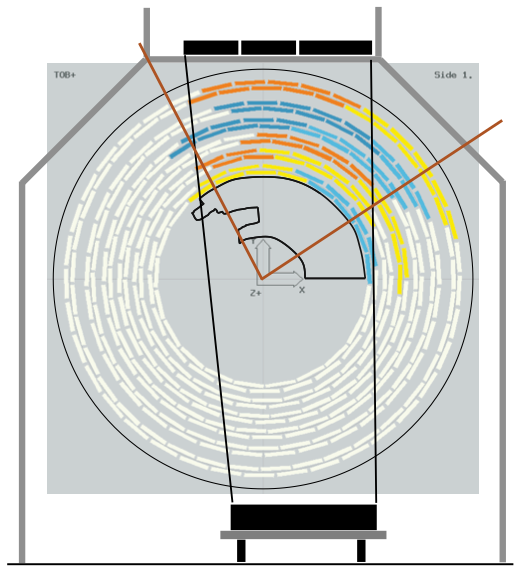}
		\includegraphics[width=9cm]{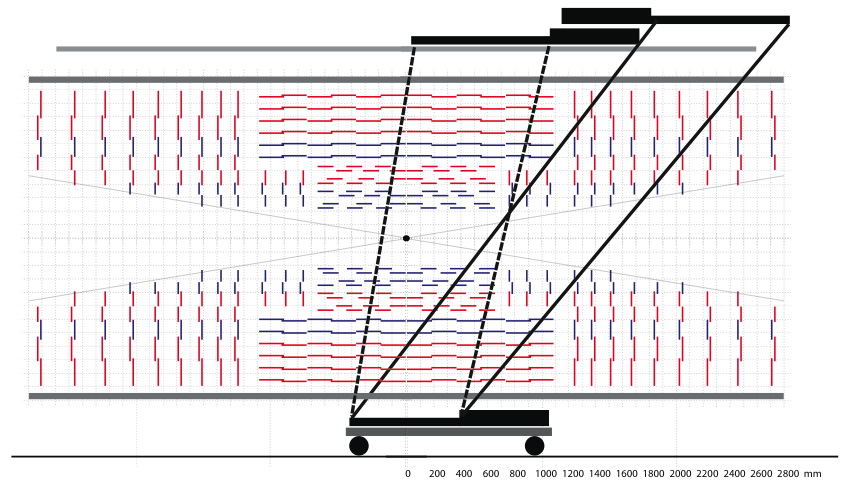} \\
		\vspace{1cm}
		(b)
		\includegraphics[width=5cm]{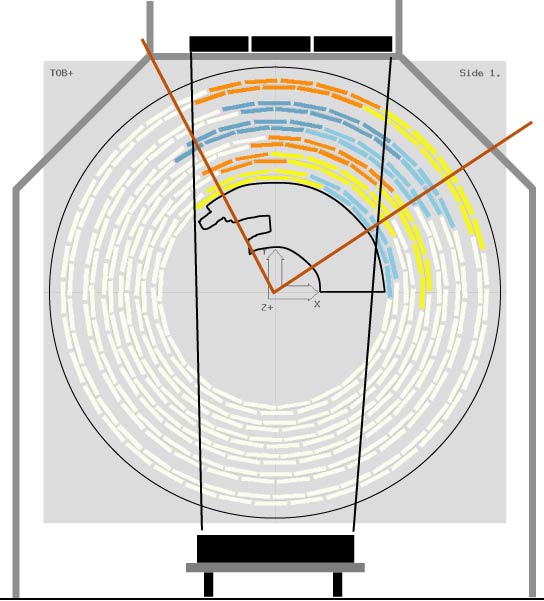}
		\includegraphics[width=9cm]{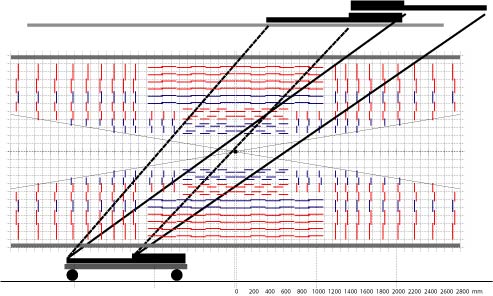} \\
		\vspace{1cm}
		(c)
		\includegraphics[width=5cm]{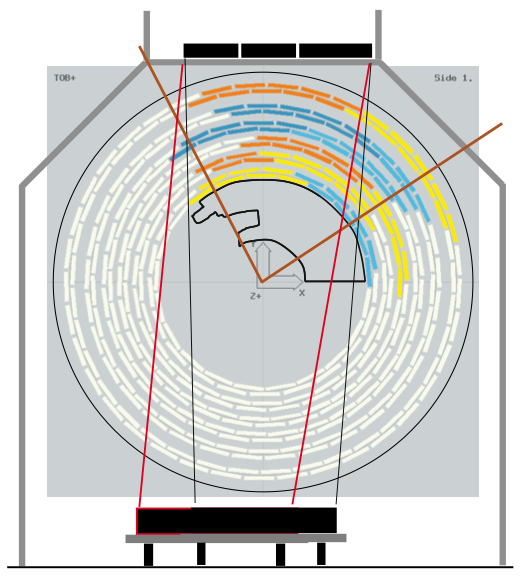}
		\includegraphics[width=9cm]{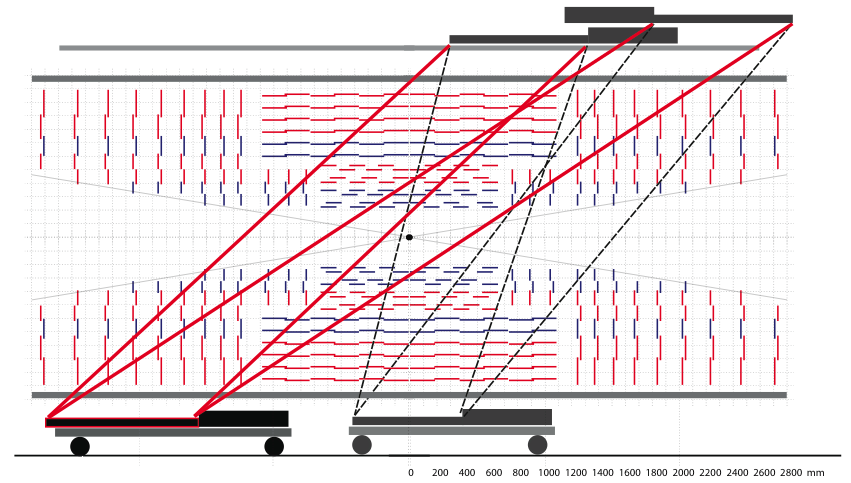}
		\caption{Layout of the various trigger scintillator positions used during the cosmic data taking at the TIF (in chronological order): (a) trigger configuration A; (b) trigger configuration B; (c) trigger configuration C. The $x$-$y$ view is shown on the left side, the $r$-$z$ view is shown on the right. The straight lines connecting the active areas of the top and bottom scintillation counters indicate the acceptance region. In the $x$-$y$~view, the active TOB modules are shown in different shading, the active TIB area is framed in black and the active part of the TEC is indicated by the radial lines.}
		\label{fig:trigger_layouts}
	\end{center}
\end{figure}

Although all components of the silicon strip tracker were commissioned at the TIF, limitations in terms of space and cost and constraints from the data acquisition and cooling systems allowed only a partial readout of the detector. 
The slice of the tracker operated at the TIF represented all four silicon strip subdetectors (TIB, TID, TOB, TEC) containing a total of 2161 modules. The pixel detector was not present. The majority of the readout modules were located in a sector defined as $z>0$ and $y>0$ as shown on the left side of Figure~\ref{fig:trigger_layouts}. With about 1.3~million electronic channels, the tracker setup consisted of nearly 15\,\% of the final silicon strip setup. The active area amounted to $\rm 24.75\,m^2$. The TIF layout is briefly summarized in Table~\ref{tab:geometry}. More information can be found in Ref.~\cite{TIFPerformanceNote}.

Cosmic muon triggering was provided by scintillation counters mounted above and below the tracker. A trigger signal was generated based on the coincidence of any top with any bottom scintillation counter. Data were recorded in various trigger layouts, which are shown in Figure~\ref{fig:trigger_layouts}. Trigger configuration~A was expected to primarily result in TIB+TOB tracks.
Trigger configuration~B was chosen to select tracks passing closer to the nominal interaction point. 
Additional availability of scintillation counters allowed for extended coverage of the trigger system and to combine trigger configurations~A and~B to form configuration~C.

A lead plate with a thickness of 5\,cm was located on top of the lower scintillation counters to avoid triggering on very low momentum tracks. This translates into a minimum cosmic energy of 200\,MeV~for the trigger system.

\begin{table}[b]
	\caption{\label{tab:geometry}Overview of the various silicon strip systems participating in the TIF cosmic data taking.}
	\begin{center}
		\begin{tabular}{|c|c|c|}
			\hline
			Silicon Strip Subdetector    & Number of Modules   & Percentage of Final System \\
			\hline
			TIB                          & 437                 & 16\,\% \\
			TID                          & 204                 & 25\,\% \\
			TOB                          & 720                 & 14\,\% \\
			TEC                          & 800                 & 13\,\% \\
			\hline
		\end{tabular}
	\end{center}
\end{table}
 
\section{\label{sec:Datasets}Data Sets and Monte Carlo Simulation}

\subsection{\label{sec:datasets_datasets}Data Samples}
A total of over 4.7~million cosmic events were recorded during the tracker commissioning at the TIF in the period from March to July~2007. 
The recorded data are grouped into different data sets, defined by active detector, trigger setup and operating temperature. An overview of the nine different data sets is presented in Table~\ref{tab:data_sample}. 

Depending on the configuration of the readout electronics the raw silicon strip detector data was written in one of two formats:

\begin{enumerate}
	\item Zero Suppression: standard operation used for proton-proton collisions (reordering to physical channel order, subtraction of strip pedestals and common-mode noise).
	\item Virgin Raw Data: used for testing, commissioning and calibration (no reordering or pedestal/common-mode subtraction is performed).
\end{enumerate}

Each run was checked using online and offline data quality monitoring tools. If a run did not meet the quality requirements, it was flagged as bad. For example, the detector readout had to be well synchronized with the passage of cosmic muons in order to guarantee an optimal signal-to-noise ratio. In addition, results from global reconstruction were used to flag bad runs. A detailed description of the different good run selection criteria can be found in Ref.~\cite{TIFPerformanceNote}. Only events from good runs are used for offline analyses, reducing the total usable data set from over 4.7~million events to about 4.2~million events.

The complete data set was processed several times using different software releases to integrate improvements in track reconstruction. The results reported in this note were obtained using the two most recent reconstruction passes.

The data sample taken at $-15\degC$ requires specific attention. Due to constraints from the cooling facility the number of powered modules had to be drastically reduced in all silicon strip sub-detectors. Although tracking is possible in this sample, the reconstructed tracks suffer from the limited setup.

\begin{table}[htb]
  \caption{\label{tab:data_sample} Overview of the TIF data samples used for the tracking analysis.}
  \begin{center}
   \begin{tabular}{|c|c|c|c|c|c|c|}
     \hline
 Active Detector  & Trigger Configuration  &  T [\degC]  & Total Events  & Good Events \tabularnewline
	\hline
 TIB+TID+TOB          & A                      & $+15$              & 703\,996       & 665\,409 \\
 TIB+TID+TOB+TEC      & A                      & $+15$              & 191\,154       & 189\,925 \\
 TIB+TID+TOB+TEC      & B                      & $+15$              & 193\,337       & 177\,768 \\
 TIB+TID+TOB+TEC      & C                      & $+15$              & 244\,450      & 241\,512 \\
 TIB+TID+TOB+TEC     & C                      & $+10$              & 992\,997       & 534\,759 \\
 TIB+TID+TOB+TEC     & C                      & $-1$              & 893\,474       & 886\,801 \\
 TIB+TID+TOB+TEC     & C                      & $-10$             & 923\,571       & 902\,881 \\
 TIB+TID+TOB+TEC     & C                      & $-15$             & 656\,923       & 655\,301 \\
 TIB+TID+TOB+TEC     & C                      & $+15$              & 112\,139       & 112\,134 \\
     \hline
    \end{tabular}
  \end{center}
\end{table}

\subsection{\label{sec:datasets_simulation}Simulation}
The simulation of a realistic cosmic muon spectrum relies on dedicated parametrizations of the energy dependence and incident angle, also accounting for the correlations between the two quantities. For comparisons with TIF data, cosmic muons have been generated starting from an ideal cylinder coincident with the CMS surface using the CMSCGEN generator~\cite{CMSCGEN}. The parametrizations have been adopted from L3CGEN~\cite{L3CGEN}, a cosmic muon generator developed for the L3+Cosmics experiment.
The original parametrization of the spectrum is based on a polynomial fitted to data above an energy of 10\,GeV and cannot be used at much lower energies.
In order to cover the range from 200\,MeV to 2\,GeV the cosmic muon spectrum is produced assuming a simple energy dependence in CMSCGEN and reweighted using the CAPRICE~\cite{CAPRICE} energy spectrum. 
Cosmic muons were generated with angles up to $88\deg$ from the vertical axis and in a time window of 25\,ns.
Since the tracker is on the surface without any iron shielding from muon stations outside, the energy loss scale factor was set equal to zero.

Before the detector and electronics response are simulated, a special filter is applied reproducing the trigger setup. The scintillation counters are modeled as virtual $1\times 1$\,m$^2$ surfaces, and the muon trajectories are extrapolated to the outside of the tracker where the intersection points with the scintillator surface are calculated. Using the intersection points, the trigger logic is applied and the cosmic muon event is either retained or discarded. Simulated events are generated separately for all three trigger layouts. 
Further details about the simulation as well as angular and momentum distributions can be found in Ref.~\cite{TIFPerformanceNote}.

\section{\label{sec:LocalReconstruction}Local Reconstruction}

The local reconstruction is done in two consecutive stages, transforming the digitized information from the silicon strips into reconstructed hits in the local coordinate of the silicon sensors. The resulting reconstructed hits are input to the various track reconstruction algorithms which are described in Section~\ref{sec:TrackReconstruction}.

The raw data coming from the readout electronics of the silicon strip detectors are unpacked and grouped according to detector modules. After the unpacking step, the raw data are commonly referred to as {\it digis}, which denotes pedestal-subtracted and zero suppressed ADC counts for individual strips. The digis are associated with a {\it detector ID}, a unique number which encodes the location of each module in the mechanical structure of the CMS tracker. Using the digis, the local reconstruction is performed in stages: cluster reconstruction and hit conversion.

\begin{enumerate}
	\item Cluster reconstruction\\
	The cluster reconstruction groups adjacent strips whose associated charge pass a set of thresholds. The thresholds depend on the noise levels characterizing the strips of the cluster. Clusters are reconstructed by searching for a seed strip with a signal-to-noise ratio (S/N) greater than~3. Neighboring strips are attached to the cluster if their signal-to-noise ratio exceeds~2. The total signal size of the cluster must exceed five times  the quadratic sum of the individual strip noises. 
The cluster reconstruction algorithm is referred to as {\it 3-Threshold} algorithm.
The signal of a cluster is based on the sum of the ADC counts of all associated strips. 
In the most recent reconstruction pass a correction for the variations in the electronic gain was applied.
These corrections had been derived from the height of digital synchronization signals \cite{TIFPerformanceNote}.

	\item Hit conversion\\
	The hit conversion associates every cluster with a hit position and corresponding errors. The hit position is determined from the centroid of the signal heights. The position resolution is parametrized as a quadratic function  of the projected track width on the sensor in the plane perpendicular to the strips~\cite{PTDR1}. Deviations from the ideal geometry (``misalignment'') are taken into account by adding an additional uncertainty on the module positions (Alignment Position Error, APE) to the hit errors. The size of the APEs was estimated from survey data~\cite{TIFAlignmentNote}. 
For the most recent reconstruction pass a first set of alignment
constants was available, and the APEs were reduced to 
about 150\,$\mu m$ in the TOB and 500 -- 600\,$\mu m$ in the other subdetectors.
\end{enumerate}

Details on the performance of the local reconstruction can be found in Ref.~\cite{TIFPerformanceNote}.

\section{\label{sec:TrackReconstruction}Track Reconstruction}

Three tracking algorithms were applied to TIF data: the two standard algorithms designed for the reconstruction of proton-proton collisions (``Combinatorial Kalman Filter'' and ``Road Search'') and one specialized algorithm for the reconstruction of single track cosmic events (``Cosmic Track Finder'').  
They use the hits described in Section~\ref{sec:LocalReconstruction}. The position estimates may depend on the local track angles. In addition, a reconstruction geometry describing the location of the modules and the distribution of passive material and condition information about the status of the different modules are needed. 

All three algorithms decompose the task of track reconstruction into three stages:
\begin{enumerate}
\item seed finding, which provides a selection of initial hits and a first estimate of parameters,
\item pattern recognition, which associates hits to a track, and
\item track fitting, which determines the best estimate of the track parameters.
\end{enumerate}

The first two items are specific to each of the algorithms while the track fit is always performed by a Kalman filter and smoother. All these software modules use some common services. In the absence of a magnetic field the tracks are extrapolated as straight lines. Material effects -- energy loss and multiple Coulomb scattering -- are estimated each time a track crosses a detector layer. The amount of material at normal incidence is obtained via the reconstruction geometry, and the same constants as for the reconstruction of proton-proton collisions are used. Since the momentum is not measured at the TIF, a constant value of $1\,\GeVc$ is assigned, close to the expected average momentum of cosmic muons.

The three track reconstruction modules have been designed (Cosmic Track Finder) or configured (Combinatorial Kalman Filter, Road Search) for the reconstruction of single track cosmic events.
They were not optimized for the reconstruction of cosmic showers.
Consequently, large multiplicity events with more than 300 reconstructed clusters were excluded from the track reconstruction.
The algorithms and the fitting procedure are described in the following subsections.

\subsection{\label{sec:CKF}Combinatorial Kalman Filter}

The Combinatorial Kalman Filter (CKF) uses the capacity of the Kalman Filter \cite{KF} for simultaneous pattern recognition and track fitting. 
Starting from an initial estimate of the track parameters the algorithm iterates through the layers of the tracker and builds a combinatorial tree of track candidates.
The CKF is identical to the one designed for proton-proton collisions except for the seed finding stage, which has been adapted for the reconstruction of cosmic tracks.

\subsubsection{Seed Finding}
In the standard tracking, {\it i.e.}, for particles coming from the interaction point, seeds are created in the innermost layers of the tracking system.
A seed is made out of a hit pair and a loose beamspot constraint or out of a hit triplet. 
The starting parameters of the trajectory are calculated from a helix passing through the three points. The selected hits must be pointing towards the interaction point and a minimum transverse momentum cut is applied. 

The situation for cosmic track reconstruction is very different with respect to proton-proton collisions, in particular:

\begin{itemize}
\item No vertex constraint can be applied, since the cosmics do not necessarily cross the tracker pointing towards the interaction point.
\item Seeds should be created also in the outer layers, because these layers have a higher acceptance for cosmic tracks.
\end{itemize}

Hence, the seed finding algorithm had to be modified to handle cosmic track reconstruction. In addition, the modified version can be used for tracking of beam halo muons. For the reconstruction of TIF data, hit triplets are used in the inner or outer parts of the barrel and hit pairs in the endcaps. Hit triplets are checked for compatibility with a straight line: the radius of the circle passing through the three hits has to exceed 5\,m. The different combinations of layers used for seeding are summarized in Table~\ref{tab:CKFseeds}.

\begin{table}[hbtp]
	\caption{\label{tab:CKFseeds}Combination of layers used for seed finding in the CKF algorithm.}
	\begin{center}
		\begin{tabular}{|l|c|}
			\hline
			Seeds & Layers \\ \hline
			Inner barrel triplets & TIB1+TIB2+TIB3 \\
			Outer barrel triplets & TOB4+TOB5+TOB6, TOB3+TOB5+TOB6, TOB3+TOB4+TOB5,  \\
			& TOB3+TOB4+TOB6, TOB2+TOB4+TOB5, TOB2+TOB4+TOB6 \\
			Endcap pairs & Any pair of adjacent TEC wheels \\ \hline
		\end{tabular}
	\end{center}
\end{table}

\subsubsection{Pattern Recognition}
The computationally most time consuming part of track reconstruction is the pattern recognition, that is, the building of a candidate trajectory by selecting its hits out of all the hits in the event. 

\begin{figure}[!ht]
        \centering\includegraphics[width=0.5\textwidth]{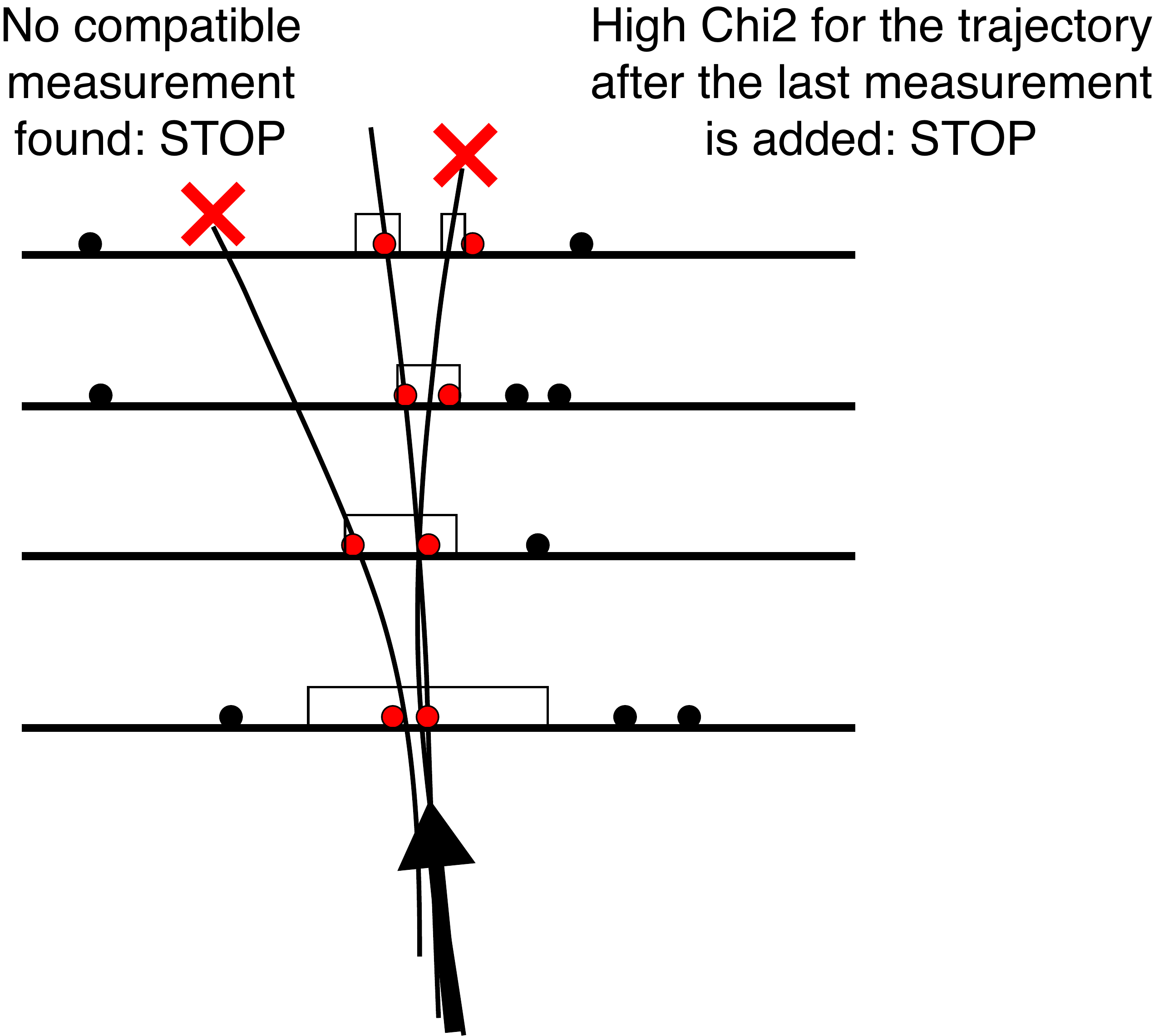}
        \caption{Schematic representation of the Kalman Filter based pattern recognition. 
The points represent hits, the curved lines track candidates and the shaded boxes the size of the search window.\label{fig:patternreco}}
\end{figure}

From each seed a propagation to the next surface is attempted. Hits are identified in a window whose width is related to the precision of the track parameters. If a hit is found within the window, it is added to the candidate trajectory and the track parameters are updated. The way in which the trajectory parameters are updated is described in some detail in Section~\ref{sec:fitting}. If several compatible hits are found, a new candidate is created for each of them. In addition, one candidate is created without adding any hit information (``lost hit''). Candidates are sorted according to quality (based on the $\chi^2$ and the number of hits) and the five best ones are retained for further propagation. As hits are added to the candidate trajectory the knowledge of the track parameters improves and the sizes of the search windows decrease. Propagation of a candidate ends if configurable cuts on the number of layers or the number of consecutive layers without a hit are exceeded. The way the Kalman filter based CKF works is visualized in Figure~\ref{fig:patternreco}.

Even if cosmic events are mainly single track events, the pattern recognition is quite difficult. The tracker is designed to be hermetic and to offer optimal module overlaps for tracks coming from the interaction point. A cosmic ray track can encounter holes as well as zones of high module overlaps. 

The cosmic ray track reconstruction has been particularly useful to test and validate the CKF trajectory builder, specifically because it was possible to test its ability to find hits on the overlap regions. When two overlapping modules are crossed, it is very important to associate both hits to the candidate trajectory to give a stronger constraint on the track parameters. Cosmics represent a suitable situation to test this, since the fraction of overlaps is higher than for tracks coming from the interaction point.

\subsection{\label{sec:CosmicTF}Cosmic Track Finder}
The Cosmic Track Finder is designed as a simple and robust algorithm, tailored to the specific task of reconstructing single tracks without imposing a region of origin, but assuming a preferred direction. It has been applied with success in the CMS magnet test~\cite{MTCCNote}. 

\subsubsection{Seed Finding}

For cosmic track reconstruction in TIF data, a dedicated seeding is used. The total number of hits in the whole tracking volume is expected to be several orders of magnitude lower than in proton-proton collisions. Hence, all the hit pairs which are geometrically compatible are considered as potential seeds. The definition of compatibility is the following: 
\begin{itemize}
\item The seed is built from a pair of hits lying in different layers (barrel) or wheels (endcaps). In the barrel, all the hit pair combinations of either the three outermost TOB layers or of the three innermost TIB layers are considered. In the endcap, hit pair combinations are accepted if the hits are separated by at most one intermediate wheel.
\item The distance between the modules along the $z$-axis is less than 30\,cm. This cut is not applied for seeds in TEC layers.
\item The distance between the hits in global $x$ is less than two times the distance in global $y$. This requirement is motivated by the small angles cosmic tracks are expected to have with respect to the vertical direction. Again, this cut is not applied for seeds in TEC layers.
\end{itemize}

The maximum number of seeds is limited to 30.

\subsubsection{Pattern Recognition}
All seeds that fulfill the previous selection criteria are considered in the pattern recognition.
For each seed, preliminary track parameters are calculated based on the line connecting the two hits. The hit-selecting algorithm is simplified with respect to the Combinatorial Kalman Filter. A seed, which comes from the previous step, can be at the top or bottom of the instrumented region of the tracker. If it is at the top (bottom), all the hits with a global $y$ coordinate lower (higher) than the hit of the seed are sorted in decreasing (increasing) order with respect to the global $y$ axis. 
A very simple procedure establishes if the hit can be selected or not:
\begin{enumerate}
\item The trajectory is propagated to the surface of the module which provided the hit.
The uncertainty from multiple scattering is considered when the track is propagated. 
\item The compatibility of the hit with the propagated trajectory is evaluated using a $\chi^2$ estimator. 
  For this analysis a cut at $\chi^2 < 40$ was chosen. 
\item If the hit is compatible, the trajectory is updated with the hit. 
\end{enumerate}
A trajectory candidate is retained if it contains at least 4 hits.
The final selection is only done after the full track fit. 
 
The fitting procedure is the same as for all the other CMS tracking algorithms and is based on the Kalman Filter. At the end of this phase several trajectories are still valid, but only one is retained since only one track per event is expected. The best trajectory is selected on the basis of the criteria below, listed in order of precedence:
\begin{enumerate}
\item the highest number of layers with hits in the trajectory,
\item the highest number of hits in the trajectory,
\item the lowest value of the $\chi^2$.
\end{enumerate}

\subsection{\label{sec:RS}Road Search}

The Road Search (RS) algorithm treats the CMS tracker in terms of {\it rings}, where a ring contains all tracker modules at a given $r$-$z$ position, spanning 360$^{\circ}$ in $\phi$.  A track will be a line in $r$-$z$, and the Road Search uses pre-defined sets of rings consistent with a line in $r$-$z$ in which it will search for a track. These pre-defined sets of rings are referred to as {\it roads}. The standard Road Search algorithm, designed for use in proton-proton collisions, had to be slightly modified in order to reconstruct cosmic muons tracks. These modifications are described in the following.

\subsubsection{Seed Finding}
For seed finding, the Road Search algorithm uses pairs of hits in seed rings. The CMS tracking system uses three different types of hits (depending on silicon module type and arrangement), and all can be used for seeding: $r\phi$, stereo, matched. While $r\phi$ and stereo hits are located on the respective modules (see Section~\ref{sec:Setup}), matched hits are virtual hits combining a compatible $r\phi$ and stereo hit in the same layer into a precise 3-dimensional measurement. All three types of hits can be used in different variations for seeding. The set of rings that composes the road will be those consistent with the linear extrapolation between the seed rings in the $r$-$z$ plane.  The seed is composed of a pair of hits in the seed rings within a maximum $\Delta \phi$, which effectively translates to a cut on the minimum transverse momentum of the track.

\begin{figure}[htb]
	\begin{center}
		\hspace*{1cm}
		\includegraphics[height=10.8cm,width=16.5cm]{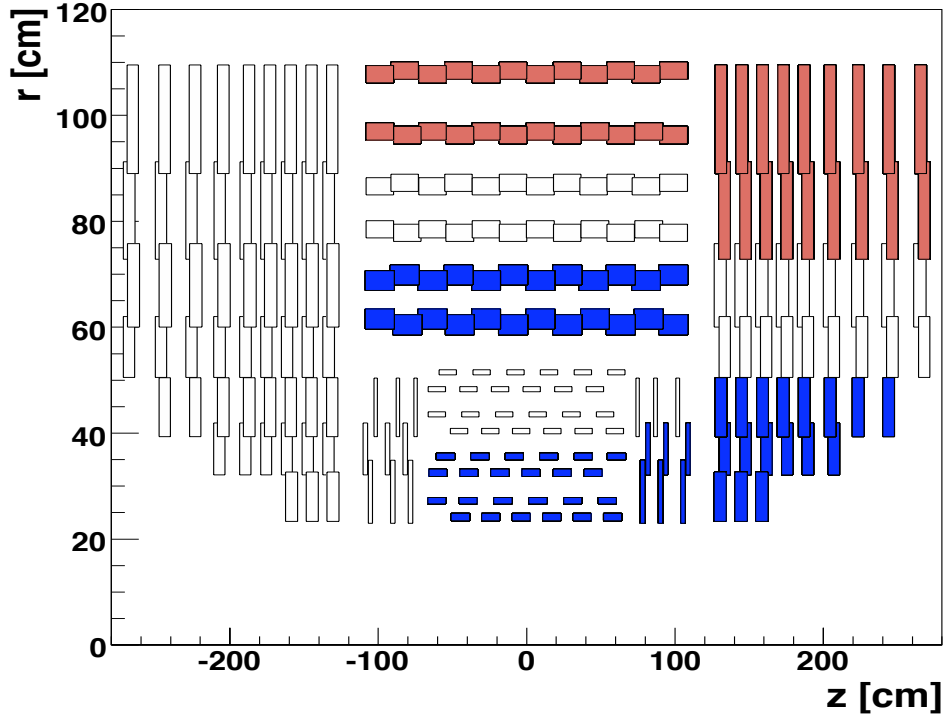}
		\caption{Overview of the ring structure of the CMS tracker. The rectangles in the plot represent the rings as defined in the Road Search algorithm. The inner seed rings are shown in dark grey (blue), the outer seed rings are shown in light grey (red). The seeding is asymmetric since modules located at $-z$ were not part of the cosmic data taking.}
		\label{fig:rs_roads_example}
	\end{center}
\end{figure}

In the standard Road Search algorithm, roads are constrained to point back to the luminous region of the beam, corresponding to roughly $|z| < 15\,\rm cm$. Cosmic rays will not point back to $z=0$, so for cosmic track reconstruction specific roads were generated, where the constraint on the extrapolation of the roads was loosened to include any pair of seed rings within the acceptance of the readout detector. An overview of the inner and outer seed rings for the TIF geometry is shown in Figure~\ref{fig:rs_roads_example}. 
The choice to use the inner layers of both TIB and TOB as inner seed rings (in addition to the inner rings of TID and TEC) was made to avoid any geometrical acceptance loss and to have one seed ring structure which fits different possible sub-detector readout combinations.

\begin{figure}[htb]
	\begin{center}
		\includegraphics[scale=0.30]{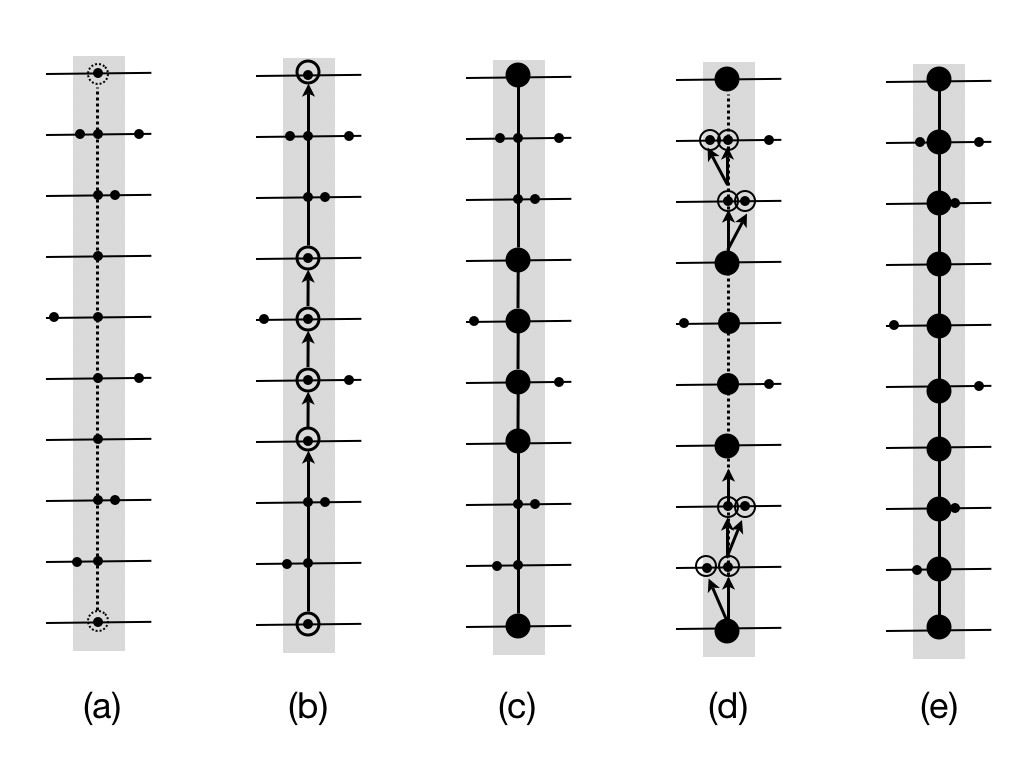}
		\caption{Schematic representation of the Road Search algorithm. In (a) a trajectory is drawn through the two circled seed hits. All hits within a window around the trajectory (shaded region) are collected in the cloud. In (b) a new trajectory is built inside-out using only hits on the low occupancy layers of the cloud, resulting in the fitted trajectory in (c). This trajectory is extrapolated back to the innermost layer and then hits on the higher occupancy layers are tested (d). The best hit on each layer is used to yield the fitted track (e).}
		\label{fig:rs_sketch}
	\end{center}
\end{figure}

\subsubsection{Pattern Recognition}
In the first part of the  pattern recognition step, an expected trajectory is determined using the two seed hits and - for the reconstruction of proton-proton collisions - the beamspot.  The trajectory is extrapolated through the other rings of the road and hits are collected inside a narrow window around the expected trajectory.  This collection of hits is referred to as a {\it cloud}. The cloud should contain all the hits of a track, along with other hits that happen to overlap and lie close to the track. In the second part of pattern recognition, the cloud is turned into a trajectory. A trajectory is first built in low occupancy layers, extrapolating inside-out. With the trajectory well-defined from the low occupancy layers, hits from the higher occupancy layers are added to the trajectory. The final track will contain at most one hit per detector module, though potentially more than one hit per layer due to detector overlaps. A schematic presentation of the pattern recognition is shown in Figure~\ref{fig:rs_sketch}.

For cosmic track reconstruction, the following modifications have been implemented: the sorting of the hits within a cloud was changed from an inside-out sorting to a sorting in $y$. In addition, all hits are compared with the expected trajectory in $r-\phi$, even for hits in the TID and TEC where the expected trajectory in $z-\phi$ is usually used.

\subsection{\label{sec:fitting}Track Fitting}

The Kalman Filter is a ``dynamic'' Least Squares Method \cite{KF}. At the intersection with a detector surface the trajectory is described by a 5-dimensional state vector
\begin{equation}
        \vec{p}=(\frac{q}{p}, \frac{p_x}{p_z}, \frac{p_y}{p_z}, x, y)
\end{equation}
whose components are the inverse signed momentum and the angles and positions in two orthogonal directions in the local coordinate system. In this system, the normal to the sensor plane defines the $z$-axis while the $y$-axis is aligned with the direction of the strips (the average direction in case of trapezoidal modules). In the absence of a magnetic field the momentum is not measured:
$\frac{q}{p}$ is set to an initial value and is not affected by the fitting procedure.

The Kalman filter proceeds in an iterative way through the list of hits established by the pattern recognition. The state vector is initialized with an estimate provided by the pattern recognition. The errors are set to high values in order to avoid any bias. Then the following three steps are repeated for each hit in the list:
\begin{enumerate}
\item The state vector and the corresponding covariance matrix are propagated to the reference surface of the hit, yielding a prediction for the state at the module.
\item The hit information is combined with the prediction to form an
  ``updated state''.
\item The $\chi^2$ of the track is increased based on the compatibility between the predicted state and the hit.
\end{enumerate}
After the last hit the full information has been included and the best estimate of the track parameters is available at the end of the trajectory.

In order to achieve maximum precision on all the surfaces (particularly the first one), a second fit is run in the opposite direction. The updated states of this backward fit are combined with the predicted states of the forward fit to obtain the final estimates (``smoothing''). The results of the smoothing can also be used to calculate the compatibility between a hit and the combination of forward and backward predicted states, {\it i.e.}, the full information provided by all other hits.

\section{\label{sec:TrackPerformance}Tracking Performance}

\subsection{Track Reconstruction Results}
The cosmic run data have been split into nine different data samples (see Section~\ref{sec:datasets_datasets}). The number of reconstructed single track events for all three track reconstruction algorithms -- without applying any track quality cuts -- is presented in Table~\ref{tab:reco_trks}. In contrast to the Cosmic Track Finder, both the CKF and the Road Search algorithm are able to reconstruct more than one track per event. Hence, it is expected that the number of single track events will be higher for the Cosmic Track Finder in comparison to the other two algorithms. Events with multiple tracks are characterized by a larger amount of wrongly reconstructed tracks and require special dedicated studies. 

\begin{figure}[p]
  \begin{center}
    \includegraphics[width=.45\textwidth]{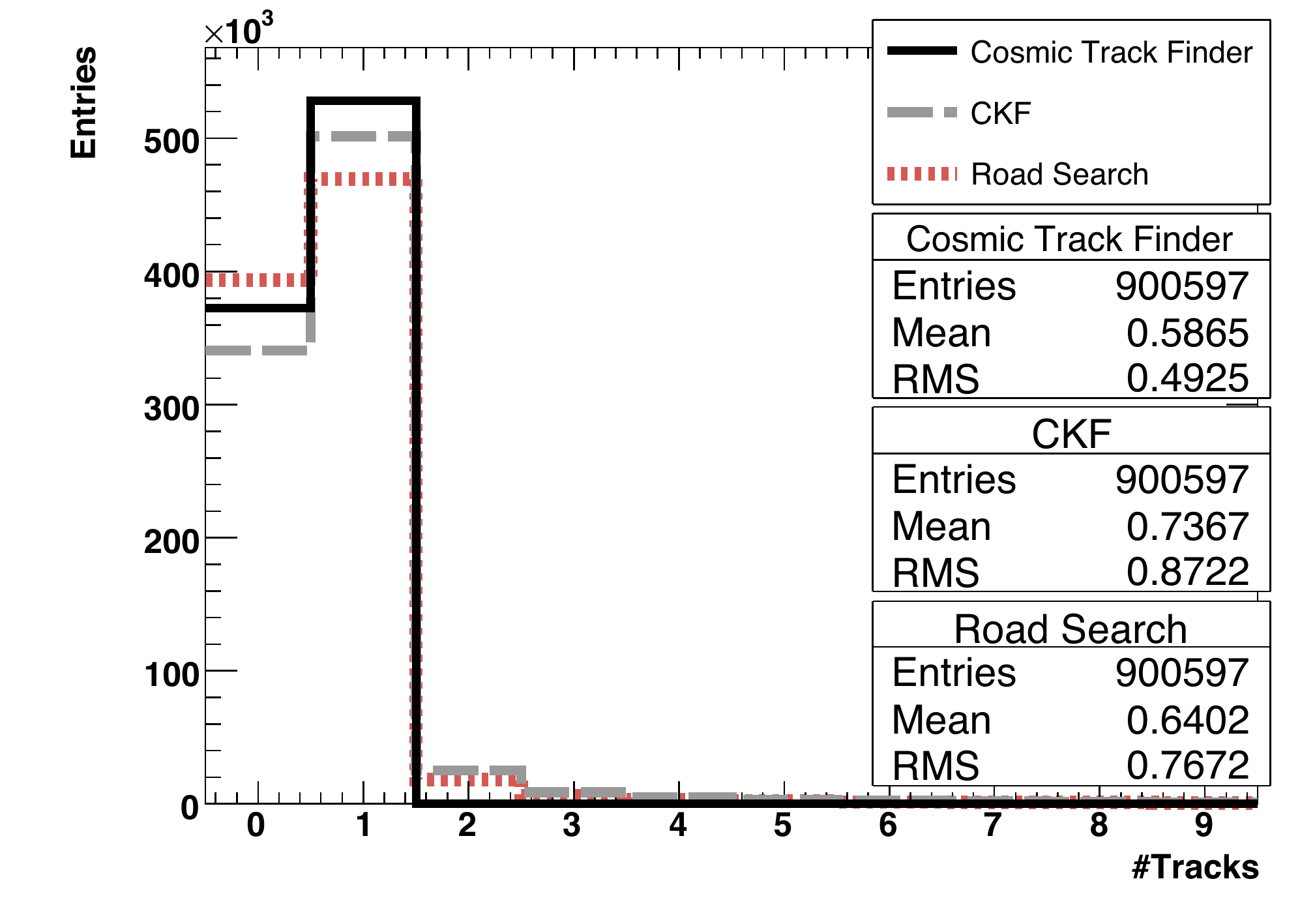}
    \includegraphics[width=.45\textwidth]{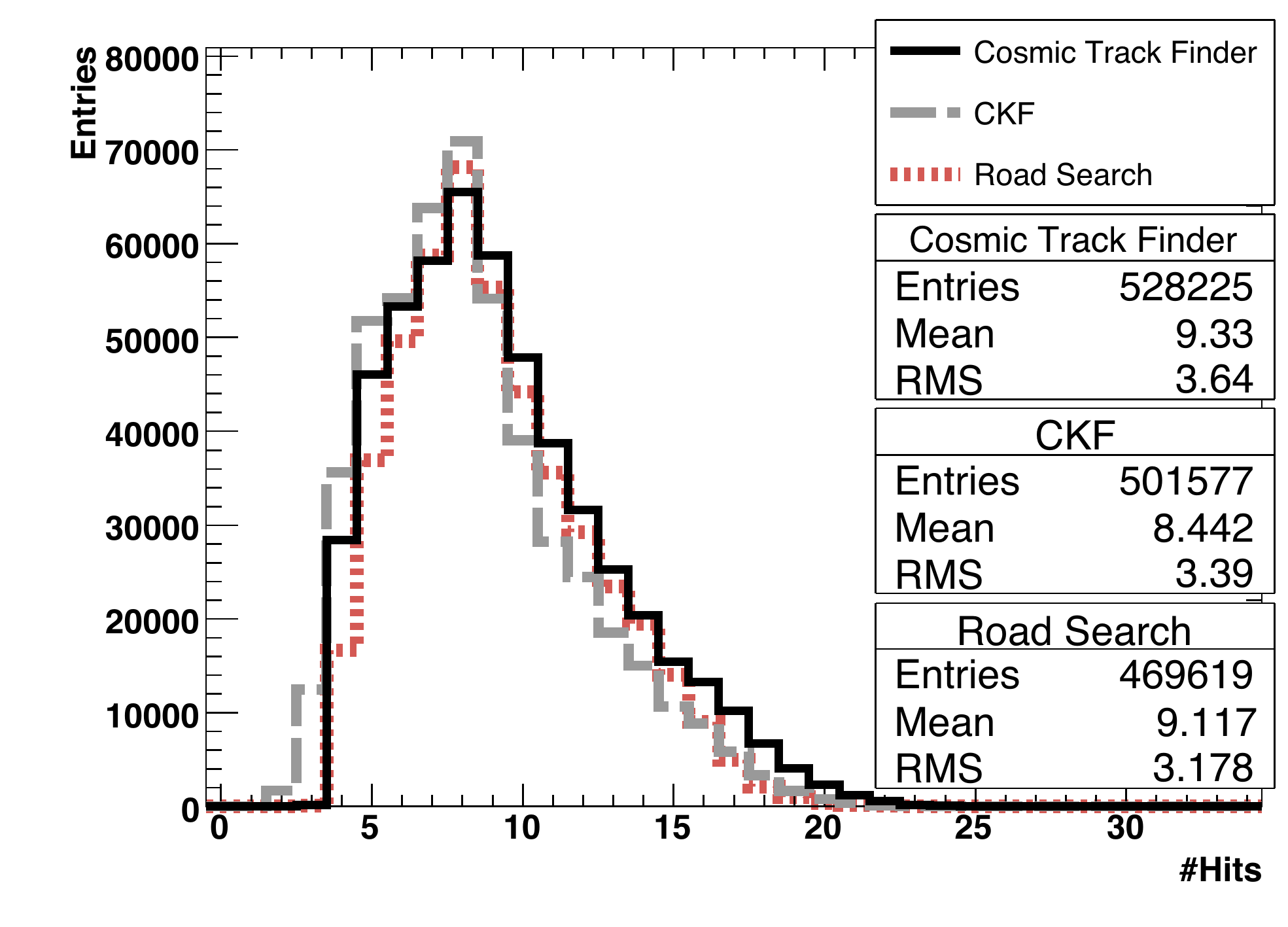}\\
    \vspace{1cm}
    \includegraphics[width=.45\textwidth]{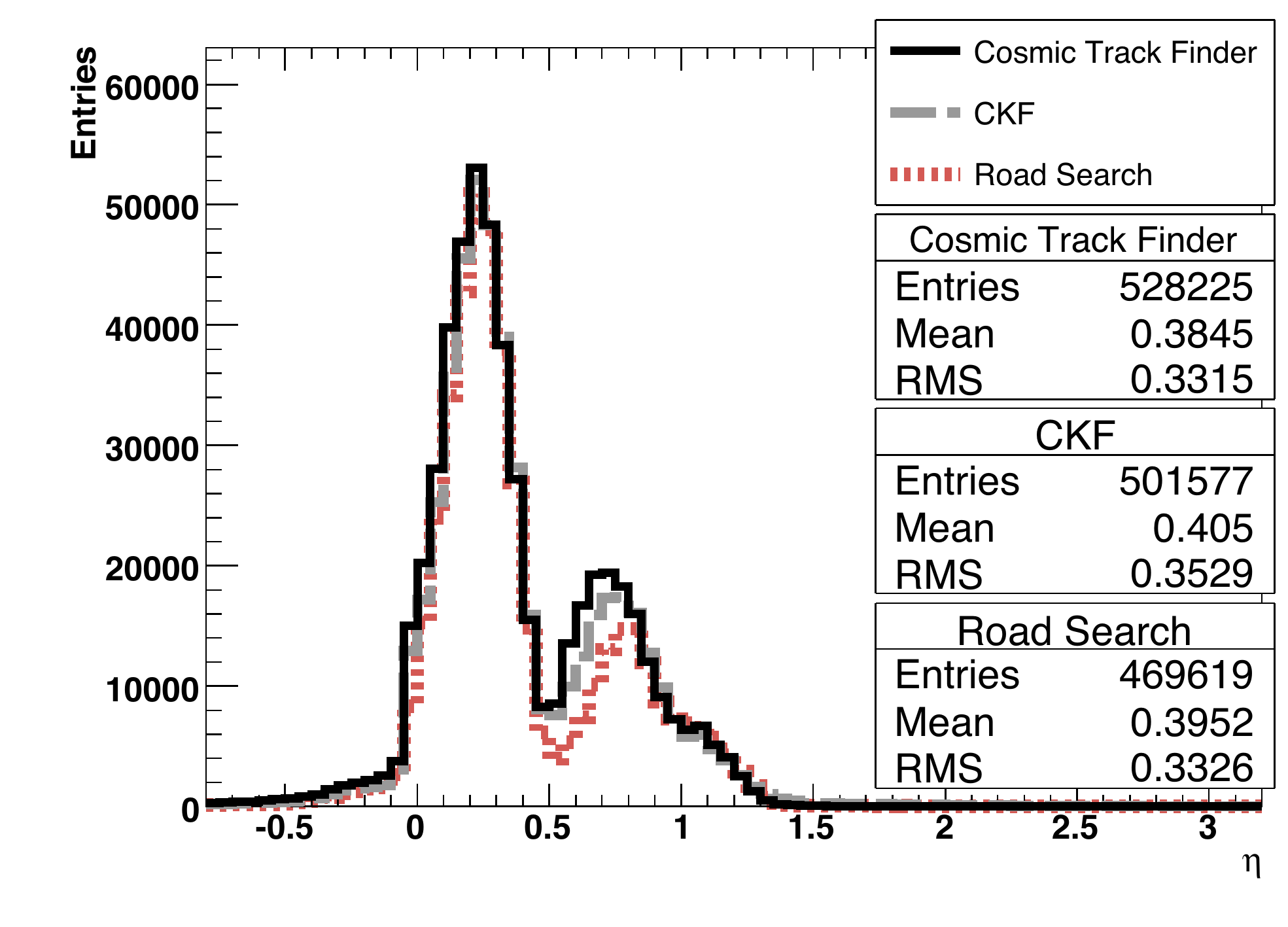}
    \includegraphics[width=.45\textwidth]{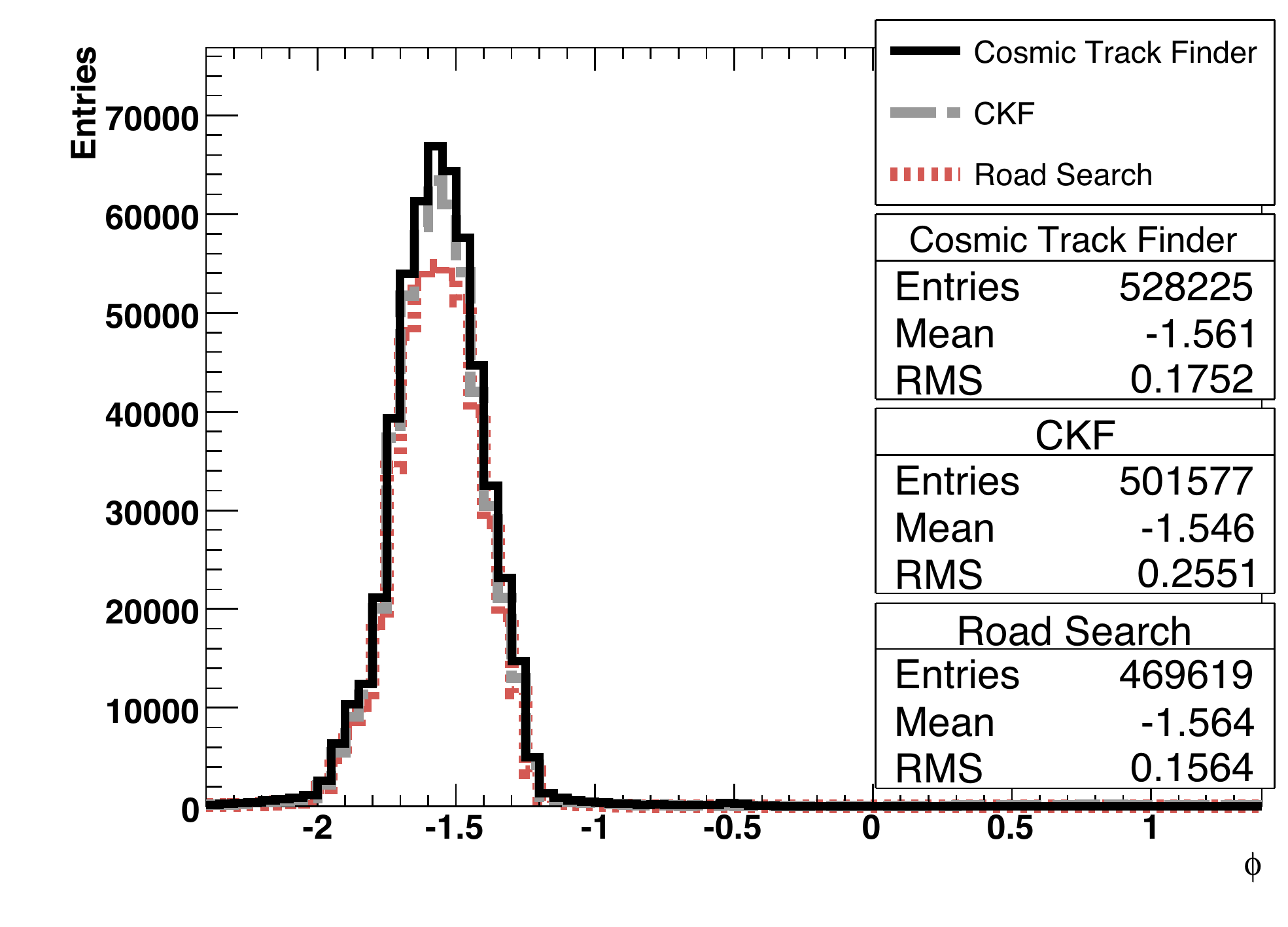}\\
    \vspace{1cm}
    \includegraphics[width=.45\textwidth]{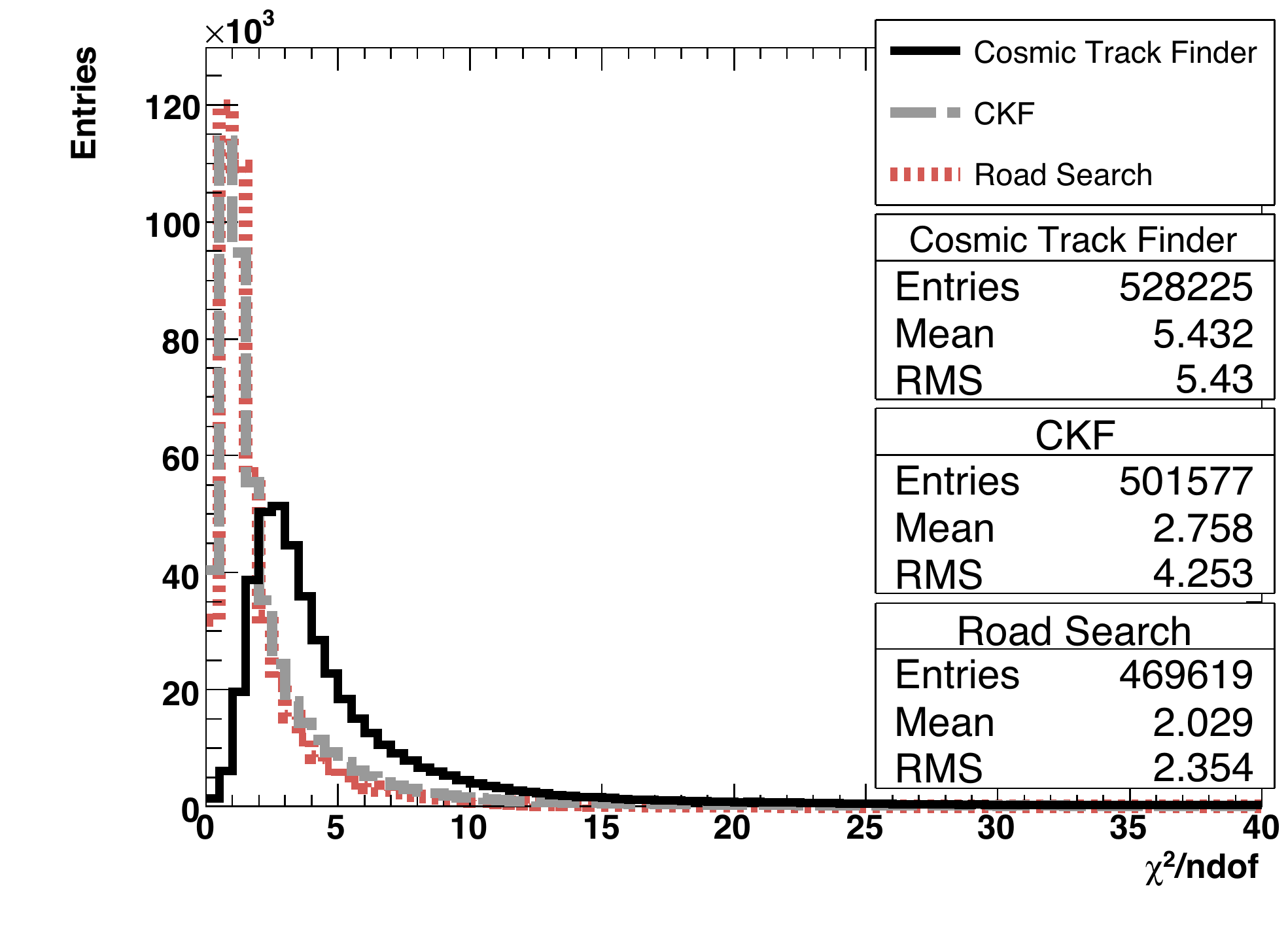}
    \includegraphics[width=.45\textwidth]{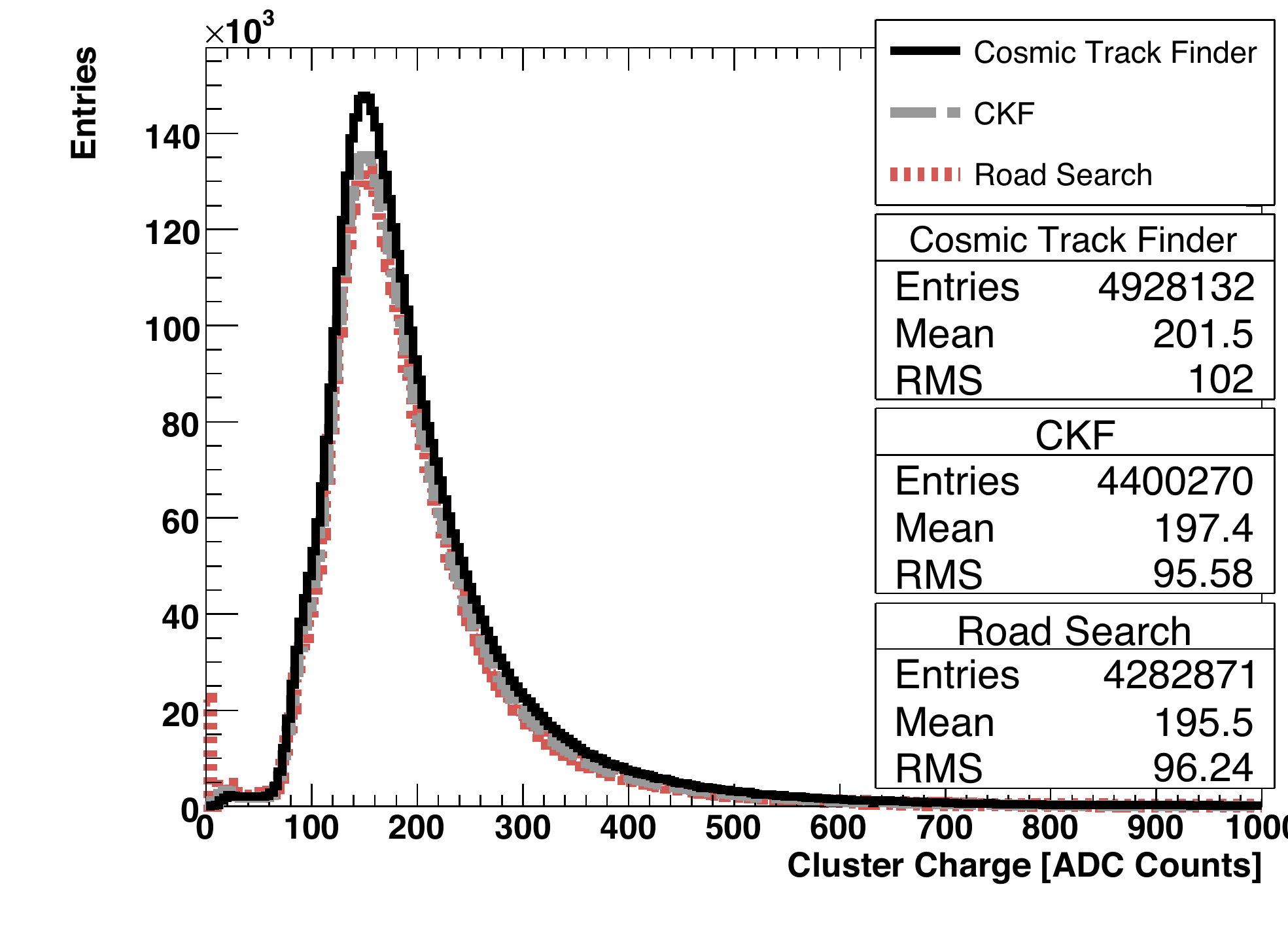}
		\caption{Number of reconstructed tracks and various track distributions in single track events using data taken at $\mathrm{T}=-10\degC$ in trigger configuration C with TIB+TOB+TEC as active detector. The results of the Cosmic Track Finder are shown as solid line, while results from the CKF are shown as a dashed line and results from the Road Search algorithm are shown as a dotted line.}
		\label{fig:trk_results_posC}
	\end{center}
\end{figure}

To visualize the standard track reconstruction results, the data sample taken at $-10\degC$ in trigger configuration~C using all silicon strip sub-detectors has been chosen. The number of reconstructed tracks for all three tracking algorithms is shown in Figure~\ref{fig:trk_results_posC}, along with various track distributions in single track events. Apart from the different numbers of reconstructed tracks, all three tracking algorithms lead to similar results. The $\phi$ distribution shows a peak around $-\pi/2$, compatible with tracks that originate from the top of the detector and travel outside in. The $\eta$ distribution is compatible with the trigger layout. 
The number of $r\phi$ and stereo hits per track shows a most probable value of 8~hits for all algorithms.
The cluster charge distribution shows that no algorithm includes a significant number of noise hits.

\begin{table}[htb]
\footnotesize
  \caption{\label{tab:reco_trks} Number of reconstructed single track events for all three tracking algorithms in the different data samples. The significant change observed for data taken at $-15\degC$ is due to the smaller number of powered modules (see section \ref{sec:TrackRecoStability}).}
  \begin{center}
   \begin{tabular}{|c|c|c|c|c|c|c|c|}
     \hline
	Active Detector  & Trigger  &  T [\degC]  & \multicolumn{3}{c|}{Number of Single Track Events} \tabularnewline
	& Configuration & & Cosmic Track Finder & CKF & Road Search \tabularnewline
	\hline
   TIB+TID+TOB         & A        & $+15$   & 502\,505 & 493\,247 & 476\,949 \\
   TIB+TID+TOB+TEC     & A        & $+15$   &  74\,116 &  71\,687 & 65\,521 \\
   TIB+TID+TOB+TEC     & B        & $+15$   & 103\,219 &  98\,085 & 92\,311 \\
   TIB+TID+TOB+TEC     & C        & $+15$   & 134\,844 &  132\,190 & 122\,528 \\
   TIB+TID+TOB+TEC     & C        & $+10$   & 309\,554 & 295\,288 & 273\,410 \\
   TIB+TID+TOB+TEC     & C        & $-1$    & 475\,467 & 452\,501 & 418\,948 \\
   TIB+TID+TOB+TEC     & C        & $-10$   & 528\,225 & 501\,577 & 469\,619 \\
   TIB+TID+TOB+TEC     & C        & $-15$   &  79\,406 &  16\,810 &  77\,181 \\
     \hline
    \end{tabular}
  \end{center}
\end{table}

\begin{figure}[htb]
	\begin{center}
		\includegraphics[scale=0.37,angle=90]{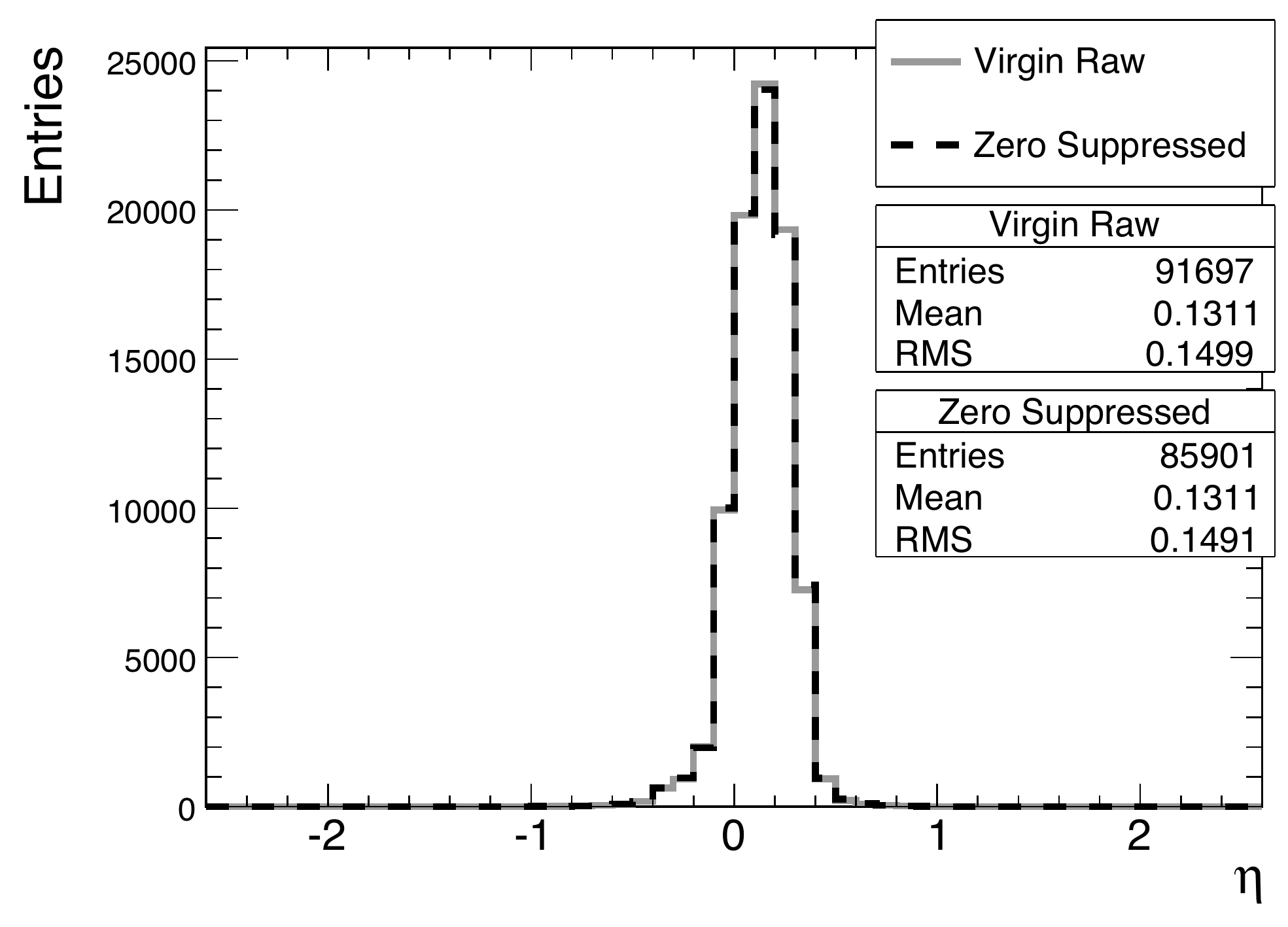}
		\includegraphics[scale=0.37,angle=90]{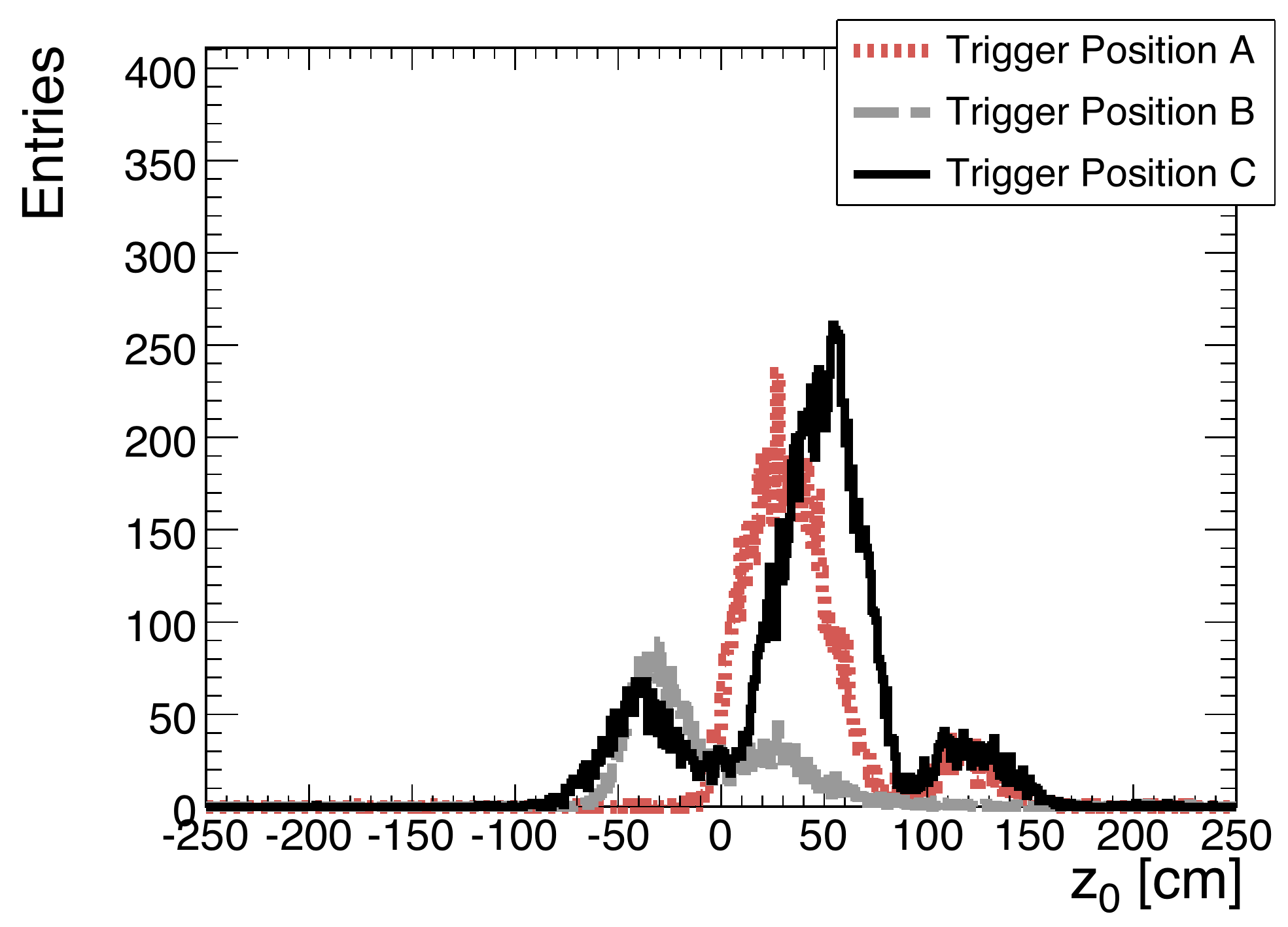}
		\caption{Comparison of $\eta$ and $z_0$ in single track events for the three different trigger layouts. All three data samples were taken at $\mathrm{T}=+15\degC$ and are normalized to each other. The tracks are reconstructed with the Road Search algorithm; the results for the Cosmic Track Finder and CKF are similar. No quality cuts have been applied.}
		\label{fig:trk_results_scintillator}
	\end{center}
\end{figure}

The effect of the different trigger scintillator positions on the track reconstruction can be seen in Figure~\ref{fig:trk_results_scintillator}. The histograms show $\eta$ and $z_0$ (measured at the point of closest approach to the origin in the transverse plane) of reconstructed tracks in single track events. The compared data samples have the same active detector and operating temperature, but differ in the position of the trigger scintillation counters as indicated in Figure~\ref{fig:trigger_layouts}. Trigger configuration~A shows a large peak around $\eta \approx 0.1$ stemming mostly from TIB+TOB tracks. A second peak is visible around $\eta \approx 0.8$, formed primarily by tracks with TEC hits. In the $z_0$ distribution, the TIB+TOB tracks give rise to the main peak at $z_0 \approx 30$\,cm, while tracks with mostly TEC hits lead to a peak at $z_0 \approx 120$\,cm. For trigger configuration~B, the lower scintillation counter was moved towards negative $z$, while the upper scintillation counters were moved slightly in the positive $z$ direction. This change led to a significant reduction in the number of TIB+TOB tracks contributing to the peak at $z_0 \approx -30$\,cm. Tracks with TEC hits can mostly be found at $z_0 \approx 25$\,cm. Trigger configuration~C is a combination of configurations~A and~B using an enhanced trigger setup. All three tracking algorithms give similar distributions, and for illustration purposes only results from the Road Search algorithm are shown.

\subsection{\label{sec:TrackRecoStability}Stability of Track Reconstruction Results}

\begin{figure}[p]
	\begin{center}
		(a)
		\includegraphics[width=.3\textwidth,angle=90]{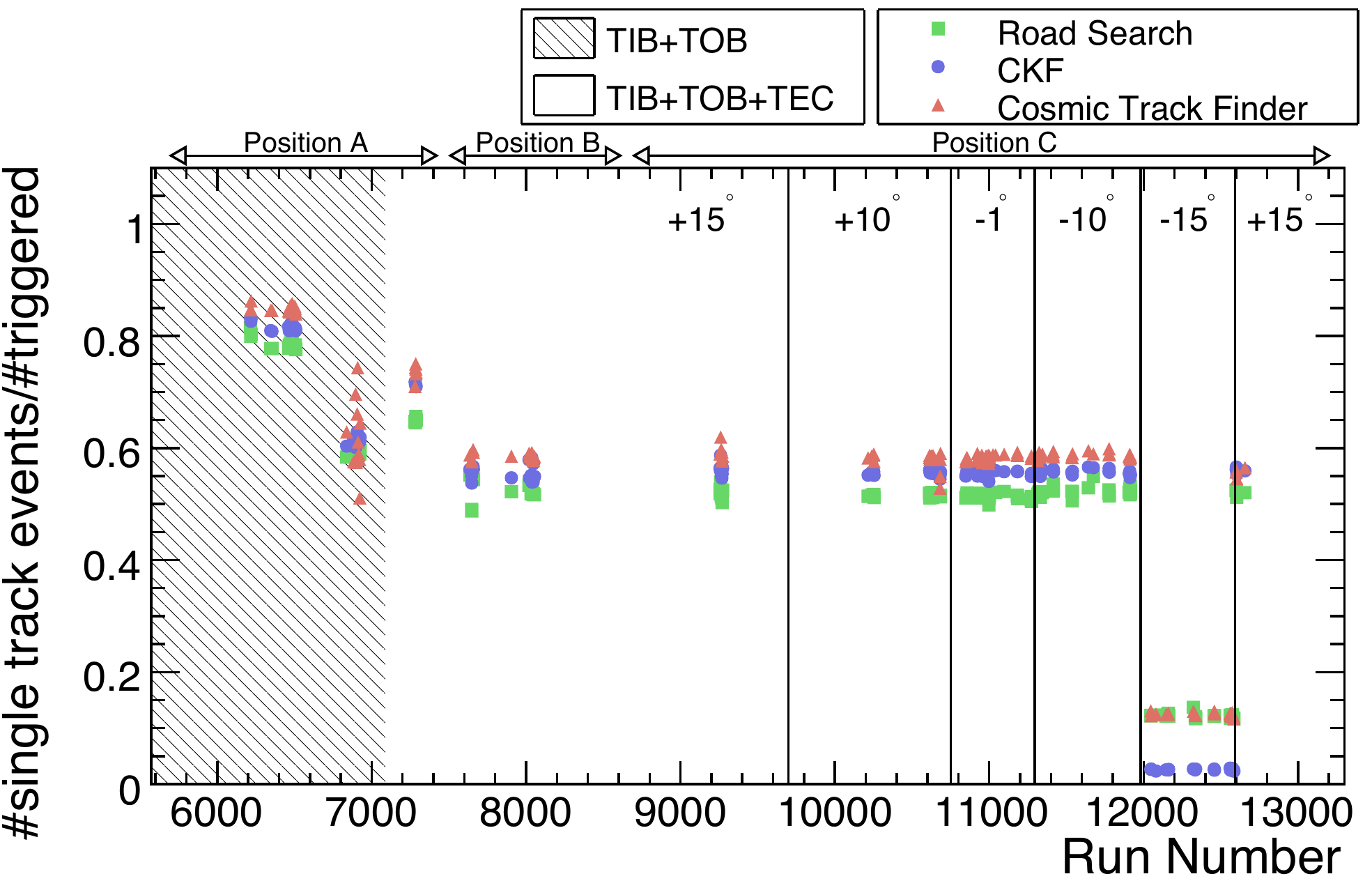}
		(b)
		\includegraphics[width=.3\textwidth,angle=90]{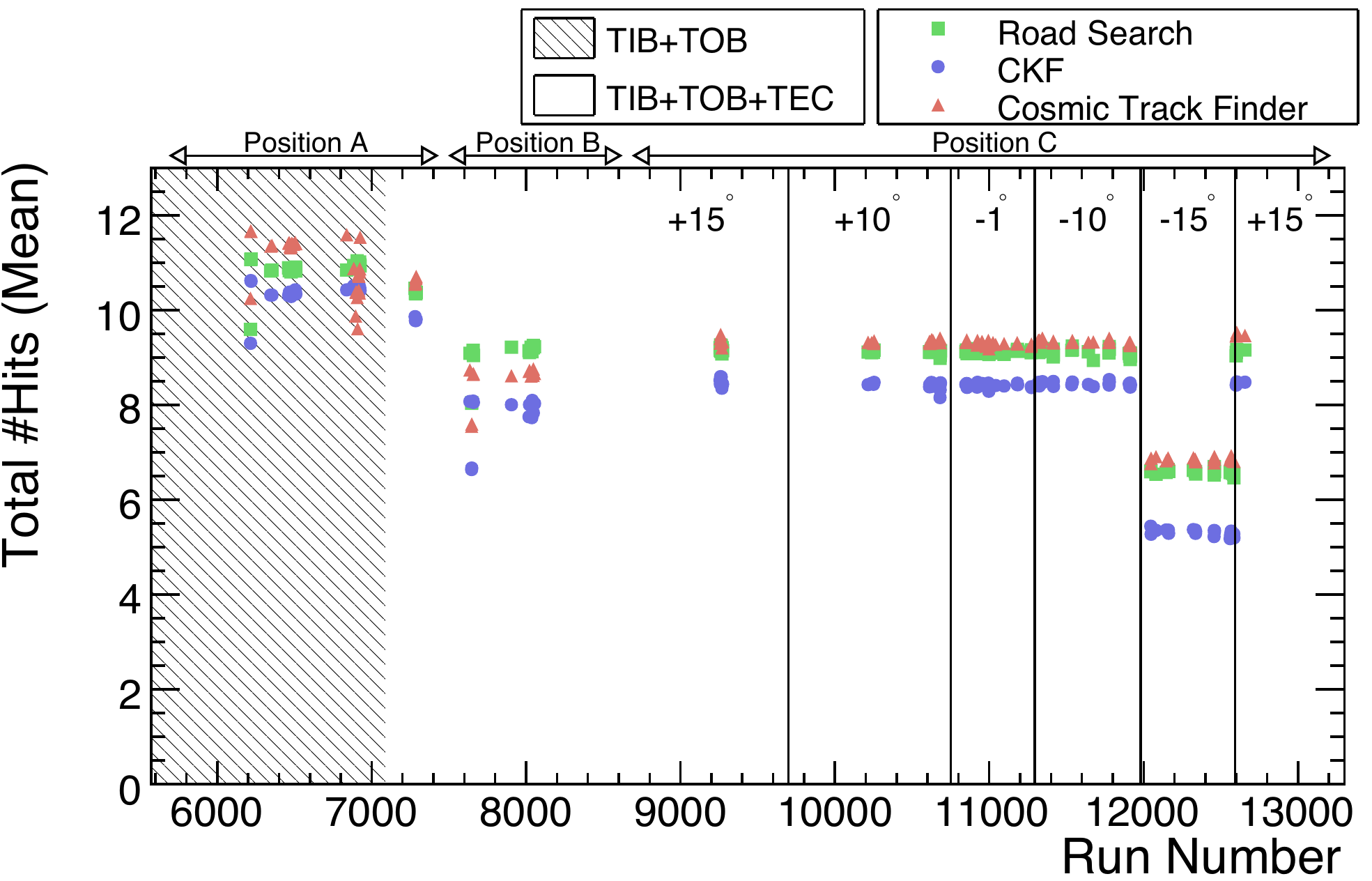} \\
		\vspace{1cm}
		(c)
		\includegraphics[width=.3\textwidth,angle=90]{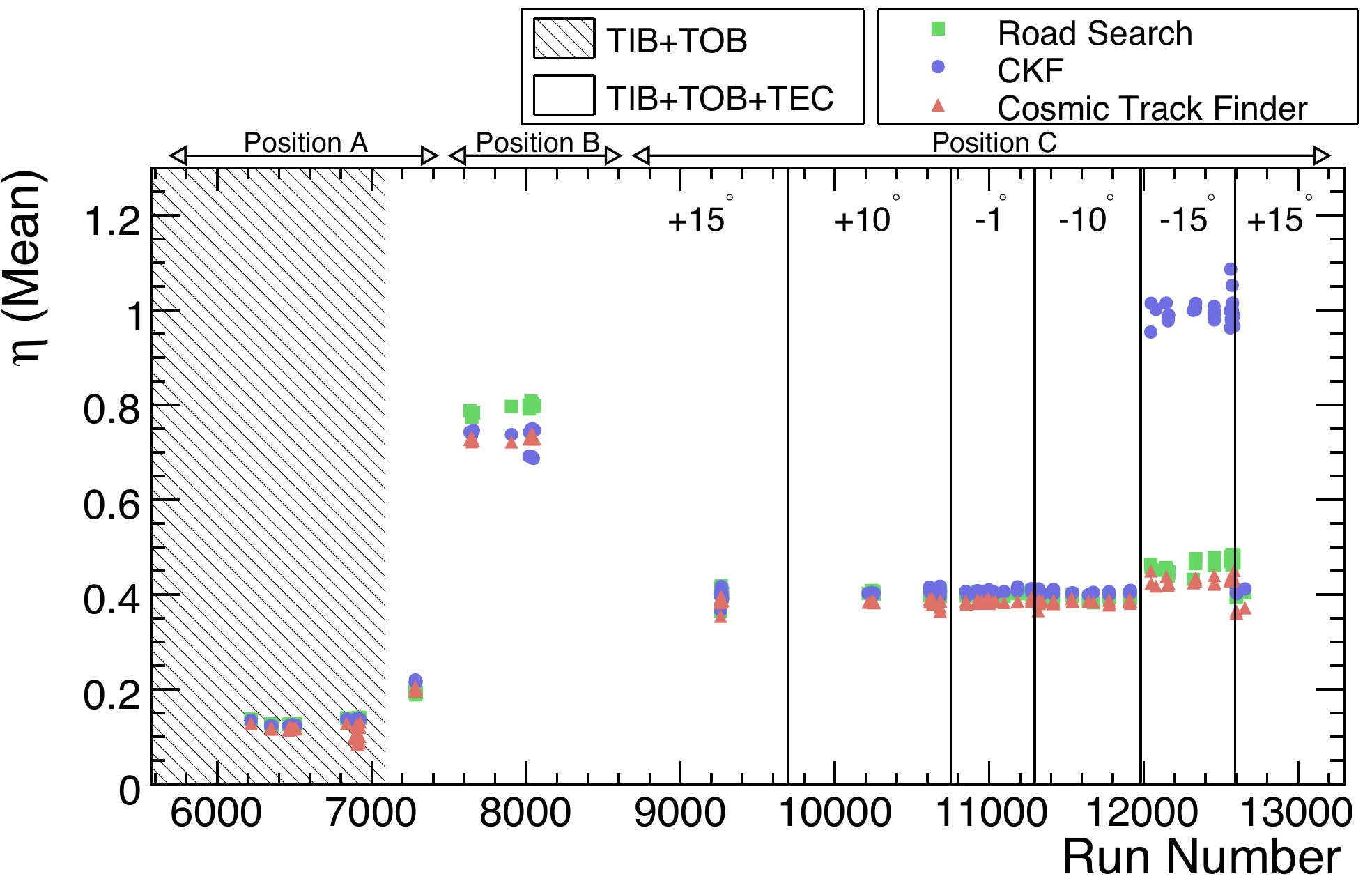}
		(d)
		\includegraphics[width=.3\textwidth,angle=90]{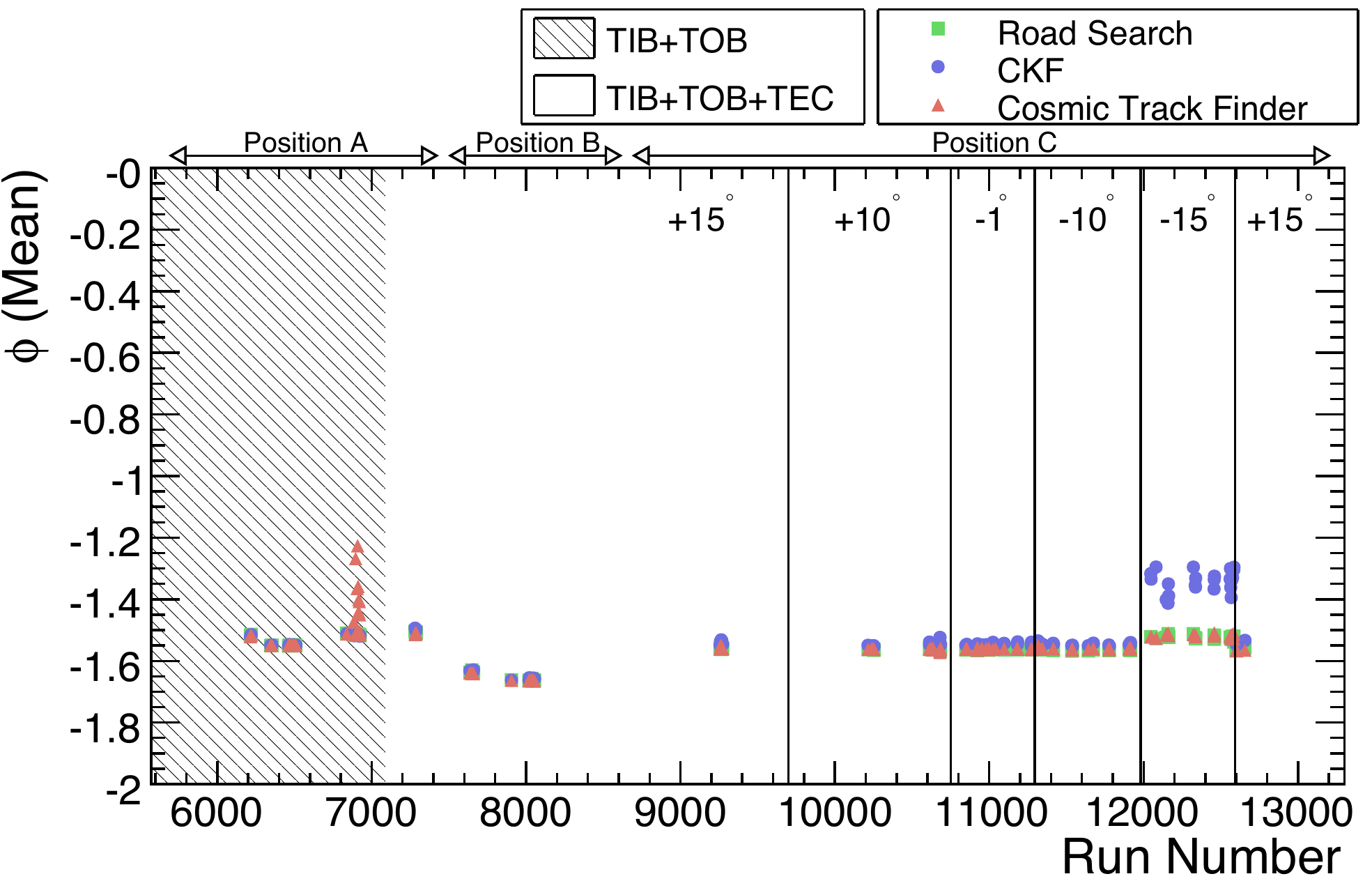} \\
		\vspace{1cm}
		(e)
		\includegraphics[width=.3\textwidth,angle=90]{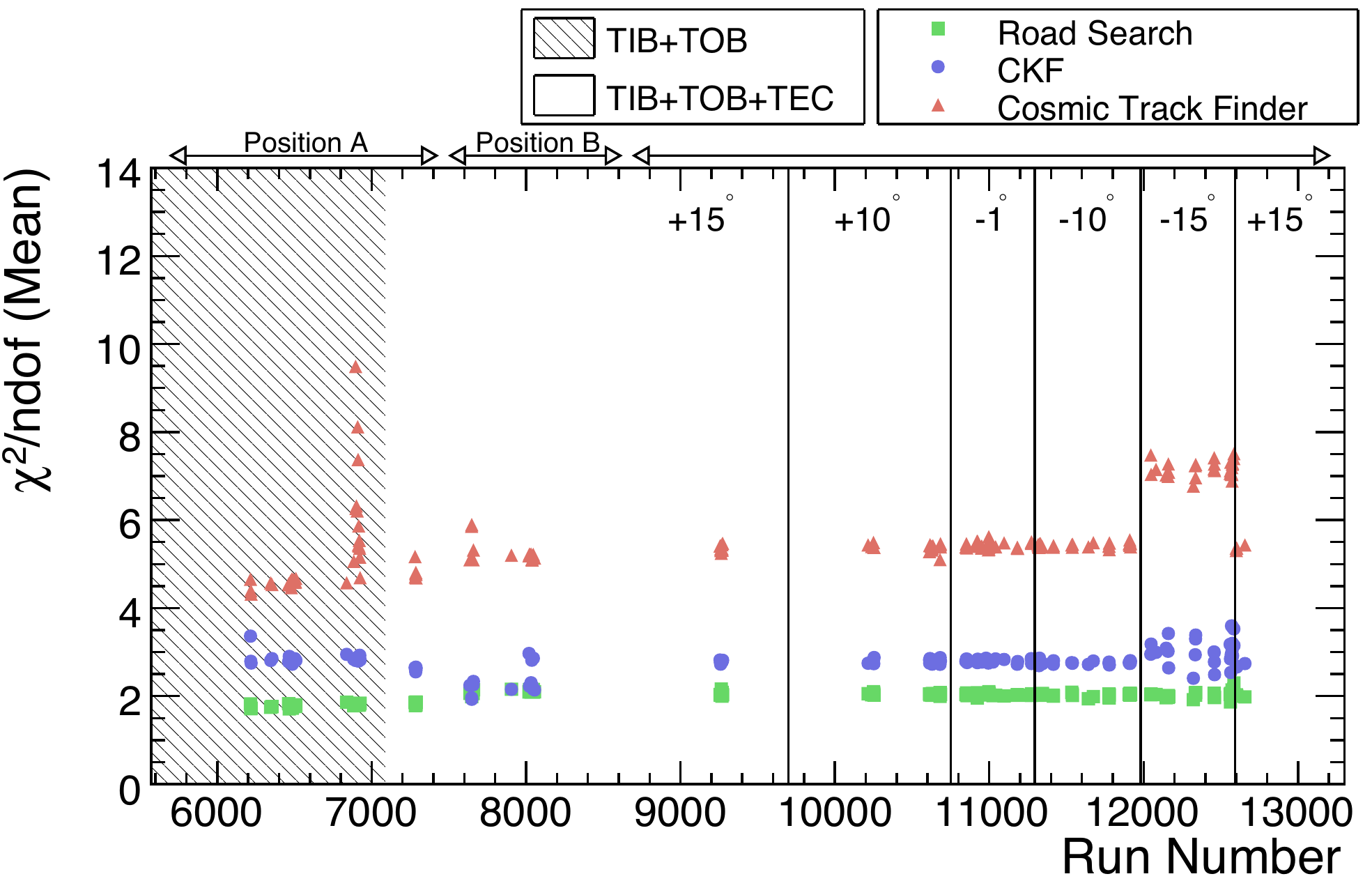}
		(f)
		\includegraphics[width=.3\textwidth,angle=90]{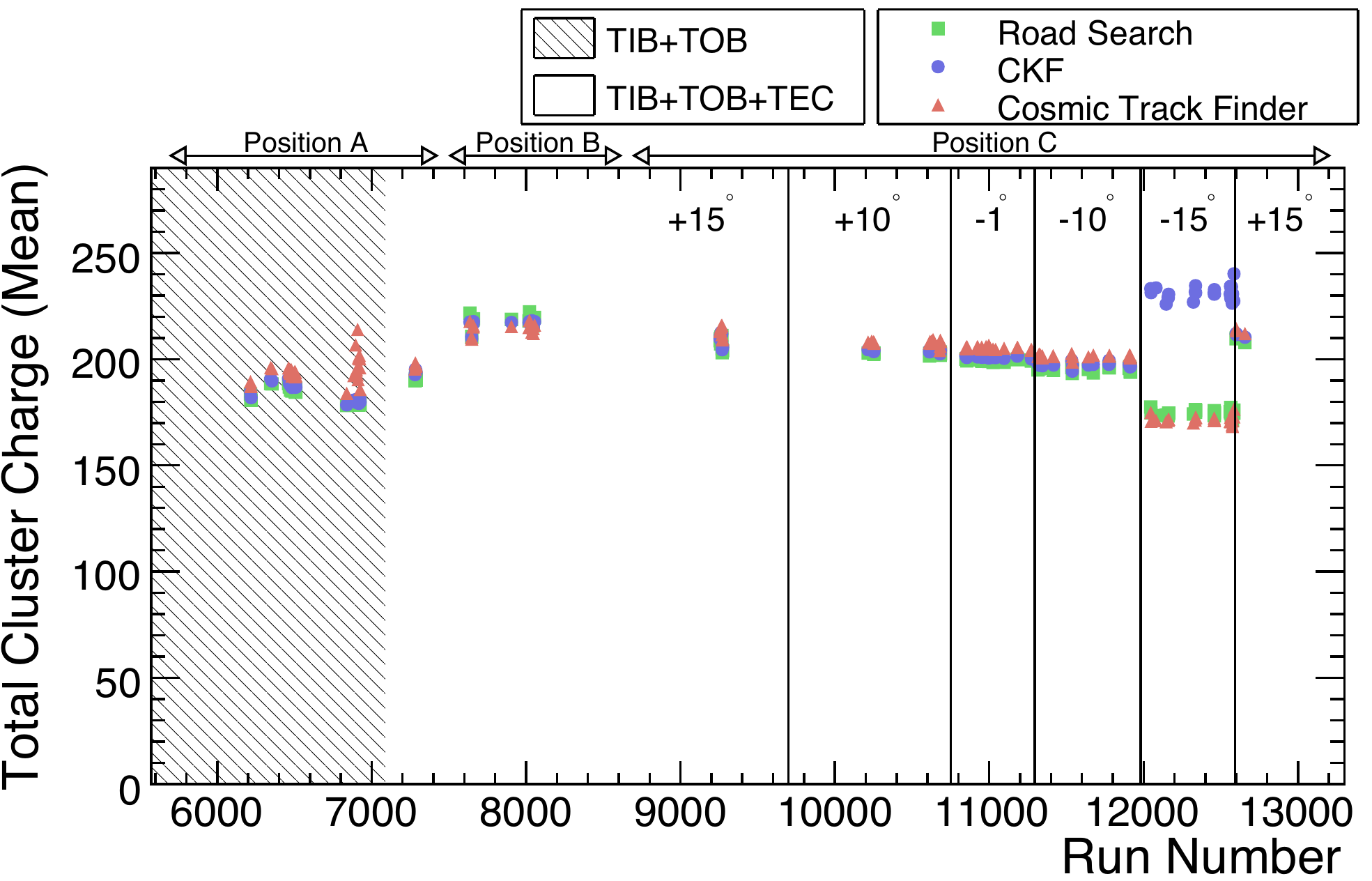}
		\caption{Control plots showing the mean of various reconstructed track parameters for all three algorithms as a function of run number. The scintillator positions and the operating temperatures (in \degC) are indicated at the top of each frame. All distributions are shown for single track events: (a) number of single track events divided by triggered events; (b) number of hits; (c) $\eta$; (d) $\phi$; (e) $\chi^2/ndof$; (f) cluster charge in ADC counts of hits belonging to the track before gain correction.
The differences observed at $-15\degC$ are due to the reduced number of active modules.
}
		\label{fig:validation_plots}
	\end{center}
\end{figure}

Control plots have been generated in order to monitor the stability of the track reconstruction results and to study the dependence on the operating temperature as a function of time. Figure~\ref{fig:validation_plots} shows the distribution of various reconstructed track parameters for single track events from all three algorithms. For each cosmic run, the mean of the parameter under study has been extracted and is plotted as a function of the run number. Each histogram indicates the active detectors, the various scintillator positions and the respective operating temperature. 

It is expected that the tracking parameters do not depend on the operating temperature, while a clear dependence on the trigger setup and the active detectors should be seen. This can be observed for trigger configuration~C, where the temperature was gradually decreased from $+15\degC$ to $-15\degC$. The reconstructed track parameters do not show a dependence on the operating temperature, except for data taken at $-15\degC$. In order to reach this operating temperature, a significant amount of modules in all four sub-detectors had to be turned off, effectively changing the active detectors.
The tracking algorithms were not retuned for this situation.
The differences in their seeding configurations translate into a different acceptance and explain the variations in the number of reconstructed tracks and the track parameter distributions.

Figure~\ref{fig:validation_plots} (f) shows a dependence of the average cluster charge before gain correction on the operating temperature.
Details of the effect of the gain corrections are shown in Ref.~\cite{TIFPerformanceNote}.

Various subtle changes can be observed for TIB+TOB in trigger configuration A, which is the result of a correction of the scintillation counter position.

\subsection{Processing Time}

The processing time of the track reconstruction algorithms has been studied in both data and simulated events. Results for the various configurations are shown in Table~\ref{tab:timing_results}. Due to the low rate of cosmic ray tracks and the expected low occupancy in the detector, no dedicated timing optimizations have been performed for any of the three tracking algorithms. However, by comparing the reconstruction times of the different algorithms with each other it is possible to detect inefficiencies in, for example, the seeding or pattern recognition step.

The timing results for simulated events show that the fraction of the overall amount of processing time spent for the tracking varies between 
2.9\% for the Cosmic Track Finder to 
8.8\% for the RS. The individual timing results of the algorithms are compatible with each other. The Cosmic Track Finder is the fastest of all three reconstruction algorithms due to its simplified structure.

A direct comparison of the timing results for data is difficult, even for runs taken in the same configuration. Nevertheless, a comparison allows general conclusions to be drawn. In Table~\ref{tab:timing_results} it can be observed that the overall processing time per event is significantly higher in virgin raw (VR) mode compared to zero suppressed (ZS) mode, except for data taken at $-15\degC$. This behavior is expected since all channels are read out in virgin raw mode and the zero suppression is done offline. The comparable reconstruction time for both virgin raw and zero suppressed data taken at $-15\degC$  is due to the limited detector geometry that is read out in this configuration. Concerning the tracking algorithms it can be observed that the processing time for both the CKF and the Road Search algorithm are comparable, while the Cosmic Track Finder is significantly faster. This is caused by the more elaborate pattern recognition stages of the CKF and Road Search algorithms.

\begin{table}[htbp] 
\small
\hspace{-1.0cm}
\caption{Reconstruction times for various configurations in both data and simulated events.}
\begin{center} 
\begin{tabular}{|c|c|c|c|c|c|c|c|} 
\hline
Trigger & T  & Readout  & \multicolumn{4}{|c|}{Time/Event [arbitrary units]}  \\
 Conf.    &   [\degC]  &  Mode  & Total    & Road Search & CKF        & Cosmic T.F. \\
\hline
\multicolumn{7}{|l|}{\it Simulated Events}\\
\hline
A   &     n/a  &    n/a   &    $0.2093\pm0.0211$  &      $0.0184\pm0.0003$  &  $0.0121\pm0.0001$ &  $0.0067\pm0.0001$ \\
C   &     n/a  &    n/a   &    $0.2033\pm0.0197$   &      $0.0179\pm0.0003$  &  $0.0127\pm0.0002$  &  $0.0059\pm0.0001$ \\
\hline 
\multicolumn{7}{|l|}{\it Data}\\
\hline
A   &      $+15$  &    VR   &    $2.2524\pm0.0105$  &      $0.0455\pm0.0049$  &  $0.0638\pm0.0035$ &  $0.0131\pm0.0002$ \\
B   &      $+15$  &    VR   &    $2.2641\pm0.0092$  &      $0.0373\pm0.0017$  &  $0.0895\pm0.0066$ &  $0.0105\pm0.0002$ \\
B   &      $+15$  &    ZS   &    $0.1931\pm0.0060$  &      $0.0341\pm0.0014$  &  $0.0744\pm0.0042$ &  $0.0091\pm0.0002$ \\
C   &      $+15$  &    VR   &    $2.3698\pm0.0111$  &      $0.1997\pm0.0048$  &  $0.0569\pm0.0030$ &  $0.0100\pm0.0002$ \\
C   &      $+10$  &    VR   &    $2.2745\pm0.0155$  &      $0.0636\pm0.0116$  &  $0.0701\pm0.0048$ &  $0.0100\pm0.0002$ \\
C   &     $-1$   &    VR   &    $2.3714\pm0.0112$  &      $0.0504\pm0.0072$  &  $0.0604\pm0.0034$ &  $0.0092\pm0.0002$ \\
C   &     $-10$  &    VR   &    $2.5425\pm0.0498$  &      $0.2257\pm0.0479$  &  $0.0576\pm0.0034$ &  $0.0096\pm0.0002$ \\
C   &     $-10$  &    ZS   &    $0.2034\pm0.0090$  &      $0.0331\pm0.0019$  &  $0.0631\pm0.0044$ &  $0.0095\pm0.0002$ \\
C   &     $-15$  &    VR   &    $0.8292\pm0.0174$  &      $0.0637\pm0.0149$  &  $0.0159\pm0.0018$ &  $0.0028\pm0.0001$ \\
C   &     $-15$  &    ZS   &    $0.1062\pm0.0089$  &      $0.0456\pm0.0058$  &  $0.0178\pm0.0016$ &  $0.0026\pm0.0001$ \\
\hline
\end{tabular} 
\end{center} 
\label{tab:timing_results}
\end{table}

\subsection{Tracking Efficiency}

TIF data were used to verify the efficiency of the tracking algorithms. The estimation of the tracking efficiency is a challenging task, due to the poor prior constraints available for cosmic tracks and the absence of an external reference other than the scintillator counters. The strategy adopted for this analysis is the reconstruction of partial tracks, using only a subset of the tracker. These track segments serve as a reference for the reconstruction in the remaining parts of the tracker. For statistical reasons, tracks in the TIB and TOB were chosen as independent subsets of the tracker.
In the following discussion, track segments reconstructed in TIB will be referred to as TIB tracks, while track segments reconstructed in the TOB are referred to as TOB tracks.

To isolate single track events, only events with less than 30 reconstructed hits were analyzed. 
TIB and TOB tracks were accepted if the normalized $\chi^2$ was smaller than 30 and if they contained hits with signal-to-noise ratio greater than 8 in at least four different layers.

Based on the selected events, the tracking efficiency in TIB (TOB) was calculated from the fraction of TOB (TIB) tracks with a matching track in the other sub-detector. This conditional efficiency can deviate from the global efficiency if the TIB and TOB acceptance differ after selection. The match between tracks was based on a comparison of the azimuthal angles. The difference was required to be smaller than five times the resolution determined from simulation. In addition, the reference track had to contain at least two hits in stereo layers and the extrapolation had to be fully contained in the active region of the other sub-detector.

\begin{table}[htbp]
	\caption{\label{tab:trk_efficiencies} Average conditional track efficiencies and corresponding statistical uncertainties for all three track reconstruction algorithms in data and Monte Carlo simulation. }
	\begin{center}
   \begin{tabular}{|c|c|c|c|c|}
	\hline
	 & \multicolumn{2}{|c|}{$\epsilon$(TIB$|$TOB) [\%]} & \multicolumn{2}{|c|}{$\epsilon$(TOB$|$TIB) [\%]} \\
	               & Data & MC & Data & MC \\
	\hline
		Combinatorial Kalman Filter 
		  & $94.0 \pm 0.2$ & $98.66 \pm 0.04$
		  & $97.7 \pm 0.1$ & $98.76 \pm 0.04$ \\
		Cosmic Track Finder 
		  & $93.1 \pm 0.2$ & $94.46 \pm 0.09$ 
		  & $96.9 \pm 0.1$ & $97.36 \pm 0.06$ \\
		Road Search 
		  & $89.9 \pm 0.2$ & $89.08 \pm 0.12$
		  & $99.0 \pm 0.1$ & $99.39 \pm 0.03$ \\ 
		\hline
    \end{tabular}
  \end{center}
\end{table}

\begin{figure}[p]
  \begin{center}
    \includegraphics[scale=0.30]{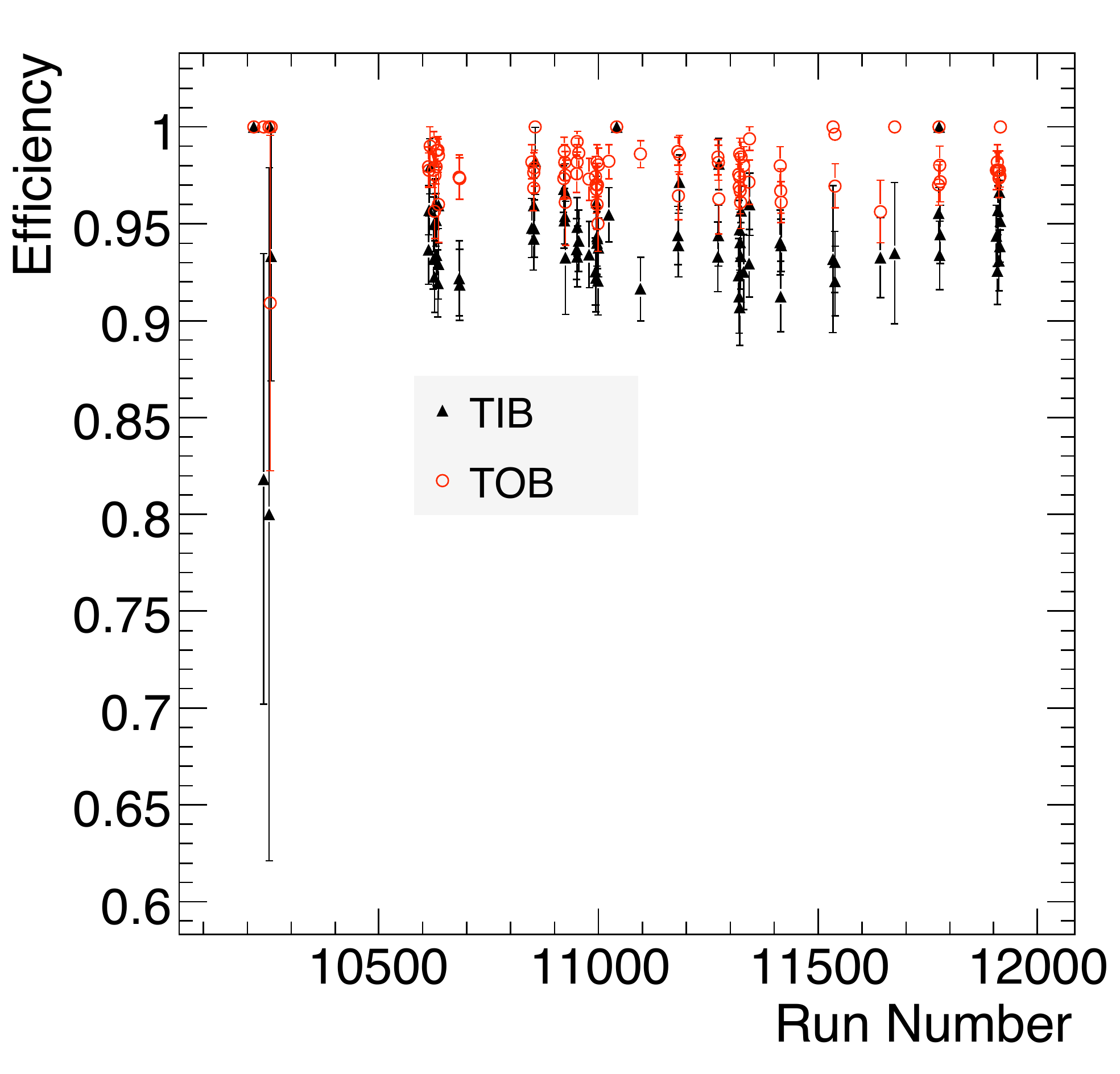}
    \caption{Conditional track reconstruction efficiencies $\epsilon$(TIB$|$TOB)~(triangles) and $\epsilon$(TOB$|$TIB)~(circles) for the CKF as a function of the run number.}
    \label{fig:trackEffVsRun}
  \end{center}
\end{figure}

The average efficiencies for all three track reconstruction algorithms are shown in Table~\ref{tab:trk_efficiencies}. The results have been obtained using data taken at  $+10\degC$, $-1\degC$ and $-10\degC$ in scintillator position~C. The corresponding values for simulated events are listed in the same table. The stability of the conditional track reconstruction efficiency is demonstrated in Figure~\ref{fig:trackEffVsRun}, which shows the distribution of $\epsilon$(TIB$|$TOB) and $\epsilon$(TOB$|$TIB) for the CKF as a function of run number.

\begin{figure}[p]
  \begin{center}
    \includegraphics[scale=0.30]{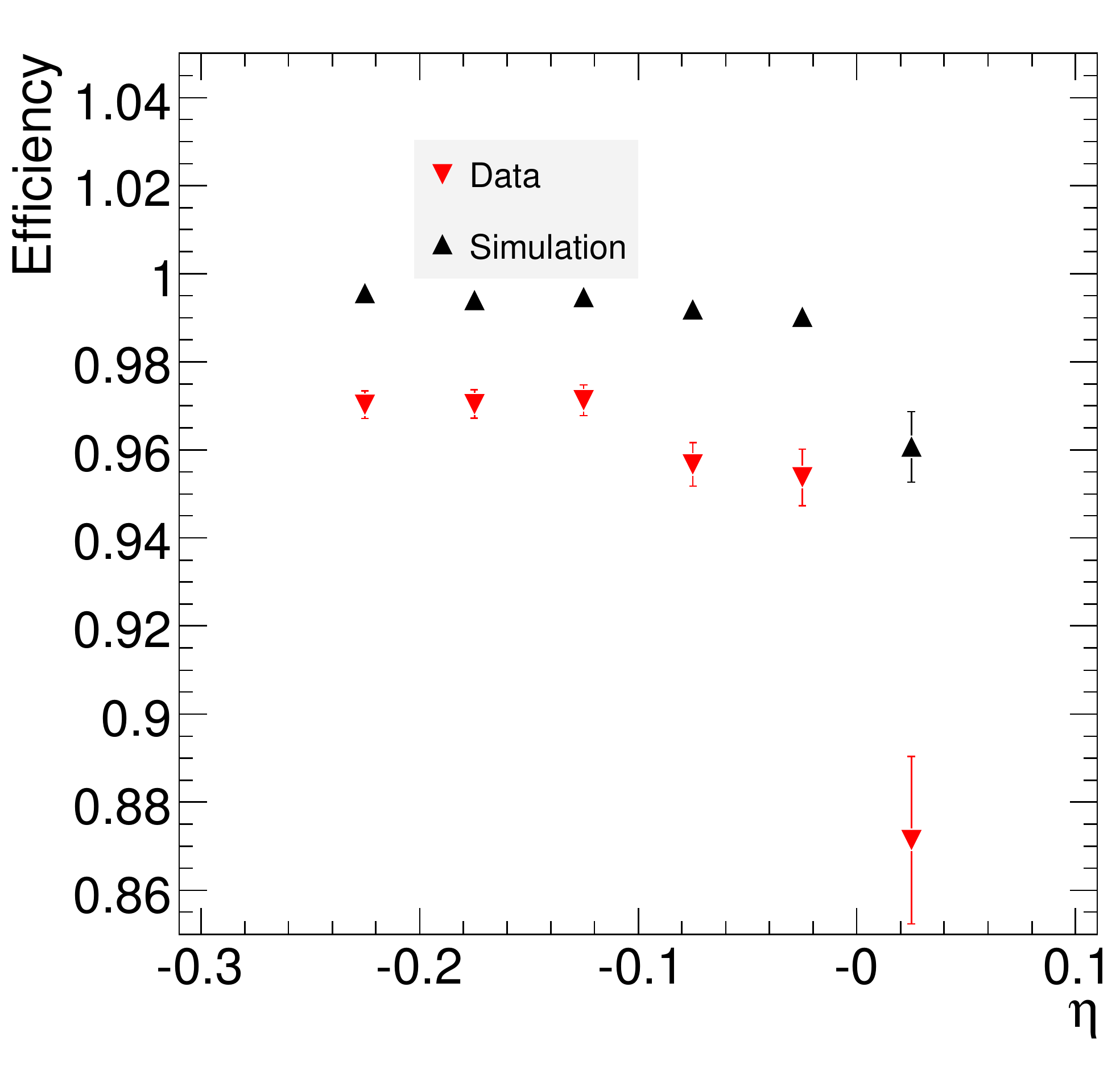}
    \hspace{0.6cm}
    \includegraphics[scale=0.30]{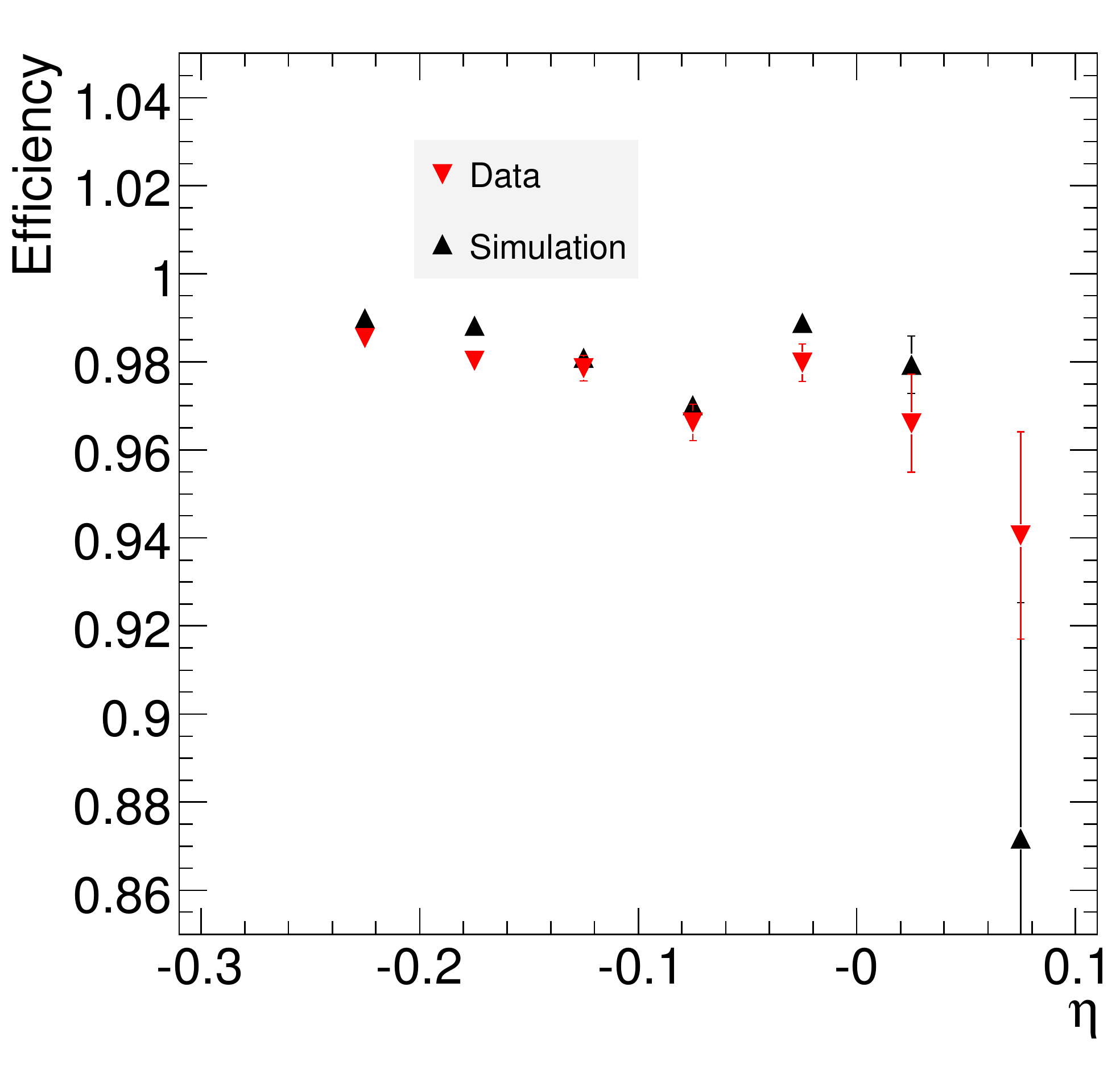}
    \caption{Conditional track reconstruction efficiency $\epsilon$(TIB$|$TOB)~(left) and $\epsilon$(TOB$|$TIB)~(right) for the CKF in data as a function of the pseudorapidity ($\eta$). Downward and upward pointing triangles indicate data and simulated events, respectively.}
    \label{fig:trackEffEta}
  \end{center}
\end{figure}

\begin{figure}[p]
  \begin{center}
  \includegraphics[scale=0.30]{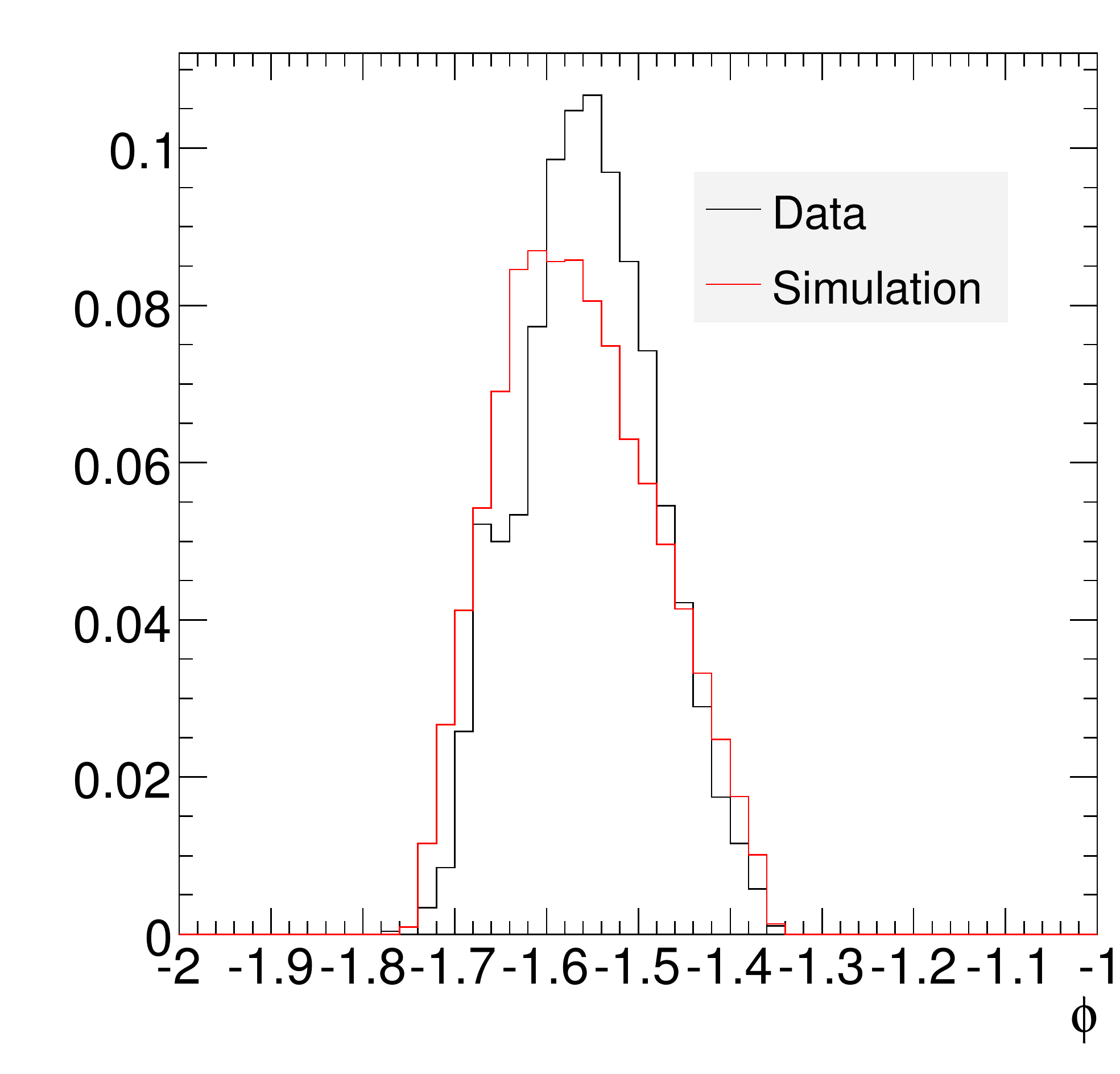} 
  \hspace{0.6cm}
  \includegraphics[scale=0.30]{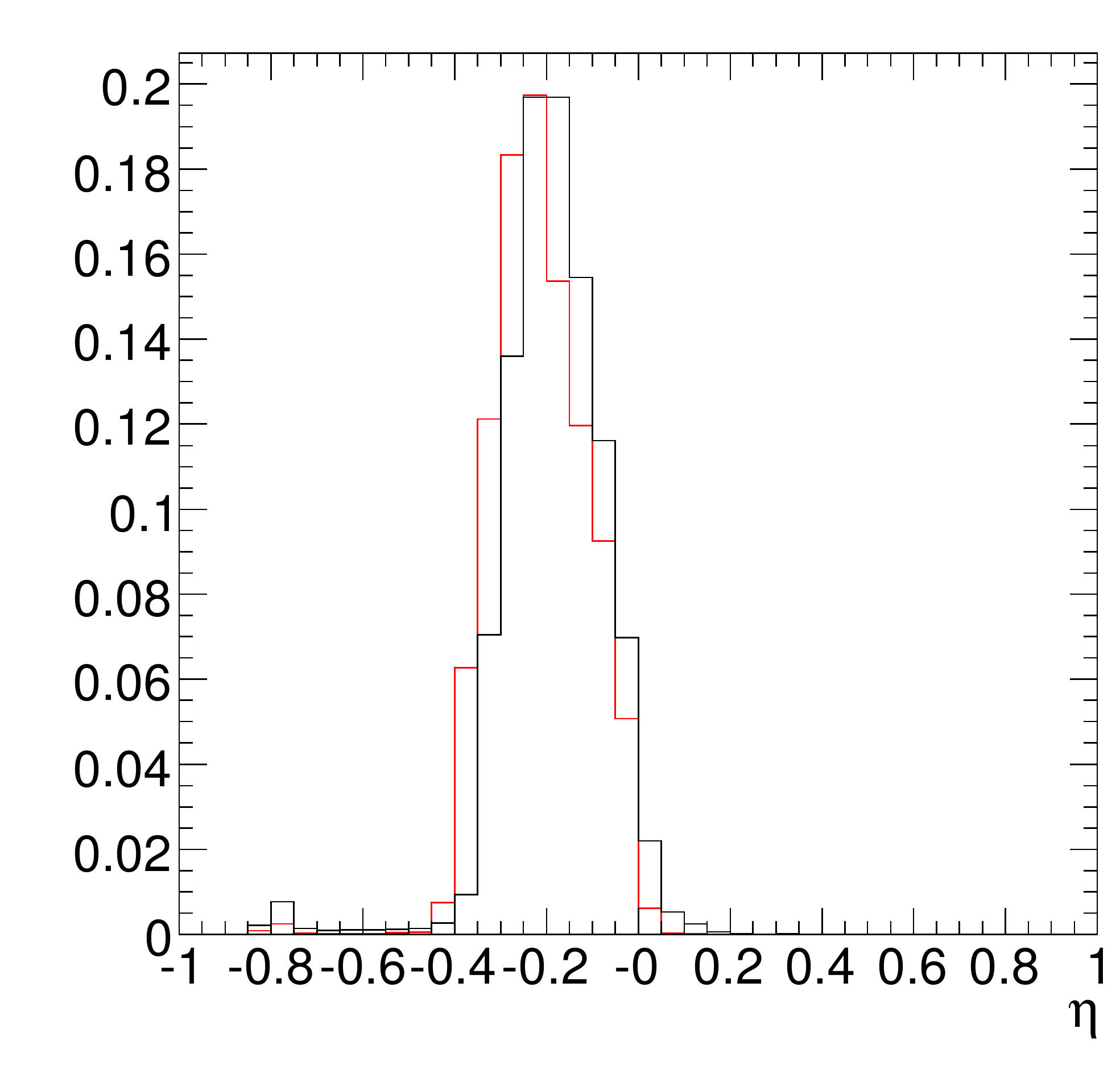} \\
  \caption{Distributions of $\phi$ (left) and $\eta$ (right) of reference tracks in the TOB. Data and simulated events are indicated by black and red lines, respectively. Distributions from the CKF algorithm are shown.}
  \label{fig:trackEffRef}
  \end{center}
\end{figure}

The dependence of the conditional track reconstruction efficiency on the pseudorapidity is shown in Figure~\ref{fig:trackEffEta} for both data and simulated events.
As an example the distributions from the Combinatorial Kalman Filter have been chosen.
Possible reasons for the slightly higher efficiency found in simulation are differences of the acceptance in data and simulation and of the resolution on the track angle, which enters via the match of TIB and TOB tracks. 
The difference in acceptance is confirmed by the angular distributions of the reference tracks, shown in Figure~\ref{fig:trackEffRef}.

\subsection{Momentum Estimation from Scattering}

Since the TIF cosmic data were taken without a magnetic field, it is difficult to select a precise, {\it i.e.} high-momentum, track sample to be used for detailed track studies. 
The majority of tracks are low momentum and dominated by multiple scattering.
In the CMS tracker, the material contribution for each layer is $x/X_0 \approx 2.5\%$. 
Only for momentum above about $20$ \GeVc~does the multiple scattering contribution fall below the intrinsic resolution.

The multiple scattering contribution can be used in order to obtain a measure of the momentum of each track. This is done by refitting the tracks with the momentum floating. The scattering contribution to the resolution of each layer is calculated from the assumed momentum and the projection distance between layers, and the momentum that makes the track fit $\chi^2$ equal to one is used as the best momentum estimate.

\begin{figure}[htb]
\begin{center}
(a) \includegraphics[width=0.45\linewidth]{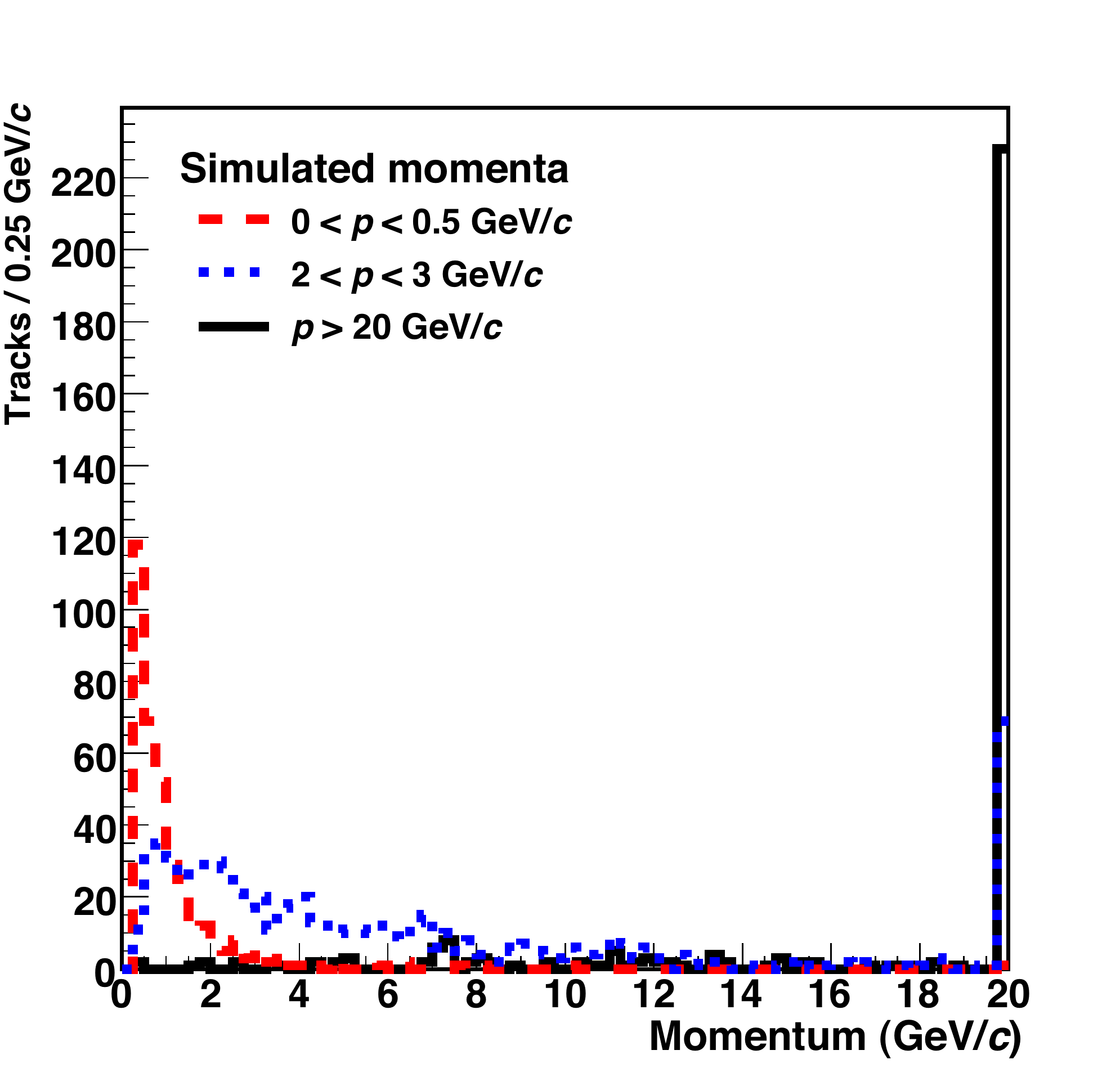} \hfil
(b) \includegraphics[width=0.45\linewidth]{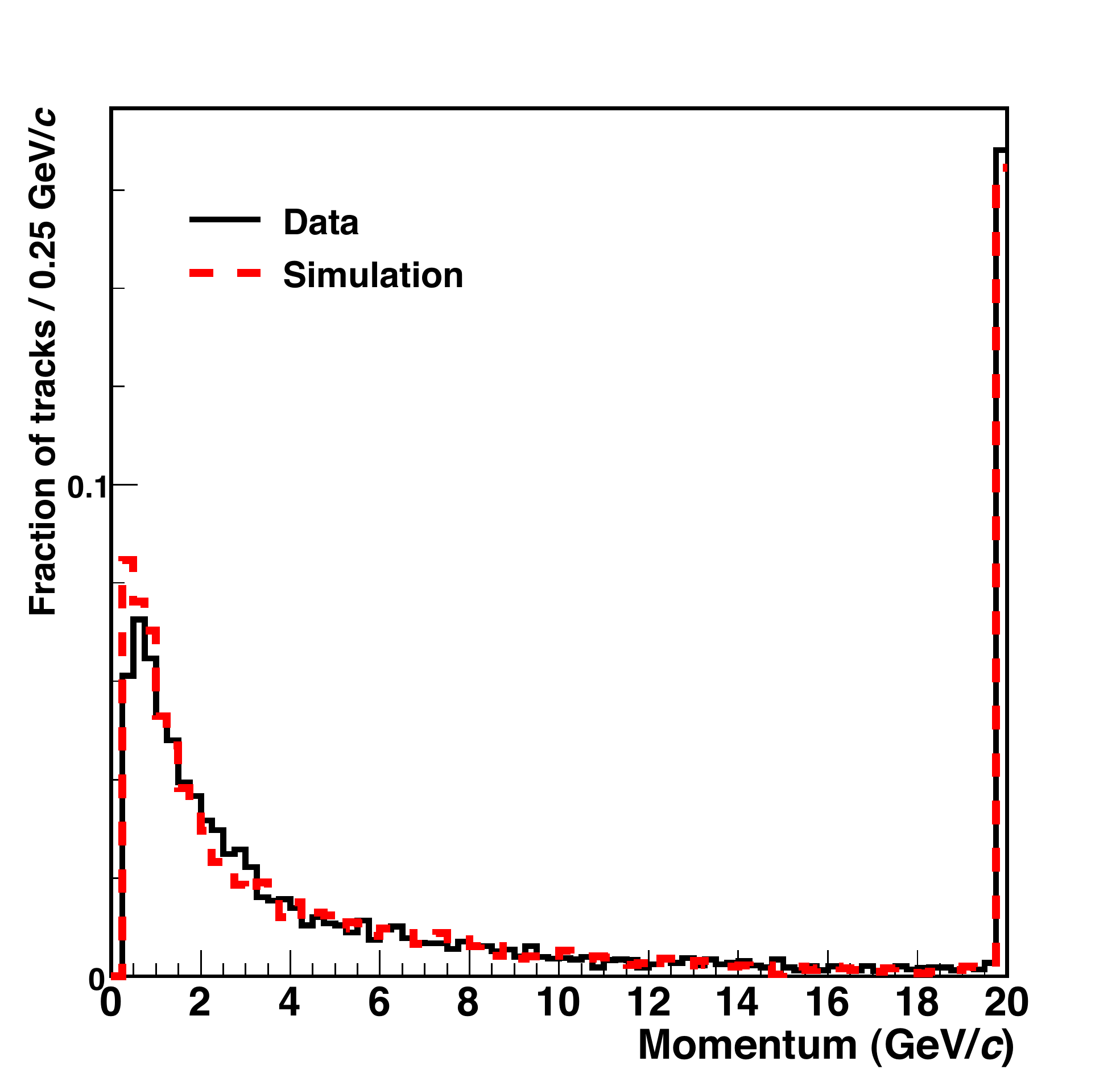}  
\caption{
(a) The ``scattering-estimated'' momentum spectrum is compared for TOB tracks in simulated cosmic ray muons with three different momentum ranges, $0<p<0.5$\,\GeVc~(dashed line), $2<p<3$\,\GeVc~(dotted line), and $p>20$\,\GeVc~(full line).
The spread of the momentum estimate is large, but the low and high momentum extremes are well separable.
(b) The ``scattering-estimated'' momentum spectrum is compared for TOB tracks in post-alignment cosmic data (full line) and simulation (dashed line). 
Reasonable agreement between simulation and data indicates that the scattering and intrinsic resolution functions are well described.
The last bin contains overflows, {\it i.e.}, tracks for which no scattering contribution is needed to obtain a good fit. 
}  
\label{fig:MCSmom}
\end{center}
\end{figure}

Since the scattering is a statistical process, this estimate does not provide a precise measure of the momentum of each track. 
However, the ensemble of tracks provides a measure of the momentum spectrum that can be compared to simulation. 
In Figure~\ref{fig:MCSmom}~(a) the momentum distribution, calculated as described above, for simulated cosmic ray muons with momenta in three different ranges are compared. Though there is large spread, the estimate follows the true momentum, and the low and high momentum extremes are well separable. 

This momentum estimate was performed in data using tracks reconstructed in the TOB. 
The TOB alignment corrections provide single layer hit resolutions of about $40~\mu$m, which is small compared to the scattering effects over most of the momentum range. The results are shown in Figure~\ref{fig:MCSmom}~(b) and compared to the momentum spectrum obtained from simulated data. There is generally good agreement.

This momentum estimate was done to validate the tracking and resolution modeling.
It also provides a track quality measure, which could be used in other studies, 
equivalent to a momentum cut in the absence of a magnetic field.

\subsection{Energy Loss Measurement for Cosmic Muons}
The signal height from a strip is related to the number of electron-hole pairs created by the traversing particle in the bulk of the silicon sensor. The analogue readout of the CMS tracking system retains this information, and each reconstructed hit is subsequently associated with a corresponding cluster charge in ADC counts. The measurement of the energy lost per unit pathlength $(dE/dx)$ by charged particles in silicon, in conjunction with the measurement of their momentum, is a valuable tool for particle identification. The $dE/dx$ of the track can provide additional information for the identification of electrons in jets -- complementing  the information from the electromagnetic calorimeter -- and most importantly allows the discrimination between different charged hadron species.

Each track crosses several micro-strip modules, and each crossing gives an independent $dE/dx$ measurement based on the deposited energy normalized to the estimated pathlength in the active part of the module.
Since the deposited energy is measured in ADC counts, a conversion factor has to be applied. In the following the default value of 250 electron-hole pairs per ADC count is used for the silicon strip tracker. This is multiplied by the energy needed to create an electron-hole pair, 3.61\,eV.  
The conversion includes also a correction for the variations in the electronic gain.
The precision of these correction factors has a direct impact on the homogeneity of the measured signal and, hence, on the resolution.

Given the trajectory of a charged particle it is possible to measure the energy deposited in the different layers of the tracking system and to attribute a single $dE/dx$ estimator to the track. The loss of energy after each traversal is negligible with respect to the momenta considered, justifying the assumption of a unique $dE/dx$ value along all the trajectory of the particle. Various most-probable-value estimators for the track $dE/dx$ have been studied in simulated events (see Ref.~\cite{EnergyLossNote}), and the two most promising methods are studied using data:

\begin{itemize}
	\item {\bf Truncated mean} \\ The highest values (in a fixed fraction of the total number of observations) are discarded, and the arithmetic mean is computed for the remaining values. The method has little sensitivity to outliers, but it introduces a bias which depends on the arbitrary cut-off parameter ({\it truncation fraction}) and it reduces the statistics. 
	\item {\bf Generalized mean} \\ The generalized mean with exponent $k$ of a variable $x$ is defined as:

	\begin{equation*}
		M_{k}(x_1,...,x_n) = \left( \frac{1}{n} \cdot \sum_{i=1}^n x^k_i \right)^{1/k}
	\end{equation*}
	A special case of the generalized mean with $k=-1$ is the harmonic mean. The harmonic mean of a set of observations is given by the number of terms divided by the sum of the reciprocals of the term. 
\end{itemize}  

Based on the results of the energy loss study in Ref.~\cite{EnergyLossNote}, a truncation fraction of 20\% and 40\% and the exponent $k=-2$  and $k=-4$ have been chosen for comparison. 

The $dE/dx$ study is performed using data taken at $-10\degC$ in trigger configuration~C. This data sample is close to the nominal tracker operation temperature and has large statistics. The following quality requirements are imposed:
\begin{itemize}
	\item Exactly one track reconstructed in the event.
	\item Number of hits $>$ 4 (with matched hits being resolved into their $r\phi$ and stereo component and counted separately).
	\item $\chi^2$/ndof $<$ 10.
\end{itemize}

Assuming that the different track reconstruction algorithms reconstruct the same track with exactly the same hits associated, the measured $dE/dx$ should be identical and independent of the chosen tracking algorithm.  
This has been confirmed by comparing events where both the CKF and the Road Search algorithm find a track that satisfies the above listed quality requirements. The resulting $dE/dx$ distributions with the corresponding means and widths are found to be exactly identical for both algorithms. Hence, to eliminate any dependence on the choice of tracking algorithm for this study, only events where both algorithms find the same track are retained for the analysis.

\begin{figure}[htb]
	\begin{center}
		\includegraphics[scale=0.34,angle=90]{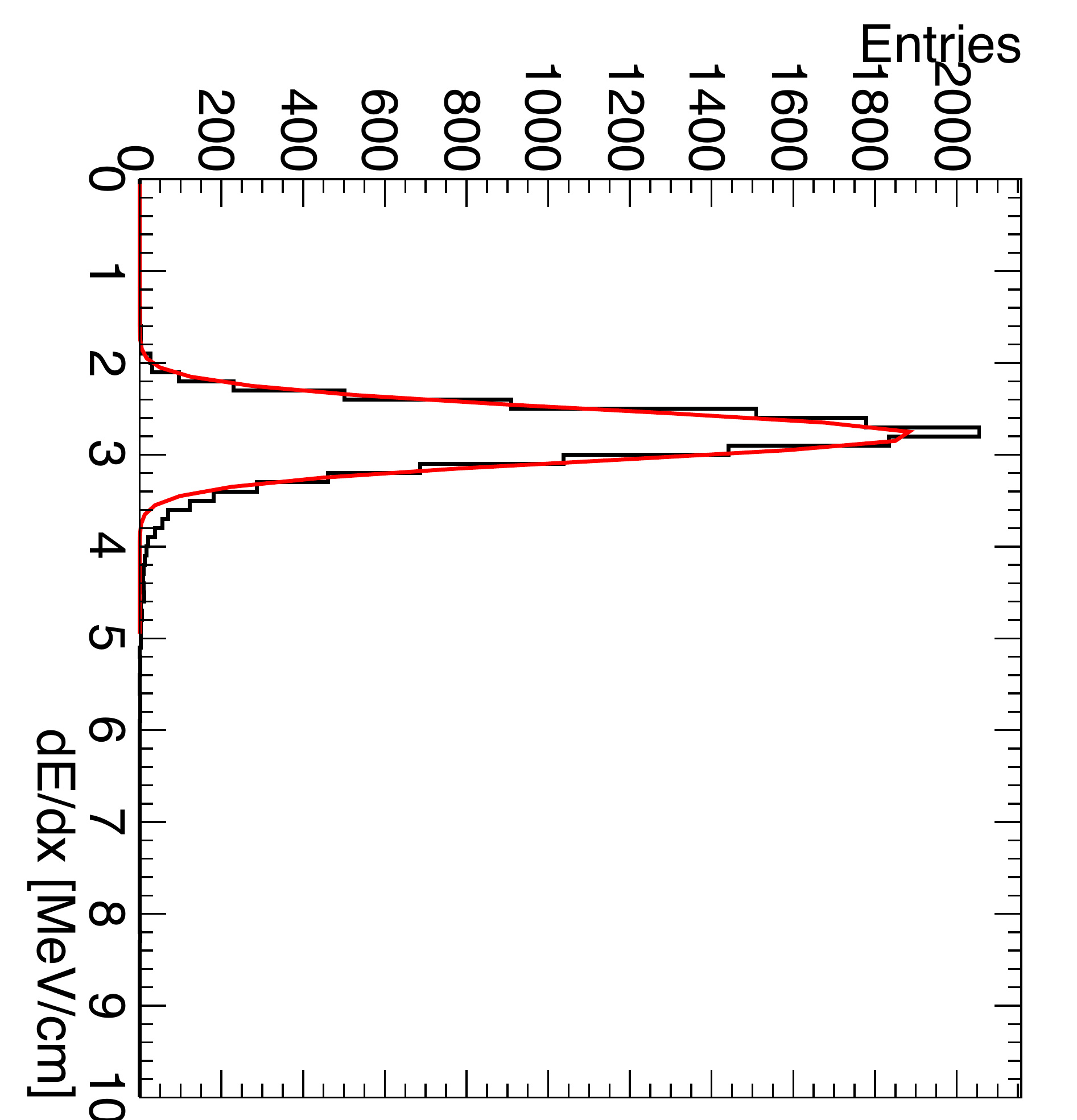} 
		\includegraphics[scale=0.34,angle=90]{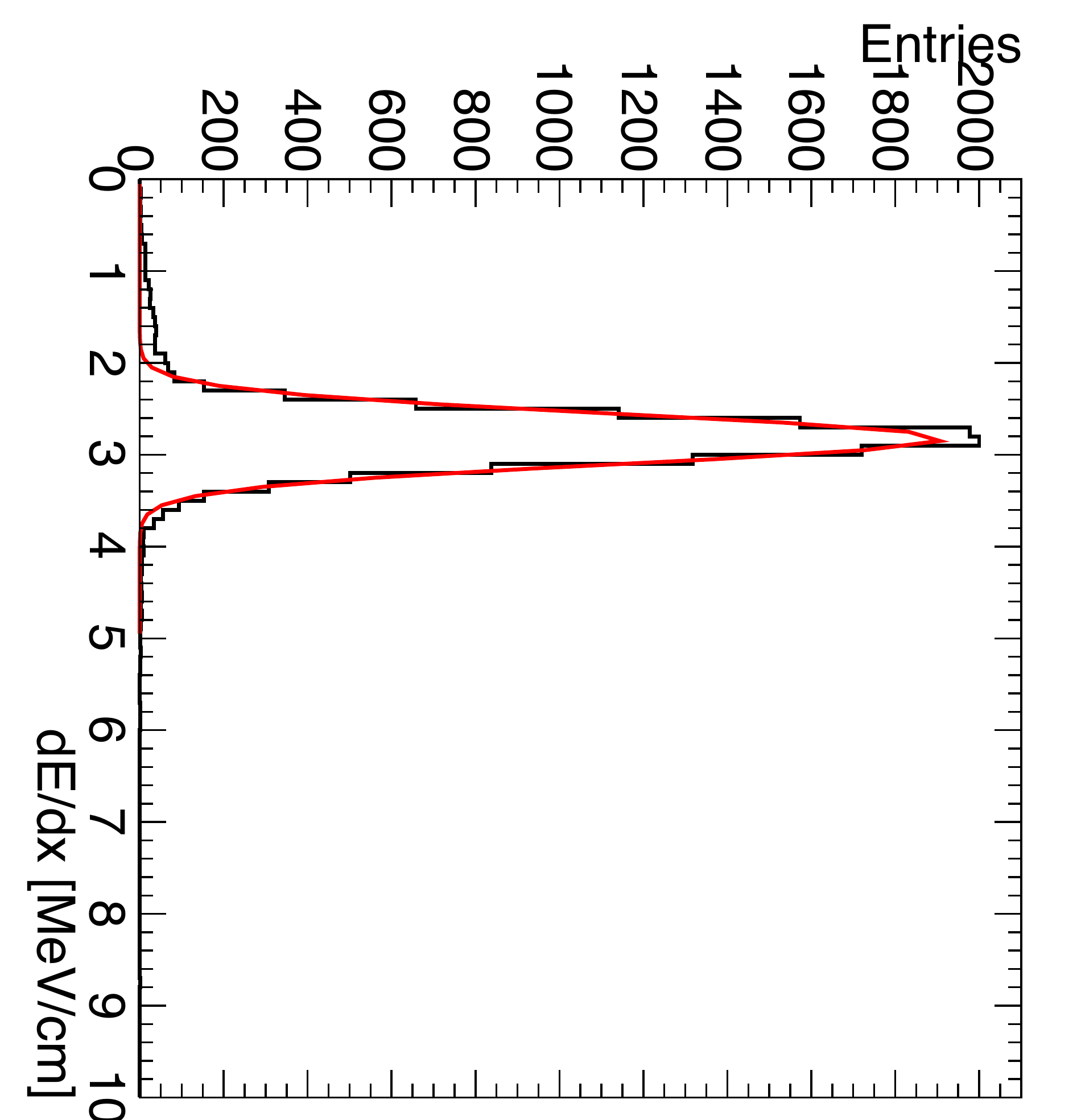}
		\caption{Results for different track $dE/dx$ estimators for TIF data taken at $-10\degC$: truncated mean with a truncation fraction of 20\% (left);  generalized mean with exponent $k=-2$ (right).}
		\label{fig:dedxestimator}
	\end{center}
\end{figure}

Results for the two types of $dE/dx$ estimators are shown in Figure~\ref{fig:dedxestimator}. Mean values and resolutions have been extracted from the measured $dE/dx$ distributions by fits to a Gaussian distribution, and the good agreement indicates that the fake rate from misidentified tracks or hits is low. The calculated $dE/dx$ resolutions are shown in Table~\ref{tab:Data_dEdx_Resolution}. It follows that the generalized mean estimator with $k=-4$ gives the best $dE/dx$ resolution of all four estimators, and that a truncation fraction of 40\% results in a better resolution for the truncated mean than a truncation fraction of 20\%. All these results are consistent with the study on simulated events in Ref.~\cite{EnergyLossNote}. 

\begin{table}[htb]
  \caption{\label{tab:Data_dEdx_Resolution}$dE/dx$ resolutions for different estimators in TIF data taken at $-10\degC$.}
  \begin{center}
	\begin{tabular}{|c|c|}\hline
		$dE/dx$ Estimator & Resolution ([\%])\\
		\hline
		Truncated mean (20\%) &  9.78$\pm$0.07\\
		Truncated mean (40\%) & 9.18$\pm$0.07 \\
		Generalized mean ($k=-2$)&  9.51$\pm$0.07\\
		Generalized mean ($k=-4)$) &  9.01$\pm$0.07\\
		\hline
	\end{tabular}
\end{center}
\end{table}

\begin{figure}[p]
	\begin{center}
		(a)
		\includegraphics[scale=0.30,angle=90]{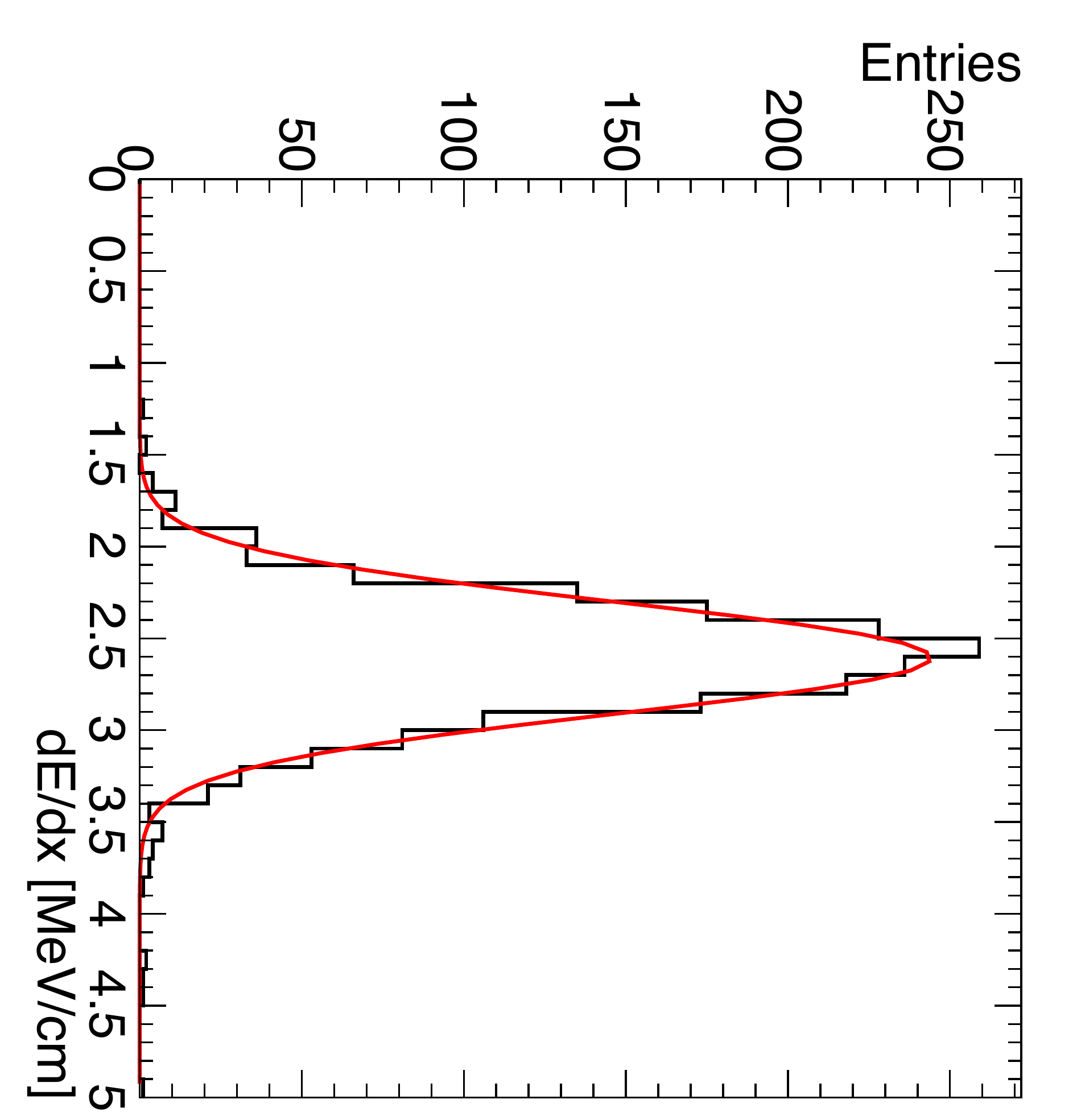}
		(b)
		\includegraphics[scale=0.30,angle=90]{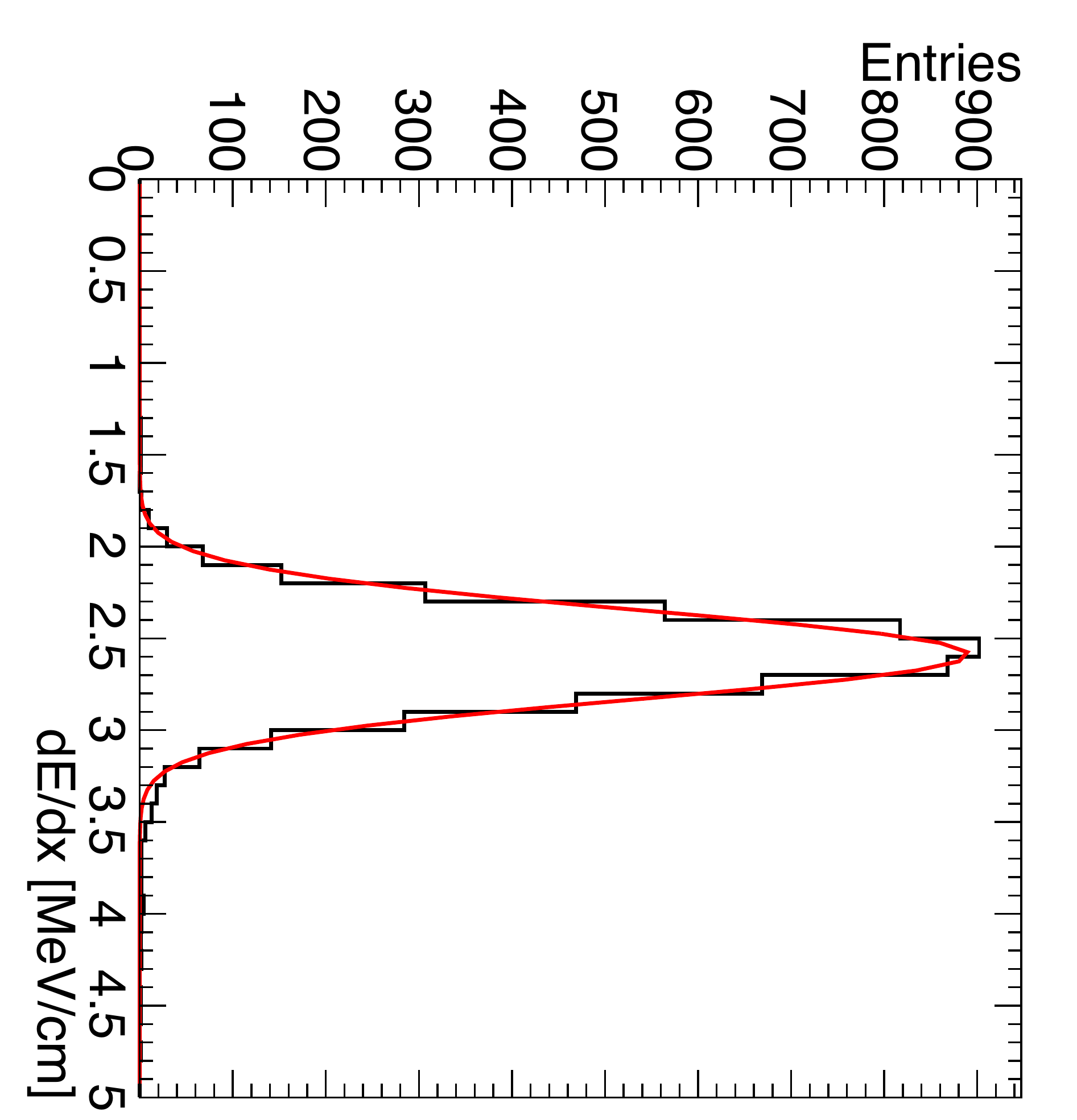} \\
		(c)
		\includegraphics[scale=0.30,angle=90]{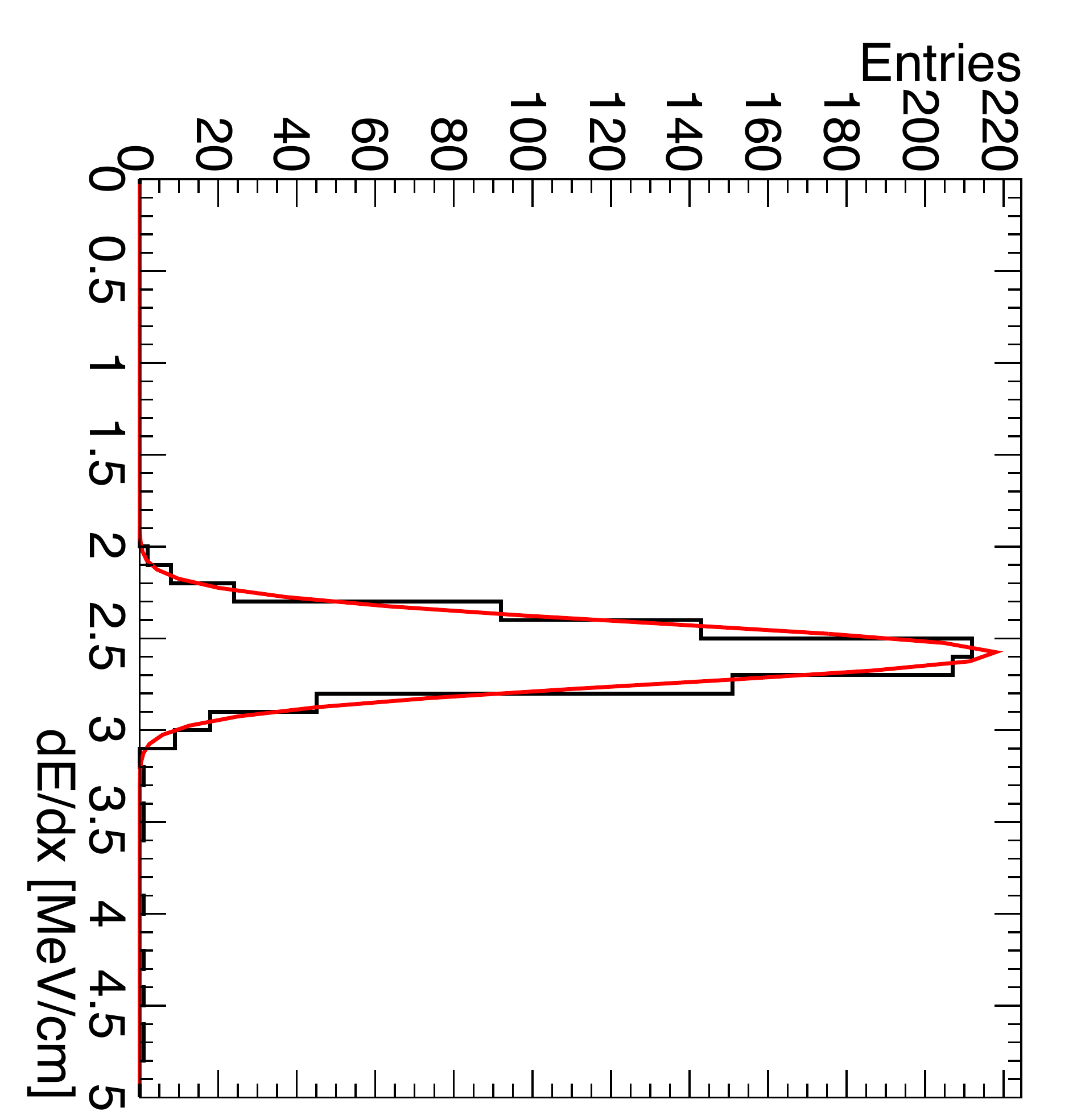}
		\caption{Energy loss estimation as a function of the number of hits per track in TIF data taken at $-10\degC$. The histograms show the truncated mean with a truncation fraction of 40\%. Tracks are required to have 5 or 6 hits (a), between 8 and 10 (b), or at least 15 hits (c).}
		\label{fig:dEdx_vs_dEdxHits}
	\end{center}
\end{figure}

It is expected that the $dE/dx$ resolution improves with an increasing number of track hits. This has been verified in data for both the generalized mean and the truncated mean by measuring $dE/dx$ in samples with a different requirement on the number of hits per track. Three samples were studied: (i) 5$\leq$~\#hits~$\leq$~6, (ii) 8$\leq$~\#hits~$\leq$~10, (iii) \#hits~$\geq$15. Figure~\ref{fig:dEdx_vs_dEdxHits} shows the $dE/dx$ for the truncated mean in all three samples. The resolutions are summarized in Table~\ref{tab:dEdx_vs_dEdxHits}. The $dE/dx$ resolution improves as expected with increasing number of track hits. The $dE/dx$ resolution obtained in the sample requiring at least 15~hits is consistent with the value obtained using simulated muon events in Ref.~\cite{EnergyLossNote}.

\begin{table}[p]
	\caption{\label{tab:dEdx_vs_dEdxHits} $dE/dx$ resolution as a function of the number of hits per track in TIF data taken at $-10\degC$.}
	\begin{center}
   \begin{tabular}{|c|c|c|}
	\hline
	Number of Hits & \multicolumn{2}{|c|}{Resolution [\%]} \\
	               & Truncated Mean & Generalized Mean \\
	\hline
	5 or 6  & 11.58$\pm$0.21 & 11.20$\pm$0.22  \\
	8 to 10  & 9.29$\pm$0.10 & 9.04$\pm$0.10 \\
	$\geq 15$ & 6.35$\pm$0.17 & 6.25$\pm$0.17 \\
	  \hline
    \end{tabular}
  \end{center}
\end{table}

Measurements at different temperatures are reported in Table~\ref{tab:dEdx_vs_temp}.
The mean $dE/dx$ shows a slight increase with increasing operating temperature which is attributed to a residual miscalibration due to changes in the operating point of the readout electronics.

\begin{table}[p]
  \caption{\label{tab:dEdx_vs_temp} $dE/dx$ measurements and resolutions for data taken at $+10\degC$, $-1\degC$ and $-10\degC$ in trigger configuration~C. The results are based on the the truncated mean method with a truncation fraction of 40\%.}
  \begin{center}
   \begin{tabular}{|c|c|c|}
     \hline
	Temperature [\degC] & $dE/dx$ [MeV/cm] & Resolution [\%]\tabularnewline
	\hline
	$-10$  & 2.596$\pm$0.002 & 9.18$\pm$0.07 \\
	$-1$  & 2.643$\pm$0.002 & 9.45$\pm$0.06 \\
	$+10$ & 2.684$\pm$0.002 & 9.35$\pm$0.06 \\
	  \hline
    \end{tabular}
  \end{center}
\end{table}

\section{\label{sec:TrkValidation}Validation of Track Reconstruction}

\subsection{Reconstructed Track Position at Scintillation Counters}

\begin{figure}[htb]
	\begin{center}
		(a)
		\includegraphics[scale=0.30,angle=90]{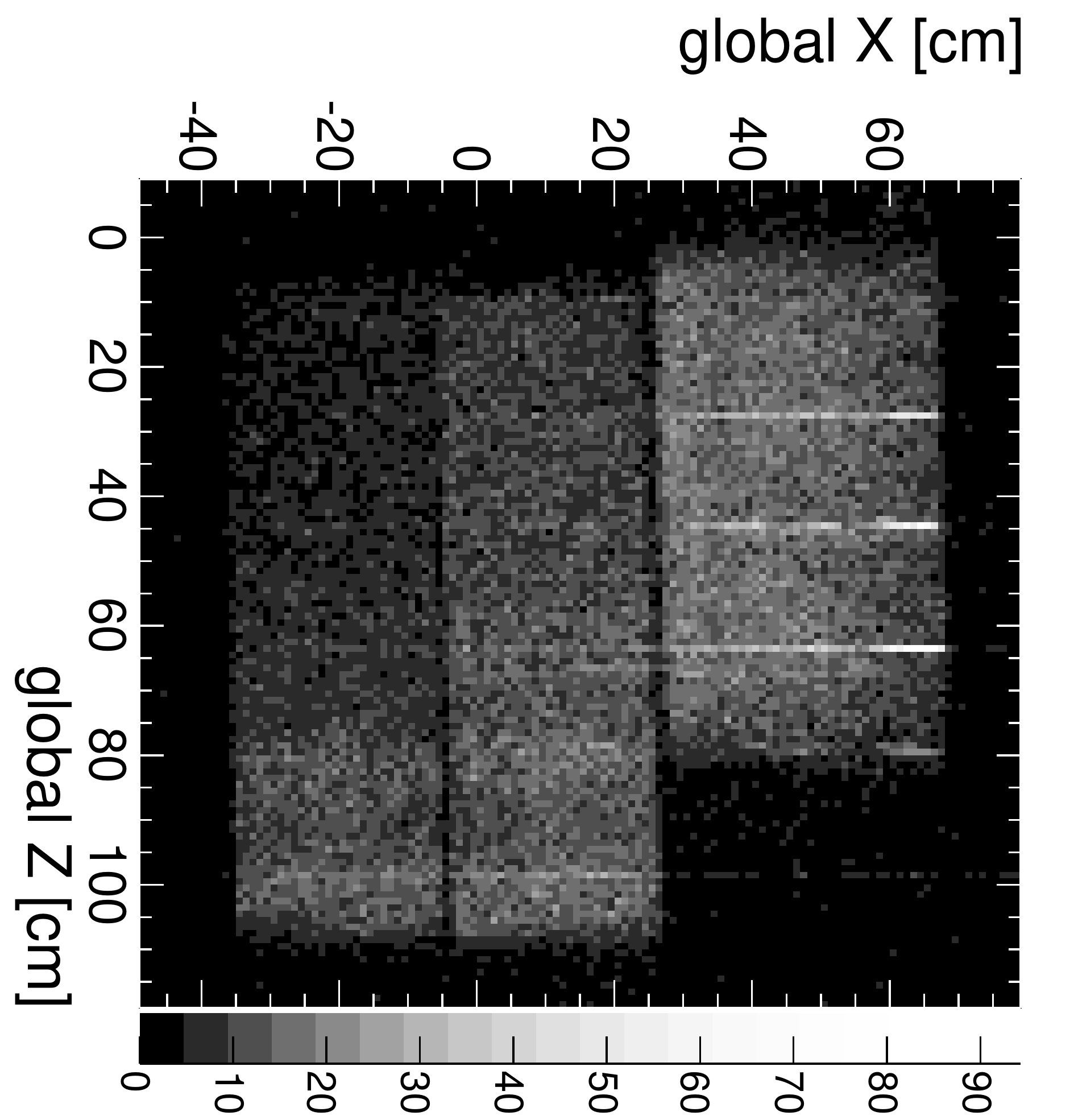}
		\hspace{1cm}
		(b)
		\includegraphics[scale=0.30,angle=90]{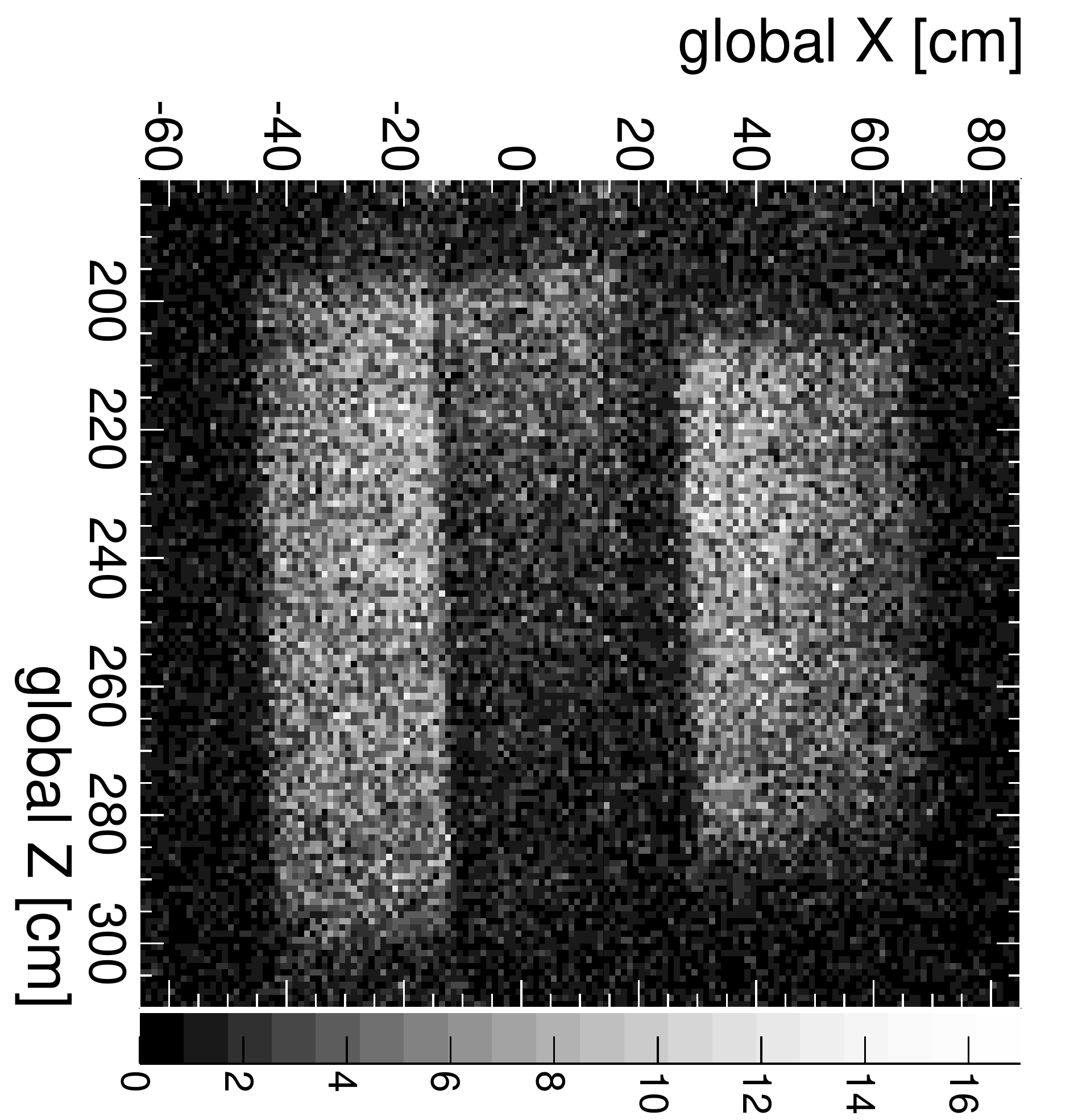}
		\caption{Reconstructed track position at the surface defined by the upper scintillation counters in position C: (a) scintillation counters for the barrel readout; (b) scintillation counters for the end-cap readout.}
		\label{fig:upper_scintillator_position}
	\end{center}
\end{figure}

For the visual validation of the tracking performance, a set of pseudo hit-maps  for the trigger scintillation counters has been compiled. Two reference planes, coplanar to the global $xz$-plane, at $y\!=\!160$\,cm and $y\!=\!-160$\,cm respectively, were used. For every reconstructed track, the trajectory state at the outermost hit was propagated to the upper reference plane using a straight-line track-model. Similarly, the trajectory states at the innermost hits were propagated the same way to the lower reference plane. 
Figure~\ref{fig:upper_scintillator_position} shows the resulting intersection points at the upper plane.

The shapes of the six scintillation counters can be clearly distinguished. 
Additionally their roughly determined positions and dimensions are in good agreement with the experimental setup. Two of the counters show a drastically decreasing light-yield for positions further away from the photocathode, as seen in Figure~\ref{fig:upper_scintillator_position} (a) and (b). This is most likely due to damages of the devices themselves, rather than from any tracking inefficiencies. The vertical lines at constant $z$ positions stem from barrel tracks that are built only out of $r\phi$ hits. Due to missing $z$ information from stereo modules these tracks are accumulated at specific positions.

\subsection{Comparison of Virgin Raw and Zero Suppressed Data}
The vast majority of the TIF data have been taken in virgin raw mode, using specific code to do the zero suppression offline. However, over 100k events have been collected in zero suppression mode, and this data have been used to compare the tracking results from both readout modes. A detailed comparison of virgin raw and zero suppressed data on the cluster reconstruction level can be found in Ref.~\cite{TIFPerformanceNote}.

\begin{figure}[p]
	\begin{center}
		\includegraphics[scale=0.35,angle=90]{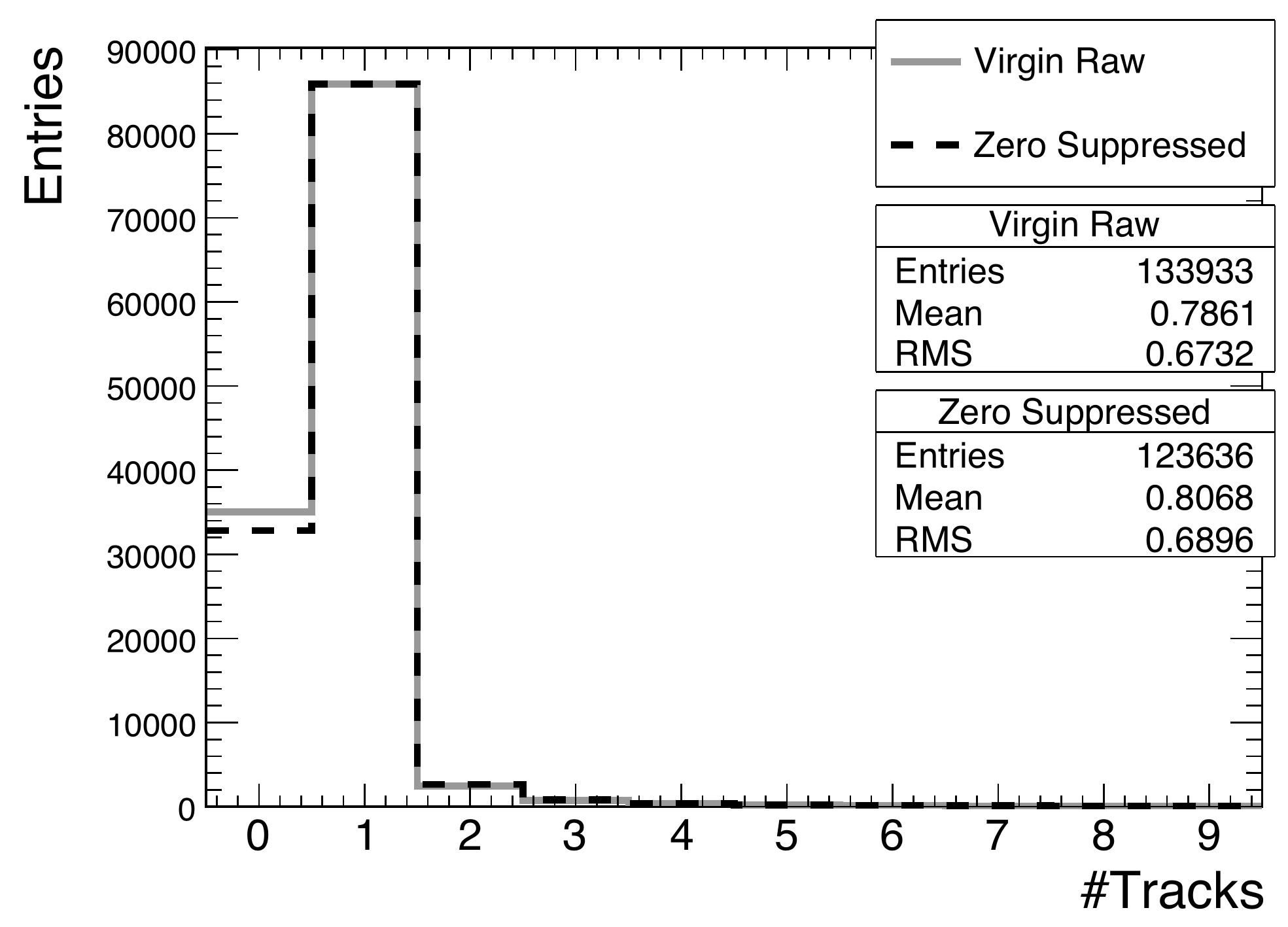}
		\hspace{1cm}
		\includegraphics[scale=0.35,angle=90]{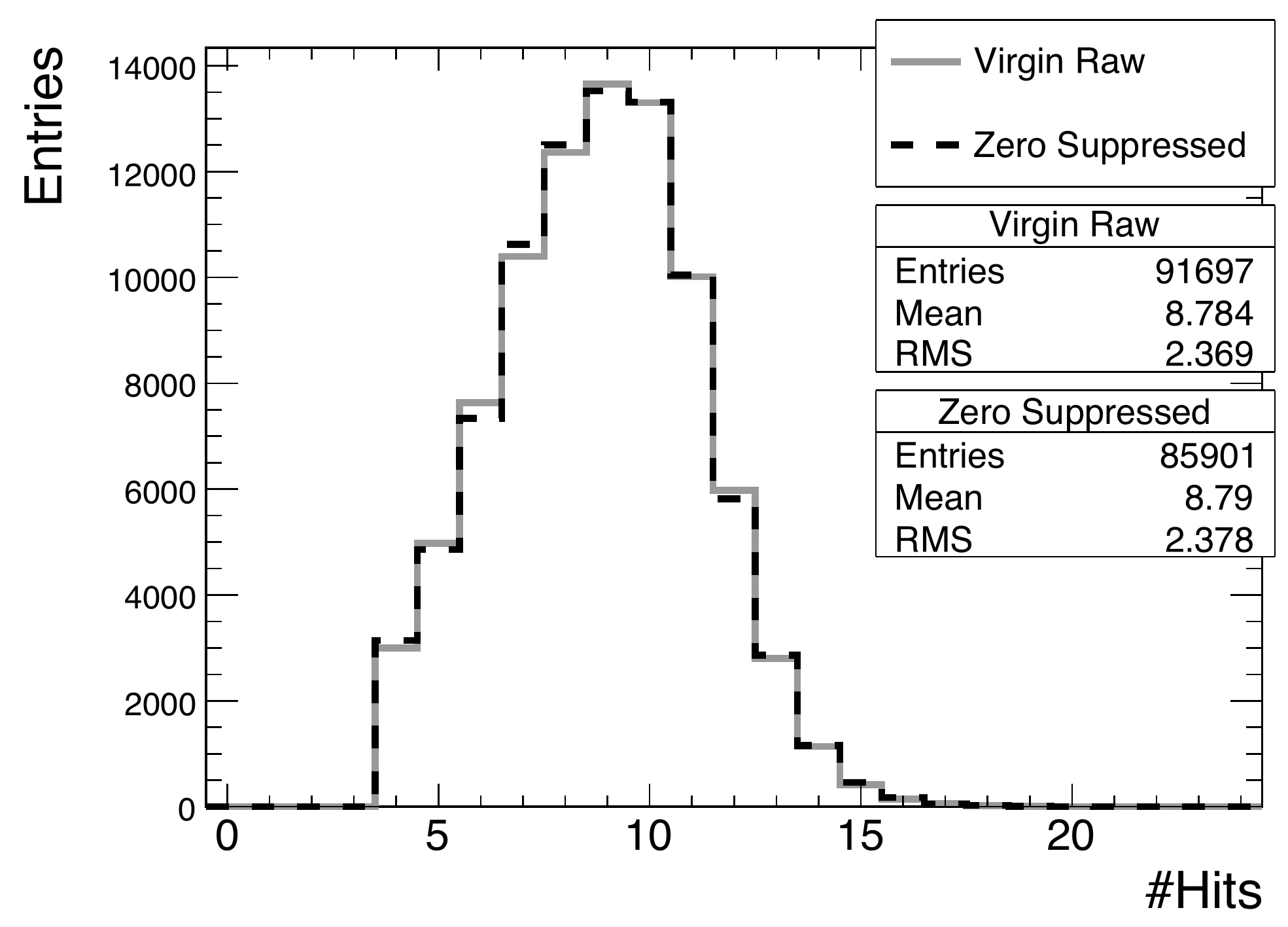} \\
		\vspace{1cm}
		\includegraphics[scale=0.35,angle=90]{Figures/eta_RS}
		\hspace{1cm}
		\includegraphics[scale=0.35,angle=90]{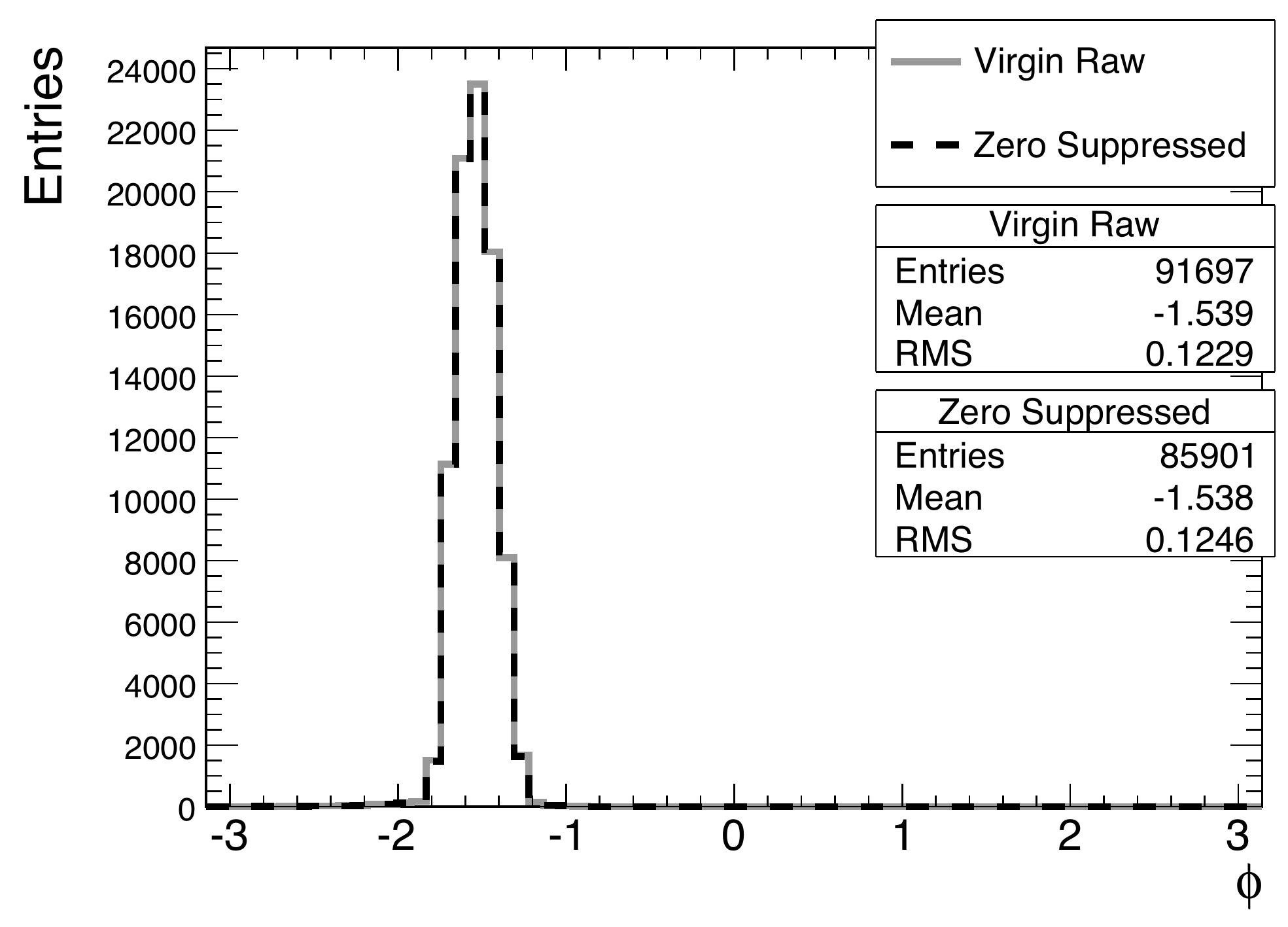}\\ 
		\vspace{1cm}
		\includegraphics[scale=0.35,angle=90]{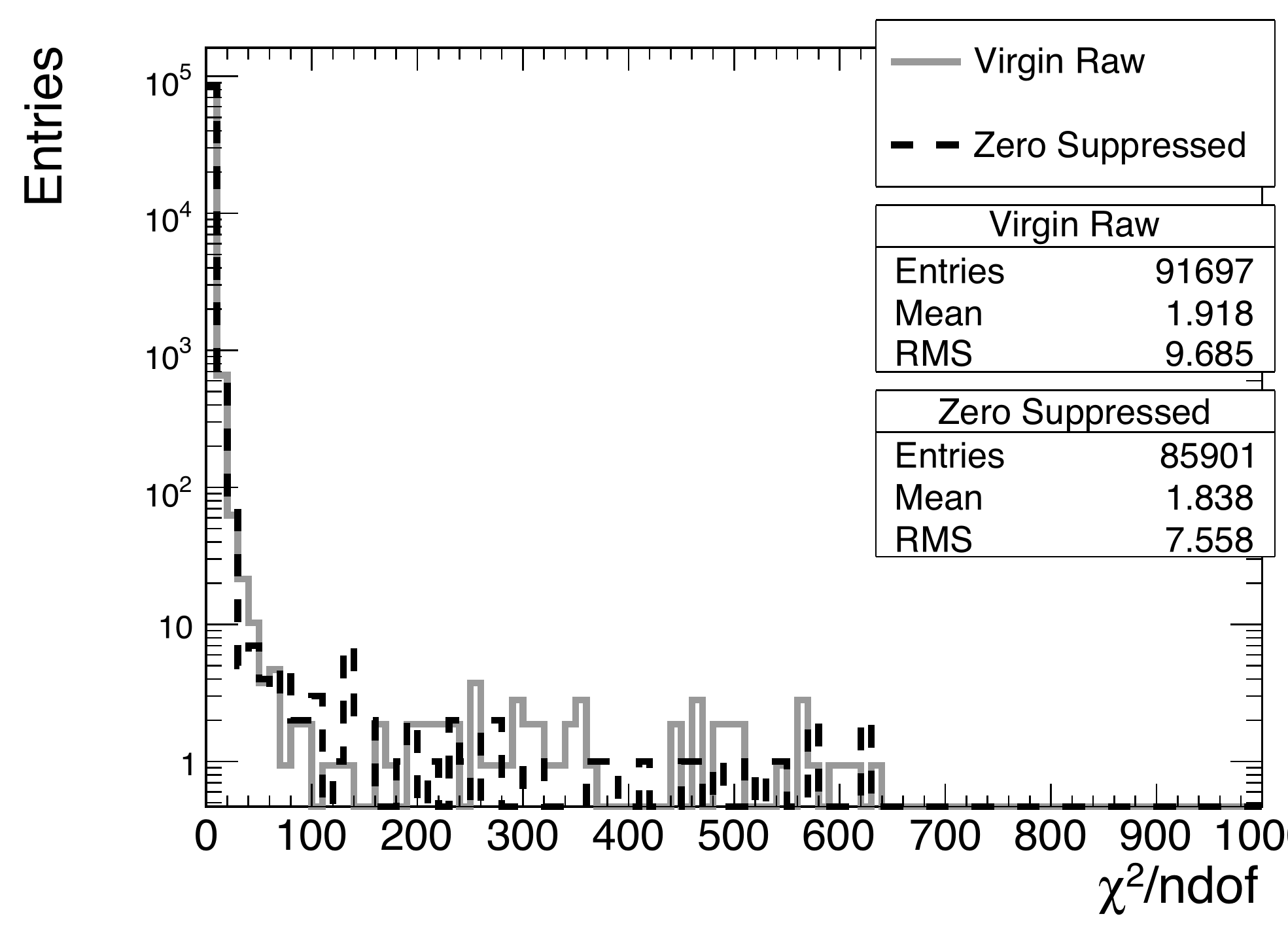}
		\hspace{1cm}
		\includegraphics[scale=0.35,angle=90]{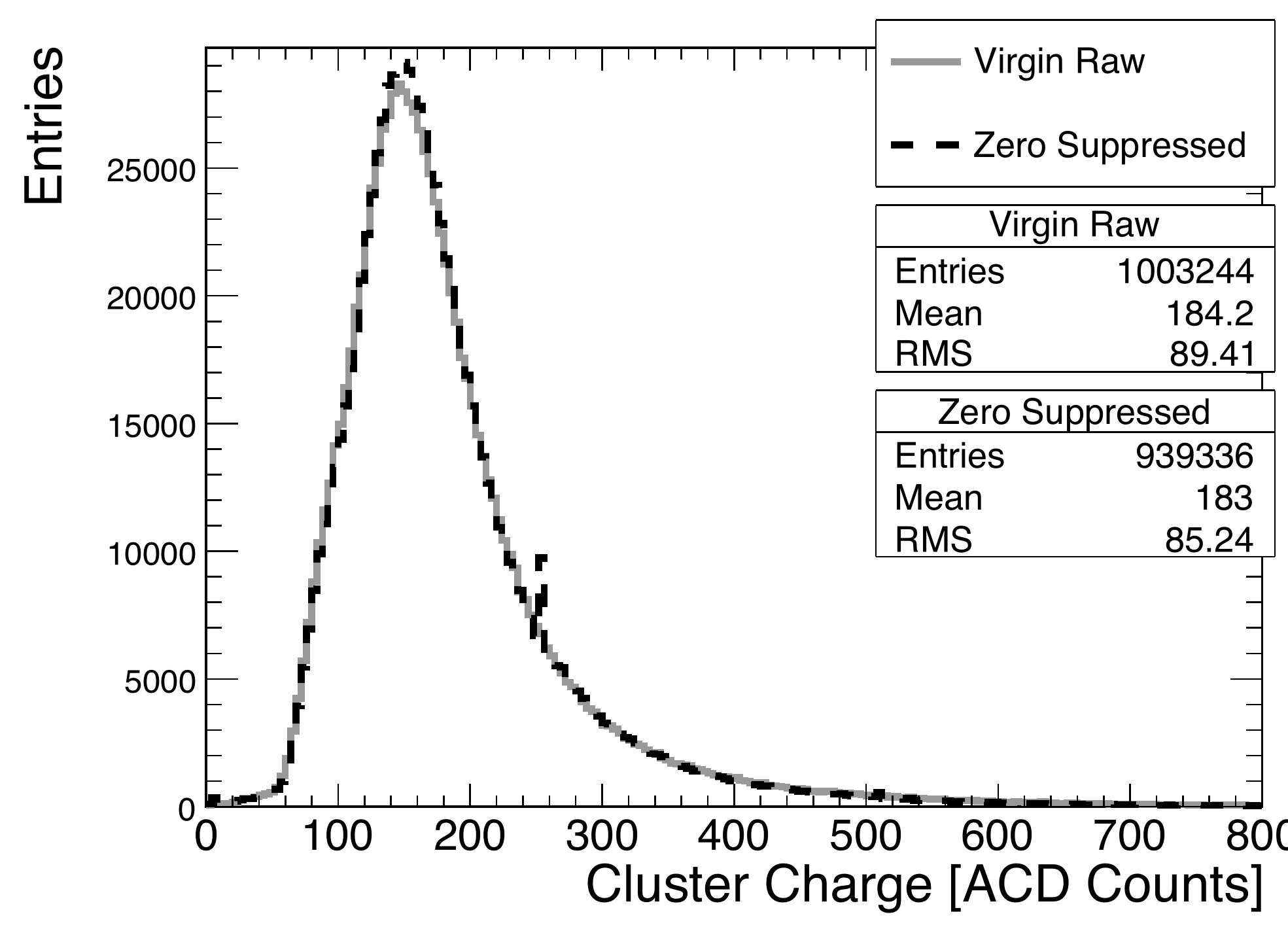}\\ 
		\caption{Comparison of tracking results obtained using data taken in zero suppressed mode and in virgin raw mode: number of tracks (top left), number of hits per track (top right), $\eta$ (middle left), $\phi$ (middle right), $\rm \chi^2/ndof$ (bottom left) and cluster charge (bottom right). As an example the distributions from the Road Search algorithm are shown -- the other algorithms show the same characteristics. The spike in the cluster charge distribution for zero suppressed mode is due to the limited dynamic range for a single channel.}
		\label{fig:vr_zs_comparison}
	\end{center}
\end{figure}

The complete zero suppressed data set has been taken at room temperature with scintillator position A. Only TIB and TOB have been included in the readout. The tracking results in the zero suppressed data set are compared to results that are obtained in a virgin raw data set taken under the same conditions. The results of the comparison are shown in Figure~\ref{fig:vr_zs_comparison} for the Road Search algorithm. Both the Cosmic Track Finder and the CKF give similar results and are not shown here. The two data sets were normalized by using the single track events. A global normalization factor has been derived by comparing the number of single track events in both samples. The normalization factor has been applied to all the distributions. The track parameters are only shown for single track events. A rather good agreement between the tracking results from zero suppressed and virgin raw data is observed. There are small discrepancies which can be attributed to the reduced resolution in the digitization in zero suppressed mode. This reduction in the dynamic range per strip generates a spike at 255 ADC counts for high-charge single strip clusters, which is observable in the cluster charge distribution.

\subsection{Regional Unpacking}

In order to speed up reconstruction for the High-Level Trigger \cite{DAQTDR}, a new software scheme has been developed. 
The new implementation of the cluster reconstruction improves the performance significantly by combining the unpacking of the raw data and the cluster finding into a single step.
Both unpacking and cluster reconstruction can now be done at the request of the pattern recognition.
Hence, only a fraction of the tracker has to be unpacked, in contrast to the currently deployed scheme where the full tracker data has to be converted for local reconstruction.

The new scheme was tested on data from one run with unpacking and cluster reconstruction triggered by the pattern recognition ({\it On Demand Mode}) and with unpacking and cluster reconstruction of all tracker data ({\it Global Mode}). For both tests the improved cluster reconstruction code has been deployed. The CKF algorithm has been chosen for track reconstruction. The resulting tracks were compared with tracks produced using the standard reconstruction and found to be identical. 

Table~\ref{tab:regionalUnpacking} shows the processing time for the different track reconstruction steps in several configurations. All three configurations use the standard seeding configuration (see Table~\ref{tab:CKFseeds}), and the full tracker data is unpacked. In the first case the standard cluster reconstruction algorithm has been used, resulting in a cluster container which is used for the subsequent seed finding. In the second case the improved cluster reconstruction algorithm is deployed, but a standard cluster container is built so that the standard track reconstruction configuration can be used. In the third case the unpacking and clusterization of the seeding layers is done in the seeding step, while the remainder is done during the pattern recognition. This avoids the use of the standard cluster container. Comparison of the time to produce clusters in the first two configurations shows a large improvement due to the improved cluster reconstruction algorithm (by a factor of 7.7). Comparing the overall track reconstruction time for the first two configurations to the third shows a further improvement.

\begin{table}[htb]
	\caption{\label{tab:regionalUnpacking} Processing time for the different steps involved in local and track reconstruction. The standard CKF seeding is used.} 
	\begin{center}
		\small
		\begin{tabular}{|c|c|c|c|}
			\hline
			& \multicolumn{3}{|c|}{Time [arbitrary units]} \\
			& Standard & New Cluster Reconstruction  & New Cluster Reconstruction \\ 
			&                & (using standard cluster container) & \\ \hline
			Unpacking \& Cluster Reconstruction & 209.7$\pm$1.7 & 27.1$\pm$0.009 & $-$ \\ 
			Seed Finding & 6.5$\pm$1.3 & 5.6$\pm$0.8 & 33.4$\pm$0.2  \\ 
			Pattern Recognition & 13.4$\pm$1.5 & 14.2$\pm$2.0 & 11.6$\pm$1.9 \\ 
			Track Fitting & 1.8$\pm$0.5 & 1.8$\pm$0.5 & 1.8$\pm$0.5 \\ \hline
			Total & 218.5$\pm$1.5 & 42.2$\pm$1.1 & 39.9$\pm$1.1 \\ \hline 
		\end{tabular}
	\end{center}
\end{table}

The standard seeding configuration uses a large fraction of the silicon strip tracker to seed (see Table~\ref{tab:CKFseeds}). Hence, all these layers have to be unpacked for the seed finding. In order to study the performance of the new software scheme in on demand mode, the seeding region was restricted to the outer three layers of TOB. This necessarily leads to a reduction in the number of reconstructed tracks, and thus the timing results can not be compared to those obtained using standard seeding. Table~\ref{tab:regionalUnpacking2} shows the processing time for two configurations that use the modified seeding layout. In the first configuration the unpacking of the three seeding layers is performed in the seed finding step, while the remaining layers are all unpacked in the pattern recognition step. In the second configuration the remainder of the strip tracker is only unpacked when requested by the pattern recognition. Comparison of these two configurations shows a further improvement in the track reconstruction when unneeded regions are not unpacked (by a factor of nearly 3). 

\begin{table}[ht]
	\caption{\label{tab:regionalUnpacking2} Processing time for the different steps involved in local and track reconstruction. The seeding region was restricted to the outer three layers of the TOB in order to compare the on demand unpacking to the global unpacking in the pattern recognition step.} 
	\begin{center}
		\begin{tabular}{|c|c|c|}
			\hline
			& \multicolumn{2}{|c|}{Time [arbitrary units]} \\
			& Global Mode & On Demand Mode \\ \hline
			Seed Finding &11.3$\pm$0.6  & 11.5$\pm$0.6  \\ 
			Pattern Recognition & 30.9$\pm$0.8  & 10.8$\pm$1.0  \\
			Track Fitting & 0.9$\pm$0.2 & 0.9$\pm$0.2 \\ \hline
			Total & 42.7$\pm$1.3 & 21.6$\pm$1.5 \\ \hline
		\end{tabular}
	\end{center}
\end{table}

\begin{figure}[htbp]
  \begin{center}
    \includegraphics[scale=0.5]{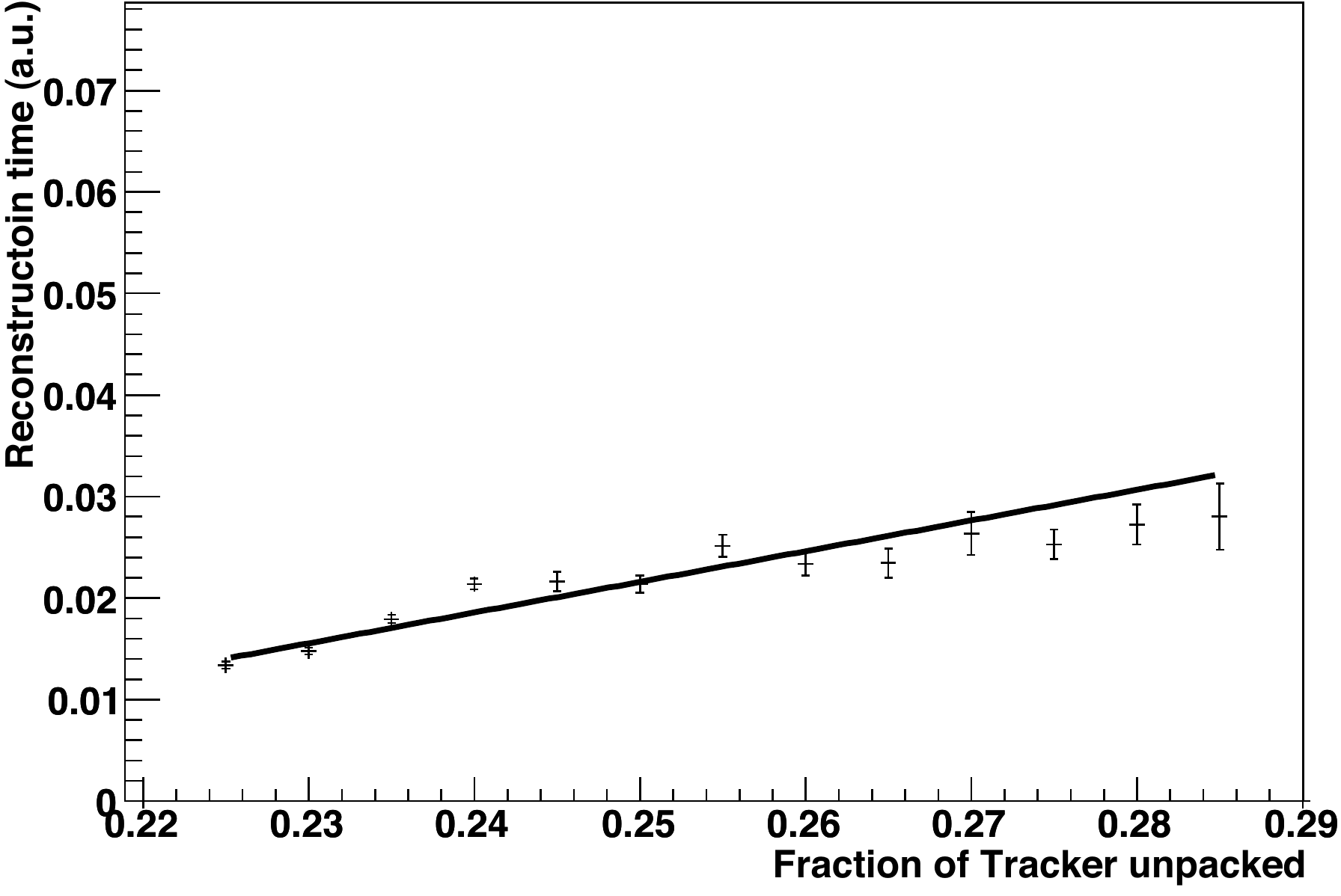}
    \caption{Total reconstruction time (in arbitrary units) from raw data to tracks versus the fraction of the active silicon strip detector being unpacked.}
    \label{fig:regionalUnpacking}
  \end{center}
\end{figure}

In some events large numbers of seeds are found. This leads to a substantial increase in tracking time due to the large number of combinations. In order to study the trend in unpacking time, these events were excluded by requiring events to have less than 3 tracks and less than 100~clusters. Figure~\ref{fig:regionalUnpacking} shows the total time taken to reconstruct tracks from raw data as a function of the fraction of the silicon strip tracker which has to be unpacked. The lower limit is due to the three TOB layers unpacked for seeding, where no seeds have been found. There appears to be a linear relationship between this fraction and the time. 

\section{\label{sec:HitPerformance}Hit Reconstruction Efficiency and Resolution}

\subsection{Hit Finding Efficiency}

The hit reconstruction efficiency for modules was estimated using the capacity of the Kalman fitter to provide optimal predictions of the track parameters on all surfaces crossed by a track. For the pattern recognition part the CKF was used and the analysis was performed for all modules of a given layer at a time. To avoid any bias due to correlations between hit and track finding efficiencies, none of the hits in the layer under consideration were used in seeding or pattern recognition. The CKF would still record intersections with modules on the tested layer as ``lost'' hits. The choice of seeding layers is summarized in Table~\ref{tab:HitEffSeeds}.

\begin{table}[hptb]
	\caption{\label{tab:HitEffSeeds}Seeding layers used for different
	layers under test.}
	\begin{center}
		\begin{tabular}{|c|c|}
			\hline
			Processed Layers & Seeding Layers \\
			\hline
			TIB 2,3,4 & TOB 4,5,6 \\
			TOB 1,2,3 & TIB 1,2,3 or TOB 4,5,6 \\
			TOB 4,5   & TIB 1,2,3 \\
			\hline
		\end{tabular}	\end{center}
\end{table}

A sample of well-reconstructed events was selected by requiring:
\begin{itemize}
  \item Exactly one track reconstructed by the CKF.
  \item One hit in the first TIB layer and one hit in one of the two
    last TOB layers.
  \item At least four reconstructed hits, with at least three matched
    hits from stereo layers.
  \item At most five ``lost'' hits and no more than three
    consecutive ones. 
\end{itemize}

These cuts reject cosmic showers, define the acceptance, ensure a correct measurement of the track in both the $r$-$\phi$ and $r$-$z$ planes and compensate for the removal of the hits in the layer under test.

The efficiency for a module was measured by asking for an intersection with the interpolated track and by checking for the presence of a hit. For increased robustness in the presence of misalignment the distance between the hit and the predicted track position was not used in the selection. An upper cut of $30^\circ$ on the angle of incidence of the track with respect to the normal to the module plane, applied in TIB layer~2, selected topologies similar to the ones expected from proton-proton collisions.

\begin{table}
	\caption{\label{tab:HitEffMargins}Width of the margins used to define
	  the fiducial area in different sub-detectors. The margins were
	  deduced from the size of the sensitive volume as described in the
	  tracker geometry.}
	\begin{center}
		\begin{tabular}{|l|c|c|c|}
			\hline
			& \multicolumn{2}{c|}{Module Edges} & {Inter-sensor Gap} \\ 
			& local x & local y & local y \\ \hline
			TIB & 3 mm & 5 mm & $-$ \\ \hline
			TOB & 5 mm & 10 mm & $\pm$ 10 mm \\ \hline
		\end{tabular}
	\end{center}
\end{table}

\begin{figure}[htb]
  \begin{center}
      \includegraphics[scale=0.27]{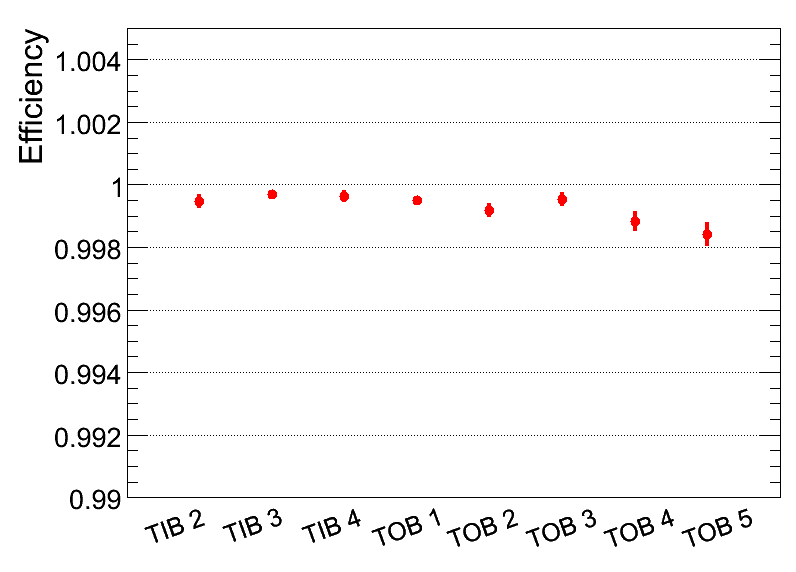} 
    \caption{Summary of the layer efficiency at room temperature.}
    \label{fig:HitEffData}
  \end{center}
\end{figure}

In order to avoid artificial inefficiencies at the edge of the sensitive region and in the central gap of double-sensor modules a fiducial area was defined (Table \ref{tab:HitEffMargins}). Intersections that were either inside the margin or passing closer than five times the predicted error from the module edge or the central gap were excluded from the analysis, as were known inactive modules. After these cuts the number of intersections per layer is of the order of 15000, both at room temperature and at $-10\degC$. 
Results for runs at room temperature are shown in Figure~\ref{fig:HitEffData}. Runs taken at $-10\degC$ show almost identical efficiencies.

The efficiency exceeds 99.8\% for all measured layers. More detailed results are reported in Ref.~\cite{TIFPerformanceNote}. The method has been validated on simulated events and data. In simulation it yields efficiencies compatible with 100\%, as expected.
In data an artificial inefficiency of 5\% was generated by randomly removing hits from the third TIB and the fourth TOB layer. 
The algorithm perfectly reproduces this inefficiency. 

\subsection{Hit Resolution}

\begin{figure}[hptb]
\begin{center}
\includegraphics[width=.6\textwidth]{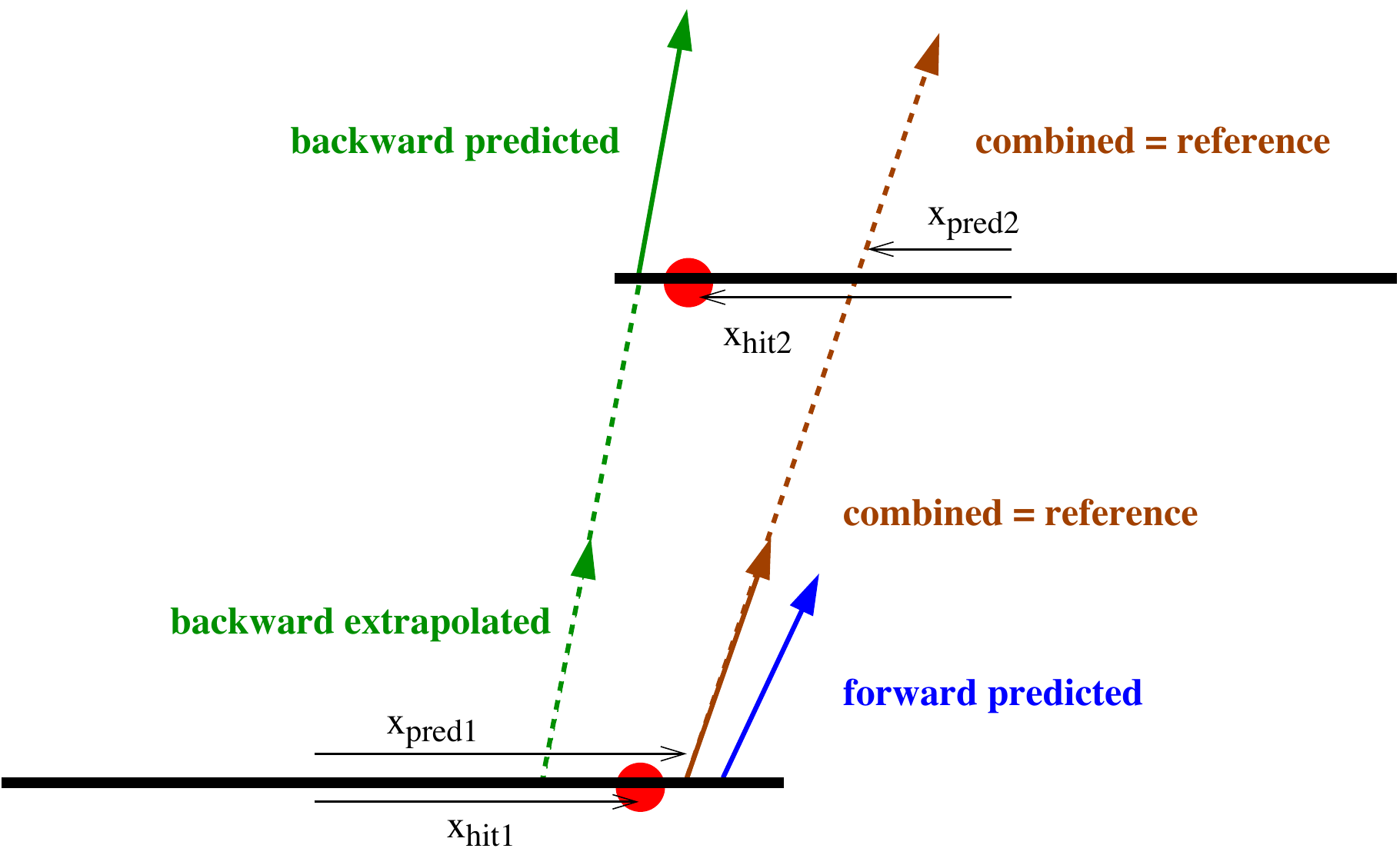}
\caption{Schematic of an overlapping pair of modules where the forward and backward predicted parameters are combined to make the best estimate of the track without the overlap layer.
$x_{hit}$ and $x_{pred}$ are measured in local $x$ coordinates, {\it i.e.}, perpendicular to the strips, with $\Delta x_{hit}\equiv x_{hit1}-x_{hit2}$ and $\Delta x_{pred}\equiv x_{pred1}-x_{pred2}$.
}
\label{fig:doublediff}
\end{center}
\end{figure}

The hit resolution was studied by considering reconstructed tracks without using the detector layer under study in the estimation of the track parameters.
To minimize the effects of any potential misalignment and reduce the effects from the need to extrapolate track position estimates between detector layers events were selected with two consecutive hits in the same layer.
Use of these so-called overlap hits also serves to reduce the potential amount of detector material between hits.
The ``backward (forward) predicted'' track parameters using layers beyond (before) the layer under study were combined to make a prediction of the track position and direction as shown in Figure~\ref{fig:doublediff}.
The difference between the local $x$-positions on the two modules in the overlap layer are compared for the actual detector hits and the prediction.
This procedure minimizes the uncertainty on the predicted position by accounting for correlations between the positions in the two modules which otherwise would dominate the measurement.

The sample of events is selected from all the data collected at scintillator position C
requiring:

\begin{itemize}
  \item Exactly one track reconstructed by the CKF.
  \item At least six reconstructed hits in the barrel region only.
  \item A normalized $\chi^2$ $\leq\ 30$.
\end{itemize}

Additionally, all overlap pairs collected from the data must have an error 
on $\Delta x_{pred} < 50 \mu m$.

In the TIB layers, the mean uncertainty for the predicted position difference for tracks passing through overlapping modules ($<\sigma(\Delta x_{pred})>$) is found to be $\approx~40\mu m$ after cuts, while the mean uncertainty for the track hit position difference ($<\sigma(\Delta x_{hit})>$) is much smaller $\approx~20\mu m$.
In these layers it is thus not possible to extract a meaningful test of the accuracy of the hit position error estimates due to the large uncertainties in the predicted positions.

In the TOB layers, however, the situation is reversed and the \,$<\sigma(\Delta x_{pred})>$ are smaller than the \,$<\sigma(\Delta x_{hit})>$.
The hit position uncertainties are calculated theoretically from the detector geometry.
The accuracy of these hit position uncertainties is compared with the resolution as measured in the data by finding the fitted width of the distribution of the hit position difference between the two modules in the overlap region compared to the predicted position difference for each track.

\begin{figure}[hptb]
\begin{center}
\includegraphics[width=\textwidth]{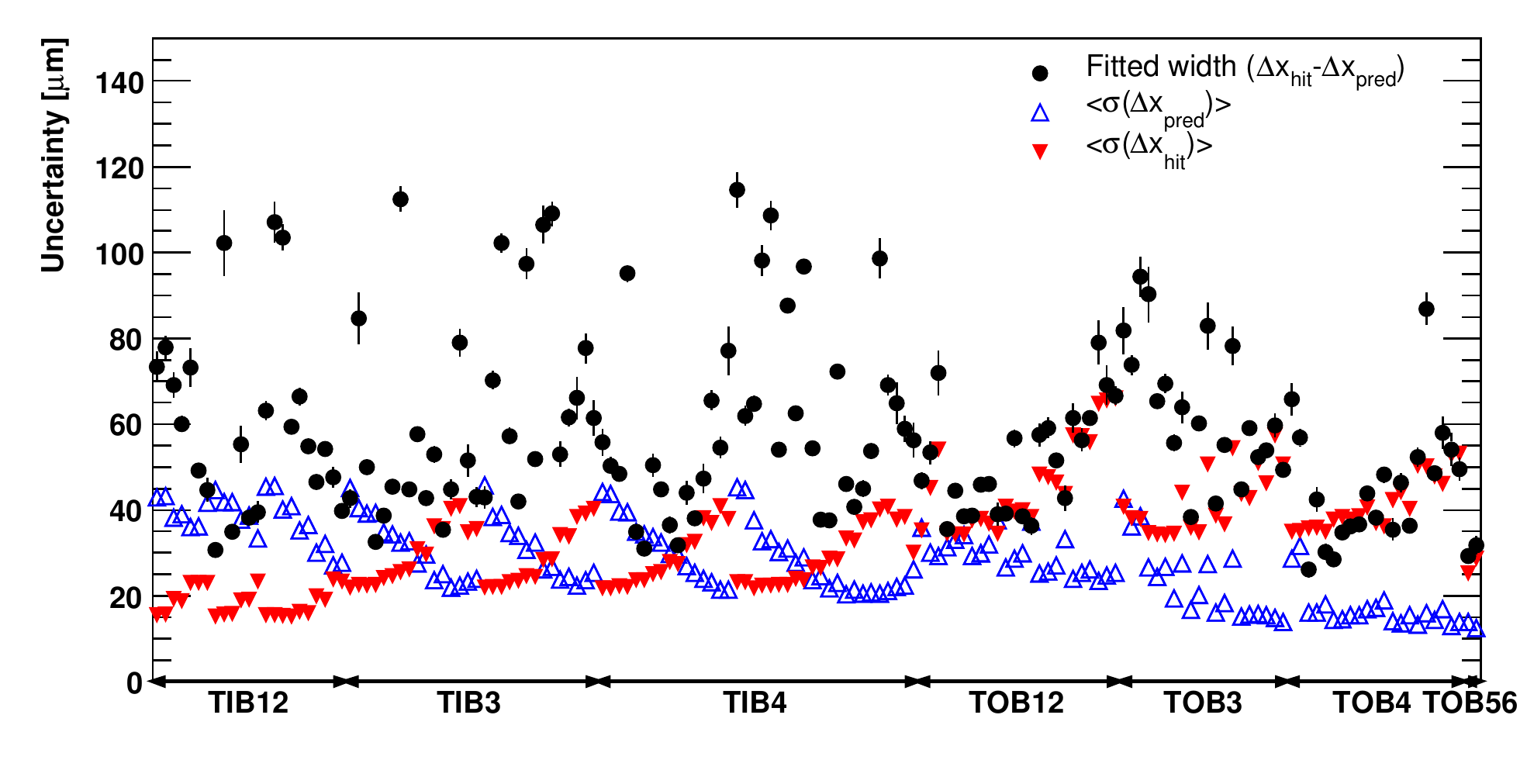}
\caption{Fitted width of the distribution of the difference between the measured and the expected $x$-position differences (circles), the mean uncertainty of the predicted position difference (upward pointing triangles) and the hit position difference (downward pointing triangles) for overlap hits in the data. Each bin represents one module pair. The uncertainties for the measured width are extracted from the fit to a Gaussian distribution. The RMS values for $\sigma(\Delta x_{hit})$ and $\sigma(\Delta x_{pred})$ are of the order of a few $\mu m$.}
\label{fig:HitRes}
\end{center}
\end{figure}

The fitted width of $\Delta x_{hit} - \Delta x_{pred}$ and the values of $<\sigma(\Delta x_{hit})>$ and $<\sigma(\Delta x_{pred})>$ for each overlap position are shown in Figure~\ref{fig:HitRes}.
The width should be a combination of the uncertainty due to the predicted position difference and the hit position difference.
In general the fitted width is close to the quadratic sum of the expected uncertainties on predicted and hit positions.
Outliers are due to residual misalignment between the modules in some of the pairs.

\begin{figure}[htb]
  \begin{center}
    \begin{tabular}{cc}
      \includegraphics[width=.49\textwidth]{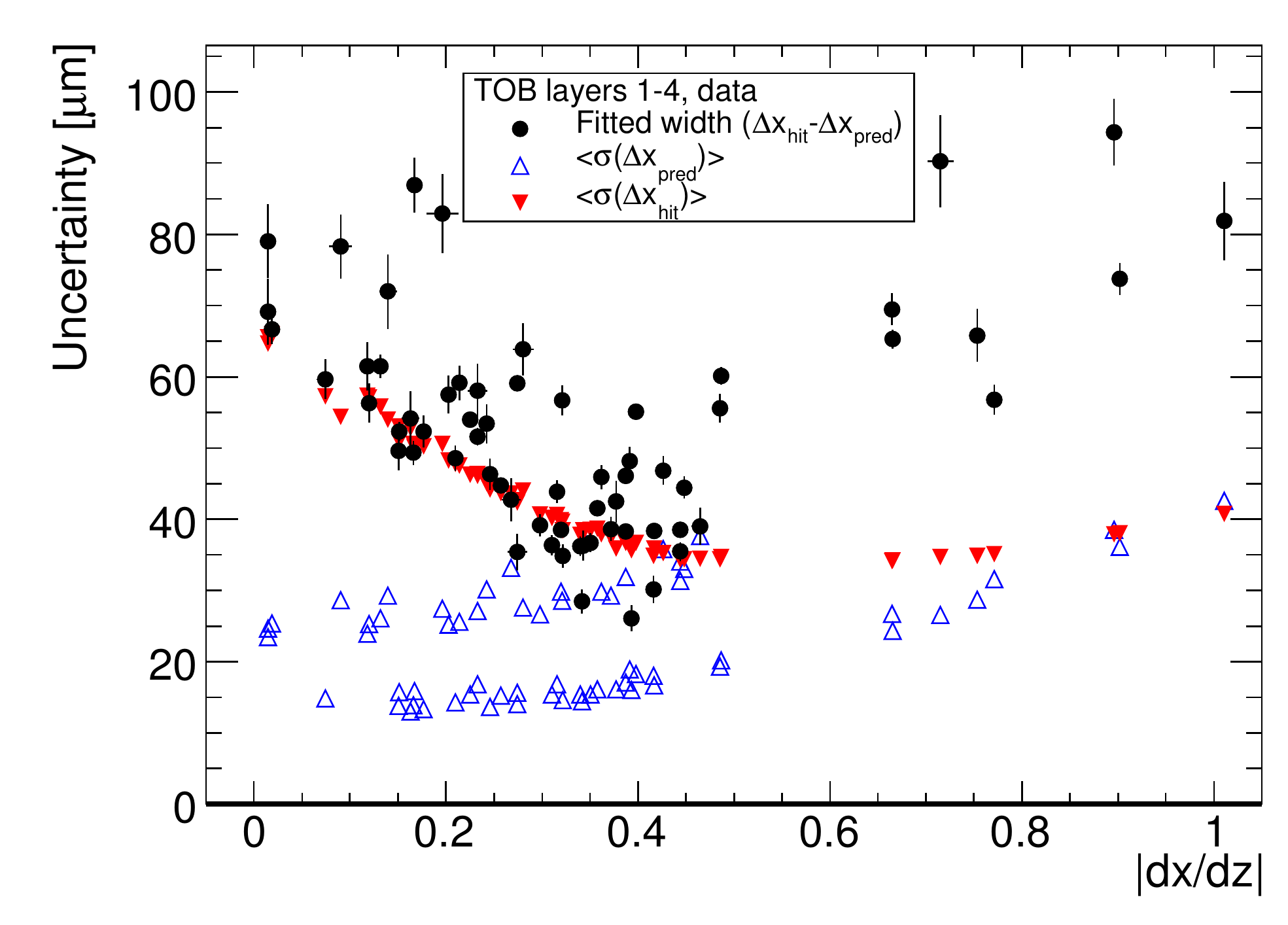} &
      \includegraphics[width=.49\textwidth]{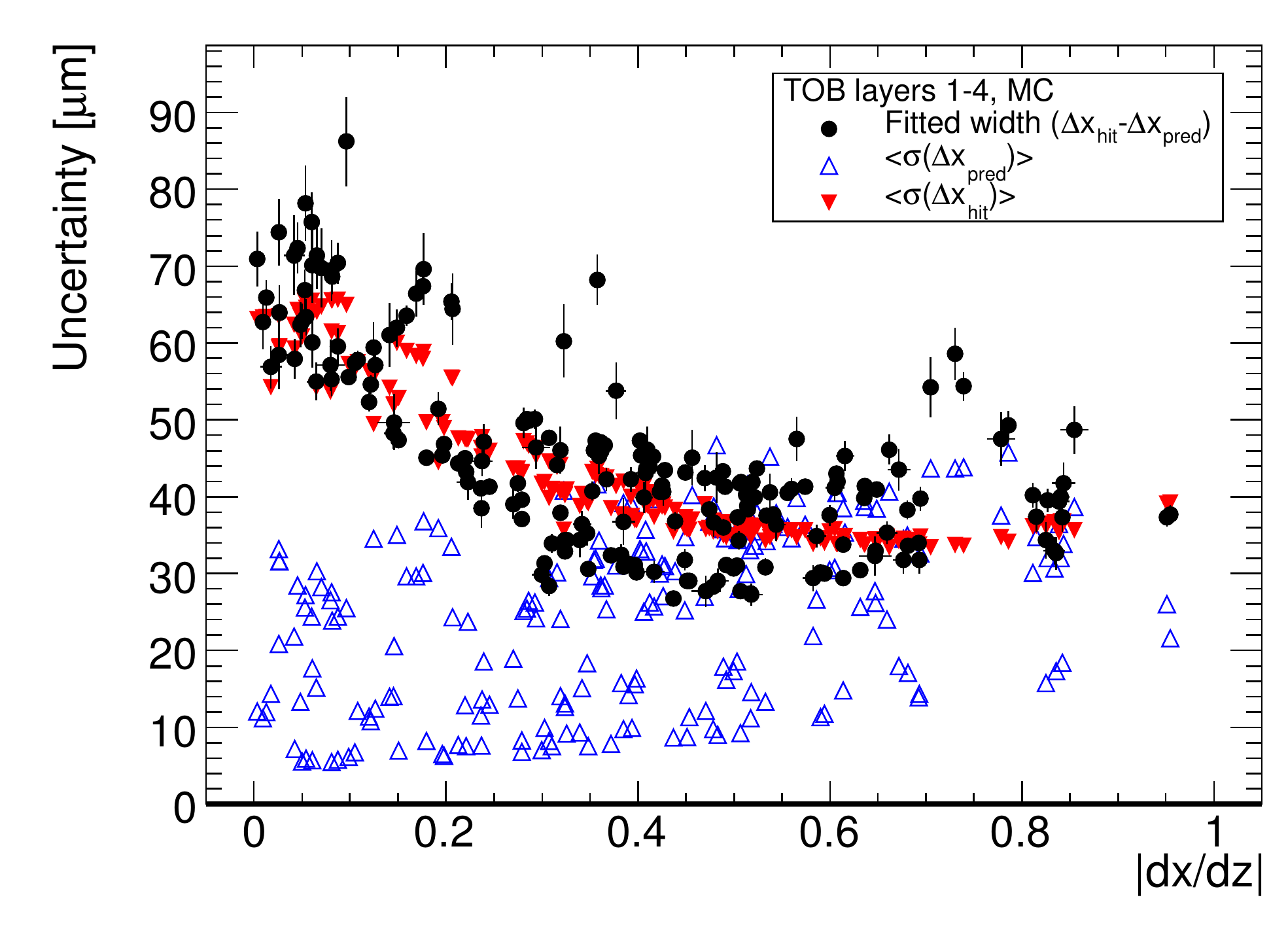}
    \end{tabular}
    \caption{Fitted width of the distribution of $\Delta x_{hit} - \Delta x_{pred}$ and the mean uncertainties for $\Delta x_{pred}$ and $\Delta x_{hit}$ for TOB layers 1 to 4 as a function of local $dx/dz$ for data (left) and simulation (right). The symbols are the same as in Figure~\ref{fig:HitRes}.}
    \label{fig:HitResVsDxdz}
  \end{center}
\end{figure}

In Figure~\ref{fig:HitRes}, within each layer, the overlap pairs are ordered by decreasing track angle with respect to the detector surface normal.
As the tracks approach normal incidence in the plane normal to the strips $<\sigma(\Delta x_{hit})>$ gets larger and $<\sigma(\Delta x_{pred})>$ gets smaller giving better precision on the uncertainty due to hit resolution.
This dependence of the resolution on the local track angle is illustrated in Figure~\ref{fig:HitResVsDxdz} for data and simulated events.
Only the module type with the highest statistics and sensitivity to the hit resolution is shown, corresponding to overlaps in TOB layers 1 to 4.

To quantify the accuracy of the hit position uncertainties, a scale factor, $\alpha$, is used on $\sigma(\Delta x_{hit})$.
In each overlapping pair of modules the variable $\alpha$ is fit according to
\begin{equation}\label{eq:alpha}
w^2=\alpha^2<\sigma(\Delta x_{hit})>^2 + <\sigma(\Delta x_{pred})>^2,
\end{equation}
where $w$ is the fitted width of $\Delta x_{hit}-\Delta x_{pred}$.
With an accurate understanding of the hit position uncertainties, the value of $\alpha$ should be consistent with one.
The best precision for $\alpha$ is reached at normal incidence, while for shallow angles the uncertainties on the prediction dominate over the hit resolution.

\begin{figure}[htb]
  \begin{center}
    \begin{tabular}{cc}
      \includegraphics[width=.49\textwidth]{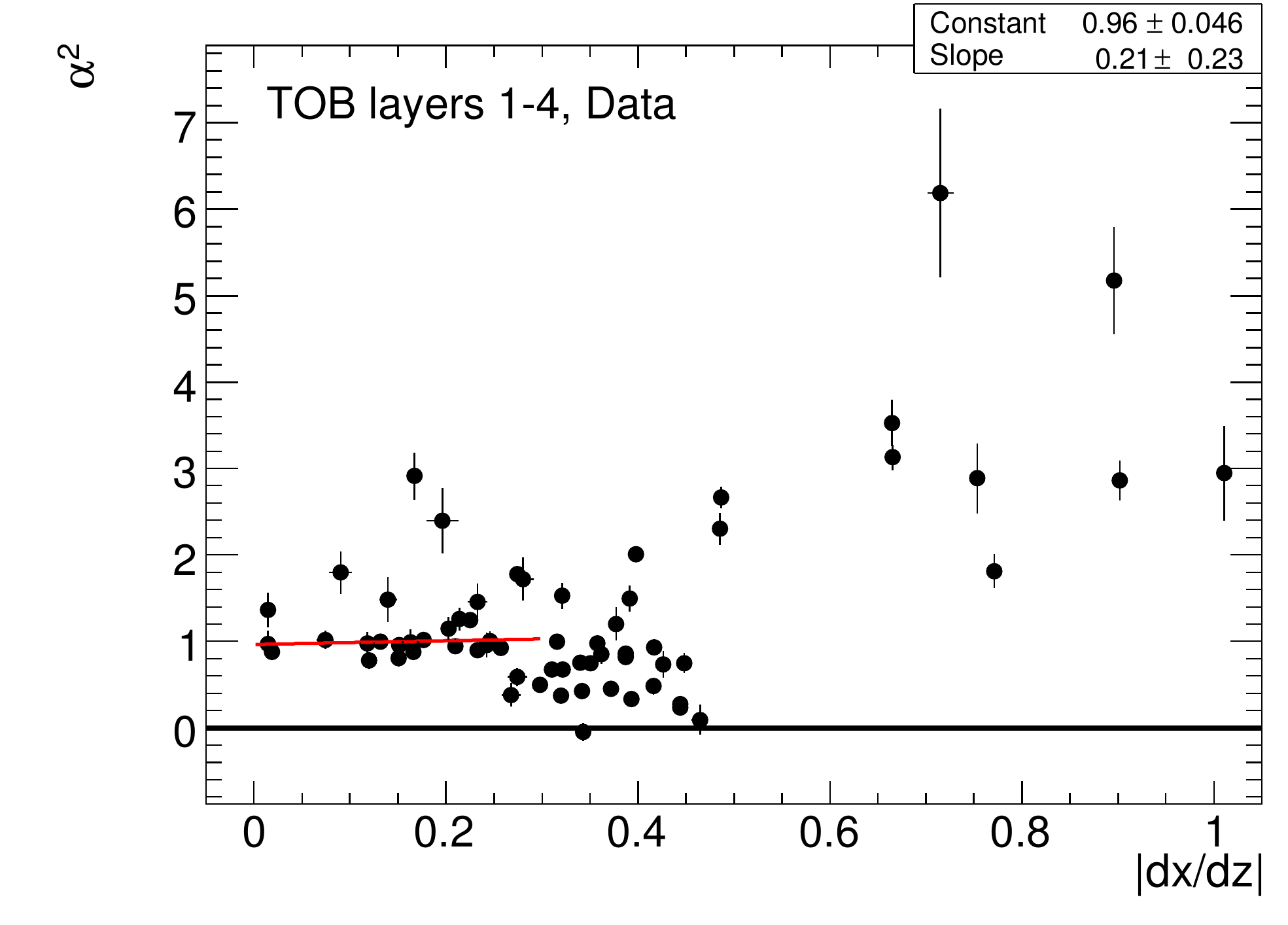} &
      \includegraphics[width=.49\textwidth]{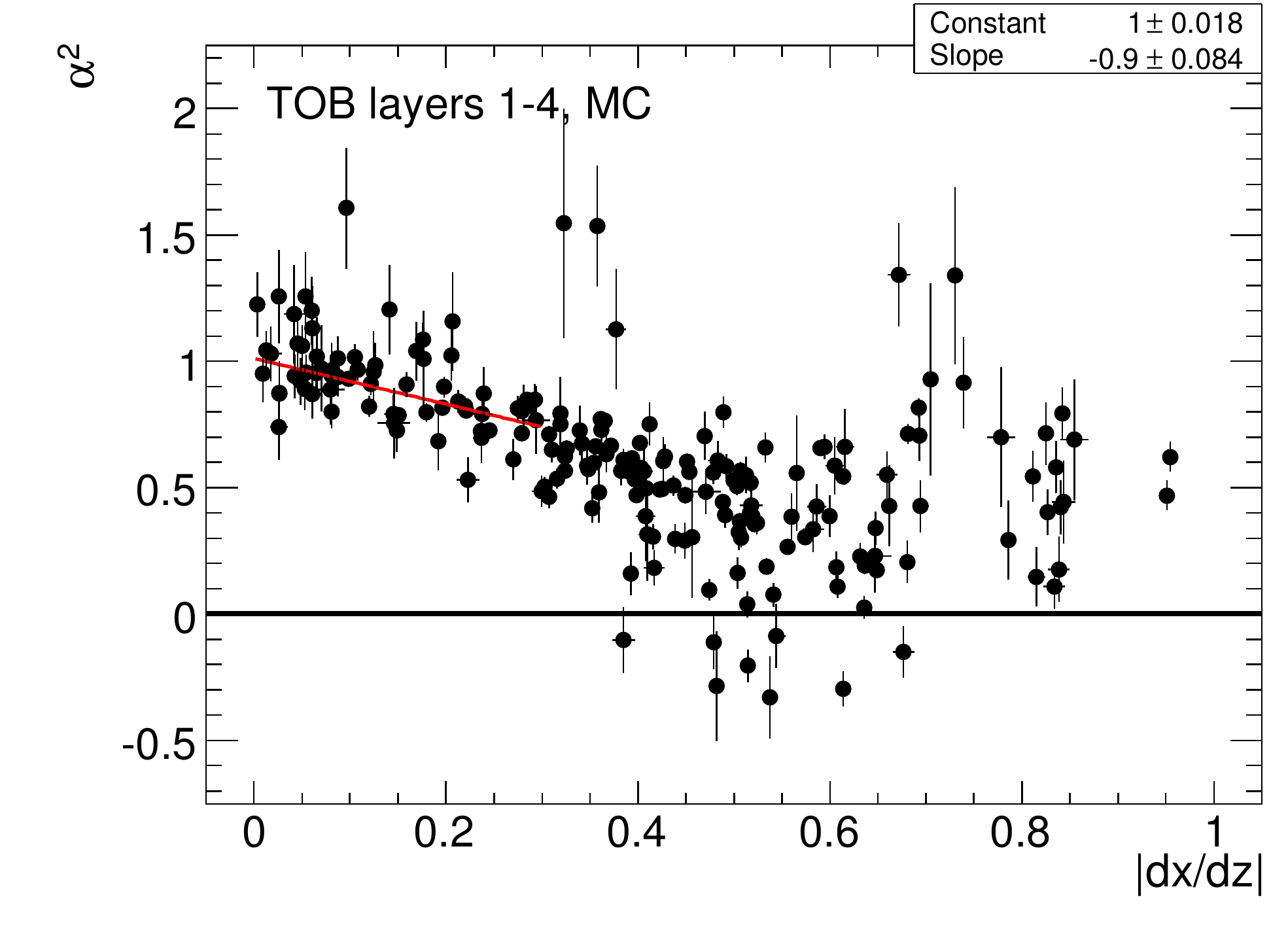}
    \end{tabular}
    \caption{Values of $\alpha^2$ per module pair as a function of local $|dx/dz|$ for TOB layers 1 to 4 in data (left) and simulation (right). A linear fit is used to estimate the value of $\alpha^2$ at $dx/dz=0$.}
    \label{fig:alpha}
  \end{center}
\end{figure}

The values of $\alpha^2$ are found to be dependent on the angle at which the track hits the detector surface as measured by $dx/dz$ in local coordinates.
Figure~\ref{fig:alpha} shows the values of $\alpha^2$ as a function of $|dx/dz|$ for TOB layers 1 to 4 in data and simulation.
An estimate of $\alpha^2$ at $dx/dz = 0$ is extracted using a linear fit at low values of $|dx/dz|$.

The average momentum of cosmic rays is low and $\sigma(\Delta x_{pred})$ is dominated by multiple scattering.
Since the momentum is not measured for individual tracks a systematic term is added.
The width of the momentum distribution in simulation is translated into an uncertainty of $45\%$ on $\sigma(\Delta x_{pred})$.
The calculation of $\alpha^2$ is repeated, varying $\sigma(\Delta x_{pred})$ by $\pm 45\%$.
Half of the difference is attributed as a systematic error, which is the dominant uncertainty.

In conclusion, the hit resolution is well understood in the inner TOB layers with the best precision at small angles.
The resolution of a particular module is dependent on the track angle. 
The uncertainties on the width of the double difference $\Delta x_{hit}-\Delta x_{pred}$ range from $20-60\mu m$.
A scaling factor for the nominal hit resolution was determined as a function of the angle of incidence.
For normal incidence, and averaging over the first four TOB layers, a scaling factor $\alpha = 0.98 \pm 0.04$ was found for a nominal resolution of $\approx 65 \mu m$. The error includes statistical uncertainties and the systematic effects due to the unknown momentum.

\section{\label{sec:Conclusions}Conclusions}

A large sample of over 4.7 million cosmic ray events was recorded at the CMS Tracker Integration Facility (TIF) in the period from November 2006 to July 2007.
These data, taken under various running conditions, were successfully processed by the tracker offline software.
This allowed for a validation of the local (module-level) and global (track) reconstruction algorithms under realistic conditions, using detector condition, alignment and calibration information derived from data.

Compared to collision data, track reconstruction at the TIF profits from the low occupancy in cosmic events.
On the other hand, the task is more complex due to the absence of constraints on the origin of the tracks.
Track momenta are low and cannot be measured directly in the absence of a magnetic field.
Three different algorithms were used to reconstruct tracks: one dedicated cosmic track finder and the two standard modules designed for LHC collision data.
The latter ones were reconfigured in order to optimize for low multiplicity cosmic events.

All three algorithms proved to be stable in different running conditions.
The results allow to assess the tracker performance at the hit and track level.
Reconstructed tracks have been used to verify the full efficiency of hit reconstruction at the module level and to provide a first measurement of the hit resolution using overlapping modules within a layer.
The efficiency exceeds $99.8\%$ for all barrel layers.
At normal incidence the hit resolution in the TOB was found to be compatible with the expected one. 
The track reconstruction efficiency was measured by comparing reconstructed segments in different parts of the detector.
Relative efficiencies were found to be in the range $90$ to $99\%$, depending on the reconstruction algorithm and the subdetector.
The momentum spectrum could be derived from the effects of multiple Coulomb scattering, and the measurement of  the energy loss confirmed the resolution of about $9\%$ expected from Monte Carlo simulation.

The results presented in this note constitute an important step in the commissioning of the tracker offline software and in the preparation for the LHC data taking.
Data taken at the TIF allowed for the first time to confirm simulation results on the performance of the tracker and the reconstruction programs on a larger scale.
The special configuration of the track reconstruction algorithms developed for TIF data will be reused in the next stage of the commissioning process, {\it i.e.}, for cosmic ray data taken in situ with the full CMS detector.
Also the analysis procedures will continue to be used for the measurement and monitoring of the tracker performance.

\section*{Acknowledgments}

We thank the administrative staff at CERN and other Tracker Institutes.
This work was supported by: 
the Austrian Federal Ministry of Science and Research; 
the Belgium Fonds de la Recherche Scientifique and Fonds voorWetenschappelijk Onderzoek; 
the Academy of Finland and Helsinki Institute of Physics; 
the Institut National de Physique Nucl\'{e}aire et de Physique des Particules / CNRS, France; 
the Bundesministerium f\"{u}r Bildung und Forschung, Germany;
the Istituto Nazionale di Fisica Nucleare, Italy; 
the Swiss Funding Agencies; 
the Science and Technology Facilities Council, UK; 
the US Department of Energy, and National Science Foundation. 
Individuals have received support from the Marie-Curie IEF program (European Union) and the A.P. Sloan Foundation.


\begin{thebibliography}{9}

\newcommand{\etal}{{\em et al.}}

\bibitem{MTCCNote} 
  W.~Adam \etal,
  {\em Tracker Operation and Performance at the Magnet Test and Cosmic Challenge}, 
  JINST {\bf 3} (2008) P07006.

\bibitem{TIFPerformanceNote} 
  The CMS Tracker collaboration,
  {\em Silicon Strip Tracker Detector Performance with Cosmic Ray Data 
  at the Tracker Integration Facility}, 
  CMS Note {\bf 2008/032}.

\bibitem{TIFAlignmentNote} 
  The CMS Tracker collaboration,
  {\em CMS Tracker Alignment at Integration Facility}, 
  CMS Note {\bf 2009/002}.

\bibitem{PTDR1} The CMS collaboration,
  {\em  Detector Performance and Software Physics Technical Design
        Report, Volume I},
  {\bf CERN/LHCC 2006-001, CMS TDR 8.1.}; \\
  The CMS collaboration, 
  {\em The CMS experiment at the CERN LHC}, JINST {\bf 3} (2008) S08004.

\bibitem{CMSCGEN}
  P.~Biallass, T.~Hebbeker, K.~H\"opfner,
  {\em Simulation of Cosmic Muons and Comparison with Data from the Cosmic Challenge using Drift Tube Chambers},
  CMS Note {\bf 2007/024}.

\bibitem{L3CGEN} T.~Hebbeker, A.~Korn, 
   {\em Simulation Programs for the L3+Cosmics Experiment}, L3-C note, 1998, \\
   {\tt http://web.physik.rwth-aachen.de/$\sim$hebbeker/l3csim.pdf} .

\bibitem{CAPRICE} M.~Boezio \etal ,
  {\em Energy spectra of atmospheric muons measured with the CAPRICE98 balloon experiment}
  Phys.~Rev.~D~{\bf 67}, 072003 (2003).

\bibitem{KF} R.~Fr\"uhwirth,
  {\em Application Of Kalman Filtering To Track And Vertex Fitting},
  Nucl.~Instrum.~Meth. A {\bf 262} (1987) 444.

\bibitem{EnergyLossNote} A.~Giammanco,
  {\em Particle Identification with Energy Loss in the CMS Silicon Strip Tracker},
  CMS Note {\bf 2008/005}.

\bibitem{DAQTDR} The CMS Collaboration,
{\em The Trigger and Data Acquisition Project, Volume II},
{\bf CERN/LHCC 2002/26}.

\end{thebibliography}
\end{document}